\documentclass[12pt]{article}

\usepackage{eucal}
\usepackage{color}
 \usepackage{ bm}
\usepackage{amsmath,bbm}
 
 \usepackage{amssymb}
 \numberwithin{equation}{section}
\usepackage{epic}

\usepackage[english]{babel}

\usepackage{graphicx}
\usepackage{float}
\usepackage{pict2e}
\usepackage{tikz}

\usepackage{epsfig}

\usepackage{lipsum}

\author{Mark Adler\thanks{2000
{\em Mathematics Subject Classification}. Primary:
60G60, 60G65, 35Q53; secondary: 60G10, 35Q58. {\em Key
words and Phrases}: Dyson's Brownian motion, Airy
process, extended kernels, random Hermitian ensembles,
coupled random matrices. \newline
 Department of Mathematics, Brandeis University,
Waltham, Mass 02454, USA. E-mail: adler@brandeis.edu.
The support of a National Science Foundation grant \#
DMS-01-00782 is gratefully acknowledged.}~~~~~~ Pierre
van Moerbeke\thanks{ Department of Mathematics,
Universit\'e de Louvain, 1348 Louvain-la-Neuve, Belgium
and Brandeis University, Waltham, Mass 02454, USA. This
work was done while PvM was a member of the Clay
Mathematics Institute, One Bow Street, Cambridge, MA
02138, USA. E-mail: vanmoerbeke@math.ucl.ac.be and
@brandeis.edu. The support of a National Science
Foundation grant \# DMS-01-00782, European Science
Foundation, Nato, FNRS and Francqui Foundation grants is
gratefully acknowledged.}}

\date{}

\newcommand{\MAT}[1]{\left(\begin{array}{*#1c}}
\newcommand{\mat}{\end{array}\right)}
\newcommand{\qed}{\leavevmode\unskip\nobreak\penalty200\hskip2pt\null
\nobreak\hfill\rule{1.1ex}{1.1ex}
\medbreak }

 \newcommand{\Om}{\Omega}
\newcommand{\AR}{{\cal A}}
\newcommand{\CR}{{\cal C}}
\newcommand{\DR}{{\cal D}}

\newcommand{\FR}{{\cal F}}
\newcommand{\GR}{{\cal G}}

\newcommand{\LR}{{\cal L}}

\newcommand{\PR}{{\cal P}}
\newcommand{\RR}{{\cal R}}

\newcommand{\BC}{{\mathbb C}}

\newcommand{\BH}{{\mathbb H}}
\newcommand{\BL}{{\mathbb L}}
\newcommand{\BP}{{\mathbb P}}

\newcommand{\BZ}{{\mathbb Z}}

\newcommand{\iy}{\infty}
\newcommand{\pl}{\partial}
\newcommand{\al}{\alpha}

\newcommand{\I}{{\rm i}}

\newenvironment{proof}{\medskip\noindent{\it Proof:\/} }{\qed}
\newenvironment
        {remark}{\medskip\noindent\underline{\it Remark:\/} }{\medbreak}

\newcommand{\om}{\omega}
\newcommand{\vp}{\varphi}
\newcommand{\la}{\langle}
\newcommand{\ra}{\rangle}
\newcommand{\ga}{\gamma}
\newcommand{\Ga}{\Gamma}

\newcommand{\Dt}{\Delta}
 
\newcommand{\sg}{\sigma}
\newcommand{\BR}{{\mathbb R}}
\newcommand{\lb}{\lambda}

\newcommand{\dis}{\displaystyle}

\newcommand{\BK}{{\mathbb K}}

\newcommand{\Id}{\mathbbm{1}}
\newcommand{\ka}{\kappa}
\newcommand{\rk}{{\frak r}}

\newcommand{\Ai}{\mbox{Ai}}

  \newcommand{\bl}{\begin{aligned}}
  \newcommand{\el}{\end{aligned}}

\def\be#1\ee{\begin{equation}#1\end{equation}}
\def\bea#1\eea{\begin{eqnarray}#1\end{eqnarray}}
\def\bean#1\eean{\begin{eqnarray*}#1\end{eqnarray*}}

 \newtheorem{definition}{Definition}[section]
 \newtheorem{theorem}[definition]{Theorem}
 \newtheorem{lemma}[definition]{Lemma}
 \newtheorem{corollary}[definition]{Corollary}
 \newtheorem{proposition}[definition]{Proposition}

\catcode `!=11

\newdimen\squaresize
\newdimen\thickness
\newdimen\Thickness
\newdimen\ll! \newdimen \uu! \newdimen\dd! \newdimen \rr! \newdimen
\temp!

\def\sq!#1#2#3#4#5{%
\ll!=#1 \uu!=#2 \dd!=#3 \rr!=#4
\setbox0=\hbox{%
 \temp!=\squaresize\advance\temp! by .5\uu!
 \rlap{\kern -.5\ll!
 \vbox{\hrule height \temp! width#1 depth .5\dd!}}%
 \temp!=\squaresize\advance\temp! by -.5\uu!
 \rlap{\raise\temp!
 \vbox{\hrule height #2 width \squaresize}}%
 \rlap{\raise -.5\dd!
 \vbox{\hrule height #3 width \squaresize}}%
 \temp!=\squaresize\advance\temp! by .5\uu!
 \rlap{\kern \squaresize \kern-.5\rr!
 \vbox{\hrule height \temp! width#4 depth .5\dd!}}%
 \rlap{\kern .5\squaresize\raise .5\squaresize
 \vbox to 0pt{\vss\hbox to 0pt{\hss $#5$\hss}\vss}}%
}
 \ht0=0pt \dp0=0pt \box0
}

\def\vsq!#1#2#3#4#5\endvsq!{\vbox to \squaresize{\hrule
width\squaresize height 0pt%
\vss\sq!{#1}{#2}{#3}{#4}{#5}}}

\newdimen \LL! \newdimen \UU! \newdimen \DD! \newdimen \RR!

\def\vvsq!{\futurelet\next\vvvsq!}
\def\vvvsq!{\relax
  \ifx     \next l\LL!=\Thickness \let\continue=\skipnexttoken!
  \else\ifx\next u\UU!=\Thickness \let\continue=\skipnexttoken!
  \else\ifx\next d\DD!=\Thickness \let\continue=\skipnexttoken!
  \else\ifx\next r\RR!=\Thickness \let\continue=\skipnexttoken!
  \else\def\continue{\vsq!\LL!\UU!\DD!\RR!}%
  \fi\fi\fi\fi
  \continue}

\def\skipnexttoken!#1{\vvsq!}

\def\place#1#2#3{\vbox to 0pt{\vss
\rlap{\kern#1\squaresize
  \raise#2\squaresize\hbox{$#3$}}
\vss}}

\catcode `!=12

\newsavebox{\foobox}

\usepackage{amsmath}

\title{Double Interlacing in Random Tiling Models
 }

\author{Mark Adler\thanks{2000
{\em Mathematics Subject Classification}. Primary:
60G60, 60G65, 35Q53; secondary: 60G10, 35Q58. {\em Key
words and Phrases}: Lozenge and domino tilings, non-convex domains, doubly interlacing sets, kernels, probability distributions. \newline
 *Department of Mathematics, Brandeis University,
Waltham, Mass 02453, USA. E-mail: adler@brandeis.edu. 
The support of a Simons Foundation Grant  
 \# 278931 is
 gratefully acknowledged. 
 }~
  ~~~~~ Pierre
van Moerbeke\thanks{ Department of Mathematics,
UCLouvain, 1348 Louvain-la-Neuve, Belgium
and Brandeis University, Waltham, Mass 02453, USA. E-mail: pierre.vanmoerbeke@uclouvain.be. The support  of a Simons Foundation Grant 
 \# 280945 is
gratefully acknowledged.  \newline The authors MA and PvM gratefully acknowledge NSF-support during their stay at MSRI, Berkeley (fall 2021). PvM thanks the Newton Institute, Cambridge (fall 2022) for its support during his stay.
 }
}

\begin{document}

\sloppy
 \maketitle
 
\hspace*{2.7cm} {\em To the memory of Freeman J. Dyson} 
 
 \vspace*{.5cm}
 
 \begin{abstract} 
 
 Random tilings of very large domains will typically lead to a solid, a liquid, and a gas phase. In the two-phase case, the solid-liquid boundary (arctic curve) is smooth, possibly with singularities. At the point of tangency of the arctic curve with the domain-boundary, the tiles of a certain shape form for large-size domains a singly interlacing set, fluctuating according to the eigenvalues of the principal minors of a GUE-matrix (Gaussian unitary ensemble). Introducing non-convexities in large domains may lead to the appearance of several interacting liquid regions: they can merely touch, leading to either a split tacnode (also called hard tacnode), with two distinct adjacent frozen phases descending into the tacnode, or a soft tacnode. For appropriate scaling of the nonconvex domains and probing about such split tacnodes, filaments of tiles of a certain type will connect the liquid patches: they evolve in a bricklike-sea of dimers of another type. Nearby,  the tiling fluctuations are governed by a discrete tacnode kernel; i.e., a determinantal point process on a doubly interlacing set of dots belonging to a discrete array of parallel lines. This  kernel enables one to compute the joint distribution of the dots along those lines. This kernel appears in two very different models: (i) domino-tilings of skew-Aztec rectangles and (ii) lozenge-tilings of hexagons with cuts along opposite edges. Soft, opposed to hard,  tacnodes  appear when two arctic curves gently touch each other amidst a bricklike sea of dimers of one type, unlike the split tacnode. We hope that this largely expository paper will provide a view on the subject and be accessible to a wider audience.

 \end{abstract}
 
  \tableofcontents


\section{Introduction}

Tiling models have been prominently present in combinatorics, probability theory and statistical mechanics. As far back as 1911, MacMahon \cite{McMa} gave us the celebrated  formula below for the enumeration of lozenge tilings of hexagons of sides $a,b,c,a,b,c$, with a more recent version by Macdonald \cite{McD},
$$
  \frac{H(a)H(b)H(c)H(a+b+c)}{H(a+b)H(a+c)H(b+c)}=
 \prod _{{i=1}}^{a}\prod _{{j=1}}^{b}\prod _{{k=1}}^{c}{\frac {i+j+k-1}{i+j+k-2}}
$$
where 
$$H(n)=(n-1)! (n-2)!\dots 1! 0!.
$$
 This result has been extended in the combinatorial community to many different shapes, including non-convex domains, and shapes with cuts and holes; see e.g., Ciucu, Fischer and Krattenthaler \cite{Ciucu,Krat2}.

Tiling models have gained a considerable interest in the physics community in the 50-60's. Indeed they have sufficient complexity to have interesting features and yet are simple enough to be tractable! In many cases tiling models can also be viewed as dimer models on certain finite or infinite bipartite graphs, thinking of the work of Kaufman and Onsager \cite{Onsager} on the spontaneous magnetization of the square-lattice Ising model, thinking of the Kac-Ward formula \cite{Kac} for the
partition function of the Ising model on planar graphs, thinking of Kasteleyn's work \cite{Kast1, Kast} on a full covering of a two-dimensional planar lattice with dimers, for which he computed the entropy. In addition think of Fisher and Temperley \cite{Temp} who counted  the number of domino tilings of a chessboard (or any other rectangular region).
  Much further work has been done on dimers on planar, bipartite and periodic graphs; see \cite{K,KO,KOS}.

 We are specifically interested in limits when the size of the tiling region grows to infinity, and in particular what happens to the tiling patterns in certain regions, viewed in different scales. To these tiling or dimer models we can also associate a height function, an integer valued function on the faces of the dimer model graph, which gives an interpretation of the dimer model as a random surface. At the global scale this height function converges to a deterministic limiting height function solving a variational problem, \cite{CKP}, and the region typically displays domains with different phases. Phases refer to the different types of local limits (limiting Gibbs measures), which can be obtained; see \cite{OR1}. It has been shown in \cite{KOS} that in a large class of dimer models these fall into three types,
frozen, liquid (or rough) and gas (or smooth). In the liquid phase, correlations are decaying polynomially with distance, whereas in the gas phase they decay  exponentially with distance. For an excellent introduction to the subject, see the books or lecture notes by Kurt Johansson \cite{Jo16}, Vadim Gorin \cite{Gorin2} and Dan Romik \cite{Romik}. 

In this paper we are only concerned with  models having no gas phase.
From \cite{AstDuse} and \cite{KO}, it appears that for a large class of dimer models in a domain, the boundary between the solid and liquid phases (arctic curves) is (asymptotically for the size going to infinity) a smooth curve possibly with singularities and some special points.  

Tiling problems have been linked to Gelfand-Zetlin cones by Cohn, Larsen, Propp \cite{Cohn}, to alternating sign matrices \cite{Elkies} and to non-intersecting paths, determinantal processes, kernels and random matrices by Johansson \cite{Jo01b,Jo02b,Jo03b}. In \cite{Jo05b}, Johansson showed that the statistics of the lozenge tilings of hexagons was governed by a kernel consisting of discrete Hahn polynomials; see also Gorin\cite{Gorin}. Johansson \cite{Jo05c} and Johansson-Nordenstam \cite{JN} show that, in appropriate limits, the tiles near the boundary between the frozen and stochastic region (arctic circle) fluctuate according to the Airy process and near the points of tangency of the arctic circle with the edge of the domain as the GUE-minor process. In the latter situation, tilings of a certain shape form a simply interlacing set, whose positions jointly correspond to the eigenvalues of the principal minors of a GUE-matrix; this will be discussed in section \ref{singleinterl}.  
 

So, the liquid region consists of patches in the domain, which may touch the boundary of the domain (turning points), which may intersect each other (leading to cusps along the arctic curve) or which may merely touch (tacnodes); touching can occur in two different ways: a soft and a split tacnode. 

At a {\em soft tacnode} we have the same frozen phase (same type of tiles) at both sides of the tacnode (see Fig. 17), whereas at a {\em split tacnode}, we have different frozen phases on each of the two sides; see the two different frozen phases, yellow and blue, appearing on either side of the split tacnode in Fig. 7. 
 
It is of great interest to study the random fluctuations around these limit shapes. Inside the liquid patches the height function is conjectured to fluctuate according to a Gaussian Free Field. This has been shown in a number of cases see e.g. \cite{OR1,Ke,BF,D,DFF,BuGo,Buf,BLR, Petrov1,Huang,Laslier}.
We can also consider the fluctuations of the dimers at the interface between the liquid and frozen phases, and, in particular, at the points where the liquid region touches the boundary or another liquid region. These type of limits fall naturally into two types depending on whether one rescales in just one direction, {\em a discrete-continuous limit}, or in both directions, {\em a continuous-continuous limit}; see \cite{Jo16} for a discussion. 

There are \underline{three basic {\em continuous-continuous limits}}; i.e. continuous in both directions: \newline (i) At {\em generic points} of the boundary of a liquid region the dimers fluctuate according to the Airy kernel process, as in \cite{PrSpohn,OR1,Jo05c,Petrov2,Duse2, Duse1}. \newline (ii) At a {\em cusp} they behave as the Pearcey process, as in \cite{OR2,ACvM}. \newline (iii) At a {\em soft tacnode} the fluctuations are described by the (continuous) tacnode kernel, as in \cite{AJvM0}. The kernel for the soft tacnode process, was given in the context of non-intersecting random walks \cite{AFvM}, nonintersecting Brownian motions \cite{Jo13} and in the context of overlapping Aztec diamonds in \cite{AJvM0}. 
When an Airy process interacts with a boundary, then a different process appears, which Ferrari-Vet\H{o} \cite{FerrariVetBr} call a hard tacnode. Yet another case is considered by Borodin-Duits \cite{BD}.

There are also \underline{three basic {\em discrete-continuous limits}}; i.e., discrete in one direction and continuous in the other:\newline  $(i)$ At {\em turning points} the natural limit process is given by the GUE-minor (or GUE-corners process), as already mentioned; this leads to a single interlacing set of dots; see \cite{JN,OR1}. \newline $(ii)$ At {\em split cusp points} with two different frozen phases inside the cusp, we get the Cusp-Airy process, as in  \cite{DJM}. \newline $(iii)$ Zooming about a {\em split tacnode} between the liquid regions one observes non-intersecting paths of dimers of definite types, leading in the scaling limit to a doubly interlacing set of dots; they form {\em filaments}  between two liquid patches evolving in a bricklike sea of dimers of another type, as in \cite{ACJvM},\cite{AJvM1},\cite{AJvM2}. Here the limiting kernel is the discrete tacnode kernel. 

All these three cases are based on the classification of the height function regularity, as mentioned. It is conjectured
in \cite{AstDuse} that the six interface processes discussed above are all that can occur in a large class of dimer models.
To be more precise, the authors in \cite{AstDuse} prove a classification of the regularity of minimizers and frozen boundaries for a natural class of polygonal  dimer model domains (simply or multiply connected). They use the fact that the asymptotic height function is the solution to a variational problem over an admissible class of Lipschitz height functions. The variational problem involves minimizing over surface tension functions which are bounded and convex in their domain and which satisfy a Monge-Amp\`ere equation. 
%
 They show the frozen boundary (the arctic curve) is the real locus of a locally convex algebraic curve (minus its isolated points); they have at most finitely many singularities, which are all first order {\em cusps, tacnodes or tangency points} of the frozen boundary with the domain boundary. The dimer models discussed here belong to the natural class required by the theorem stated in  \cite{AstDuse} and the height function used in our paper is affinely equivalent to their height function. Therefore the singularities of the frozen boundary must have the types mentioned above. 
 
 In this review paper, we will be considering  tilings of non-convex shapes: tilings with dominos or lozenges. These were already considered by Okounkov-Reshetikin\cite{OR1,OR2} and Kenyon-Okounkov\cite{KO} and then more specifically by Petrov \cite{Petrov2}, who considered lozenge tilings  of hexagons, where one side only has a finite number of cuts (say along the top), opposed to Fig. 10(b), where cuts appear at opposite sides. Petrov computes the kernel for the resulting interlacing set of red tiles, starting from the bottom, when the size of the hexagon and the cuts become very large. He also computes the statistical behavior of the tiles in the bulk, which Okounkov-Reshetikhin \cite{OR1} and Petrov \cite{Petrov2} shows satisfy the incomplete beta-kernel. 
 
 Many of these processes are believed to be {\em universal}, since they appear in a number of different models. These computations depend on the explicit models. As mentioned, the GUE-minor process has appeared in lozenge tilings of hexagons and domino tilings of Aztec diamonds; see \cite{JN,OR2}. Aggarwal and Gorin \cite{Ag-Go} were able to extend some of the results to lozenge tilings of generic domains, by localizing the problem. To be precise, when the boundary of fairly generic domains 
 contains, in the limit, three straight segments, with $2\pi/3$ angles, then the GUE-minor process appears asymptotically along this line-segment; even, more generally, if the tilings have an embedded trapezoid, formed by the sides of the lozenges of the tiling and with the same angle requirements, then also the GUE-minor process will appear. Moreover, Aggarwal and  Huang \cite{AH} prove a universality result for the Airy-statistics of the tiles near the arctic boundary between the solid and liquid phases for a very general class of shapes, again by localizing. Similar results are shown by Amol Aggarwal in \cite{Ag}.

In this paper, we will be concerned with domino tilings of so-called {\em skew-Aztec rectangles} \cite{AJvMskewAzt} in section 5 and lozenge tilings of {\em non-convex hexagons} \cite{AJvM1,AJvM2} in section 6; in the former situation, the skew property implies a (very slight) non-convexity along opposite sides of the rectangle. In the appropriate scaling limit when the size of the domains, rectangles or hexagons, gets very large, one finds, as mentioned in $(iii)$, two liquid regions in the domain, connected with filaments; 
 the filaments evolve within a strip bounded by two parallel lines; see Fig. 7. The width $\rho$ of the strip $\{\rho\}$ and the number of filaments $r$ will be important parameters in the problem.  The existence of those filaments is tantamount to saying that those tiles form a determinantal point process of a doubly interlacing set of dots (along parallel lines to the strip) in a neighborhood of the strip. Looked at from a distance, the two liquid regions appear to merely touching each other.

It turns out that, in the appropriate scaling limit, both models lead to tile fluctuations governed by the so-called {\em discrete tacnode kernel} (\ref{Ldtac}), a kernel consisting of a Heaviside part and a sum of four multiple integrals; when $\rho=r$, the kernel simplifies considerably, as explained in Section 5.2. The form of the kernels will appear in section 3. The fact that the  discrete tacnode kernel appears in these two very distinct models, having nevertheless some features in common, provides strong evidence that the discrete tacnode kernel is a universal limiting kernel that naturally occurs whenever we have double interlacing pattern of the type encountered here. 

As the GUE-minor statistics occurs for a single interlacing system, so do we have another statistics for double interlacing systems \cite{AvM1,AvM2}. This follows from computing the probability by means  of the the discrete tacnode kernel, as will be discussed in section 4. When $\rho=r$, the double interlacing is equivalent to two single interlacing systems, put side-by-side and pointing in opposite directions. \cite{Krat,McMa1} are references for the basic combinatorics. 

This paper uses basic combinatorics which can be found in McDonald \cite{McD} and Stanley \cite{Stanley}.







We thank Sunil Chhita for having made the very insightful simulations in this paper.
 


\section{GUE-matrices, single interlacing and Aztec diamonds}\label{singleinterl}

GUE-matrices are given by $n\times n$ Hermitian matrices $X$, with  distribution function,
$$dP(X)=\frac1{Z_n}e^{-\frac{\mbox{\tiny Tr }X^2}2}dX.$$
This induces a distribution on the eigenvalues $\{x_j^{(\tau)}\}_{1\leq j\leq\tau\leq n}$, of all  $\tau^{th}$ upper-left principal minors $X^{(\tau)}$, viewed jointly . The consecutive sets of eigenvalues ${\bf x}^{(\tau)}=(x_1^{(\tau)},\dots,x_\tau^{(\tau)})$ are known to interlace; i.e., {$x^{(\tau+1)}_1\leq x^{(\tau)}_1\leq x^{(\tau+1)}_2\leq\dots\leq x^{(\tau+1)}_{\tau } \leq x^{(\tau)}_\tau\leq x^{(\tau+1)}_{\tau+1}$,} which is denoted by
\be\label{interlace}{\bf x}^{(1)}\prec {\bf x}^{(2)}\prec \ldots \prec {\bf x}^{(n)}={\bf x}\ee
 This GUE-minor point process is also a determinantal point process with correlation kernel,
 \be\label{Lminor}\bl
  {\mathbb L}^{\mbox{\tiny GUEminor}}
     ( \tau_1, y_1 ;\tau_2, y_2)  
  := - &
{\mathbb H}^{\tau_1-\tau_2} ( y_1-y_2)   
,~~~~~~~(\tau_i,y_i)\in \BZ\times\BR
\\& +\oint_{\Ga_0}\frac{du} {(2\pi\I)^2}\oint_{  L_{0+}}  \frac{ dv}{v-u}\frac{v^{  \tau_2}}{u^{  \tau_1}}
\frac{e^{-\frac{u^2}2  + y_1   u  }}
{e^{-\frac{v^2}2  + y_2    v  }},\el\ee
where $ L_{0+}$ is an up going vertical line in $\BC$ to the right of $0$ and where
\be \bl\BH^{m}(z)  :=\frac{z^{m-1}}{(m-1)!}\Id _{z\geq 0}\Id_{m\geq 1},~~~~(\mbox{Heaviside function}).
 \label{Heaviside}\el\ee

 Consider uniform measure 
 $$
 d\mu_{{\bf x}^{(\tau)}}=\prod_1^{\tau-1} d{\bf x}^{(k)}\Id_{{\bf x}^{(k+1)}\succ {\bf x}^{(k)} },\mbox{~for }d{\bf x}^{(k)}=\prod_i^kdx_i^{(k)}$$ on the cone ${\mathcal C}_{\bf x}$ for ${\bf x}\in \BR^\tau$

 $${\mathcal C}_{\bf x}=\{({\bf x}^{(\tau-1)}, \ldots , {\bf x}^{(1)}) \in \BR^{\tau(\tau-1)/2}\mbox{ such that }{\bf x} \succ {\bf x}^{(\tau-1)}\succ \ldots \succ {\bf x}^{(1)}\}.
 $$
 According to Baryshnikov \cite{Bar}, upon fixing the spectrum ${\bf x}$ of the GUE-matrix $X^{(\tau)}$, the spectra
  ${\bf x}^{(k)}$ of the minors $X^{(k)}$ for $1\leq k\leq \tau-1$ are uniformly distributed, with the volume expressed in terms of the Vandermonde determinant,
%
\be\begin{aligned}
\BP^{\mbox{\tiny GUE}}\left(\bigcap_{k=1}^{\tau-1}\{{\bf x}^{(k)}\in d{\bf x}^{(k)}\}~\Bigr|~
{\bf x}^{(\tau)}={\bf x}
 \right)=
  \frac{d\mu_{\bf x} }{\mbox{Vol}( {\mathcal C}_{\bf x}  ) } 
, 
 \mbox{with }\mbox{Vol}( {\mathcal C}_{\bf x} )=\frac{ \Dt_\tau({\bf x})}{\prod_1^{\tau-1}k!}, 
   \end{aligned}\label{Bary1}\ee 
   and thus we have, setting ${\bf x}:= {\bf x}^{(\tau)}$,
   \be\begin{aligned}
\BP^{\mbox{\tiny GUE}}\Bigl(\bigcap_{k=1}^\tau\{{\bf x}^{(k)}\in d{\bf x}^{(k)}\}
 \Bigr)=
 \rho_\tau^{ \mbox{\tiny GUE}}({\bf x}^{ })d{\bf x}^{ }
  \frac{d\mu_{\bf x}}{\mbox{Vol}( {\mathcal C}_{\bf x}  ) } = \rho_\tau^{ \mbox{\tiny minor}}({\bf x}^{ })d{\bf x}^{ }d\mu_{\bf x}
,
  \end{aligned}\label{Bary1'}\ee
in terms of the densities
\be\label{GUE-dens}\rho_\tau^{\mbox{\tiny GUE}}({\bf x}  )d{\bf x}: = \frac{\Dt_\tau^2({\bf x})} 
 {\prod_{k=1}^{\tau-1}k!}\prod_{j=1}^\tau  \frac{e^{-\frac{x_j^2}2}dx_j}{ \sqrt{2\pi}},~~~
 \rho_\tau^{ \mbox{\tiny minor}} :=  \Dt_\tau ({\bf x}) \prod_{j=1}^\tau  \frac{e^{-\frac{x_j^2}2}dx_j}{ \sqrt{2\pi}}
 .\ee

 
  {\em Aztec diamonds and domino tilings}. Consider next an Aztec diamond of size $n$ as in Fig. 1(a), which is a union of squares, alternatively blue and white, with lattice points in the region $-1\leq \xi,\eta\leq  2n+1$ of the plane in the $(\xi,\eta)$-coordinates as depicted in Fig. 1(a) for size $n=8$.   
  This diamond can be covered with $4$ different $2\times 1 $ vertical and horizontal domino's (as in Fig. 1(b)), where the blue square can be on the left (color blue), on the right (color red), on top (color green) or at the bottom (color yellow); this covering is depicted in the simulation of Fig 1(c) for $n=20$. The cover of the diamond with dominos is such that the white (blue) square of the domino can only be put on the white (blue) square of the diamond; this can be done in $2^{n(n+1)/2}$ ways. See \cite{JPS,Jo02b,Jo05c} and Johansson's Harvard lectures \cite{Jo16}.
 
  The lines $\xi=2\tau-1$, for $1\leq \tau\leq n$  pass through exactly $\tau$ blue or green tiles 
   at locations $\eta =2j$, with $0\leq j\leq  \tau$; in the simulation Fig 1.(c), the line $\xi=1$ passes through one green domino, the line $\xi=3$ through one blue and one green, the line $\xi=5$ through two blue and one green, etc... Put a red dot in the middle of the left (upper) square of the blue (green) tile in question, as in Fig. 1(b). One shows that the vectors $\eta^{(\tau)}=(\eta^{(\tau)}_1,\dots,\eta^{(\tau)}_\tau)\in ( \BZ_{\tiny\mbox{even}})^\tau$,  giving the positions of the red dots (white dots in Fig. 1(c)) on the lines $\xi=2\tau-1$, interlace (see \ref{interlace}):
   %
   $$ \eta^{(1)}\prec \dots \prec \eta^{(\tau)}\prec 
\eta^{(\tau+1)}\prec\dots\prec \eta^{(n)}
 .$$%
  The interlacing ensemble $
     \eta^{(\tau)}_j $ is a determinantal point process. In the limit when $n\to \iy$, Johansson and Nordenstam \cite{JN} show the following joint limit statement around the point $n $, for fixed levels $\xi=2\tau-1$ with $1\leq \tau\leq N$ and any $N$ fixed:
 $$  \lim_{n\to \iy}\left\{ \frac{\eta^{(\tau)}_j-n}{\sqrt{n}}\right\}_{1\leq j\leq \tau\leq N}\to
 \left\{  x_j^\tau\right\}_{1\leq j\leq \tau\leq N}$$
  More precisely for each continuous function $\phi:{\mathbb N}\times \BR\to \BR, ~0\leq \phi $ of compact support, we have for fixed $N$, 
 $$ {\mathbb E}\left(\prod_{1\leq j\leq \tau\leq N} (1-\phi\Bigl(\tau,\frac{\eta^{(\tau)}_j-n}{\sqrt{n}}\Bigr)
 \right)\to 
 {\mathbb E}\left( \prod_{1\leq j\leq \tau\leq N} (1-\phi(\tau,x^{(\tau)}_j))\right)
 $$

  \newpage
\setlength{\unitlength}{0.015in}
  \begin{picture}(0,0)
   \put(178,-180)  
 {{\makebox(0,0){\includegraphics[width=190mm,height=238mm]{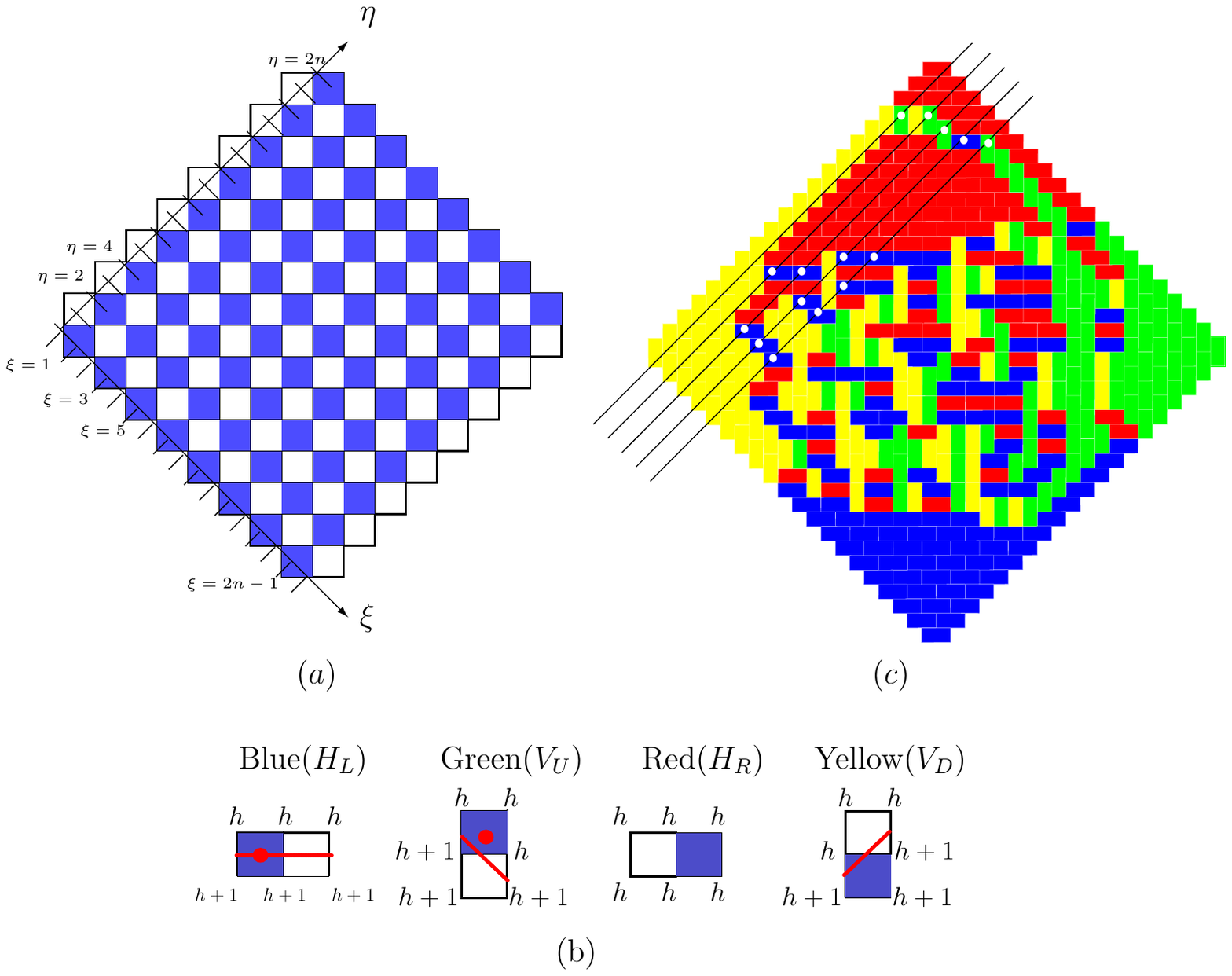}}}}
 
\end{picture}

 \vspace{7.5cm}
 
 {\small Figure 1: An Aztec diamond (a) with coordinates $(\xi,\eta)$, four kind of dominos (b) and a tiling by these dominos, vertical $V_U,V_D$ and horizontal ones $H_L,H_R$, with the blue square on the upper- or lower-side or to the left- or right-side; see simulation (c). Each of the dominos (b) is equipped with a level surface, whose height along the boundary is given by $h\in \BZ_{\geq 0}$; also with a level curve of height $h+1/2$ and with a red dot in the appropriate square. Covering the diamond with these 4 types of dominos leads to a level curve of height $h+\frac 12$. The colors, blue, green, red yellow, correspond to the colors in the simulations. For the sake of visibility the red dots are replaced by white dots in Fig. 1(c).}
 
 \vspace*{ 1cm}




Johansson \cite{Jo02b} then shows that in the limit when $n\to \iy$, the tiling tends to a configuration which is an inscribed circle (the arctic circle) tangent to the $4$ sides of the Aztec diamond, with four points of tangency  half way along the sides; inside the circle the tiling is chaotic and outside it is bricklike as in Fig. 1(c). If horizontal tiles are more likely than vertical tiles, that is when $0<a\leq 1$ in the probability (\ref{Prob}), then the arctic circle becomes an ellipse, also having four tangency points with the boundary.
 In \cite{JN} analogous results are shown for lozenge tilings of hexagons.
 
 The point process of red dots in the Aztec diamond 
  is a determinantal point process with correlation kernel, given by (see \cite{JN}):  (in the case discussed here $a=1$)
 \be\label{2.7}\bl\BK^{(a)}(2r,x;2s,y)=& \int_{L_{0+} } \frac{dz} {(2\pi \I)^2}\oint_{\Gamma_0}
 \frac{dw}{z-w}\frac{w^{y-1}}{z^{x}}\frac{(1-aw)^{n-s}(1+\frac aw)^{ s }}{(1-az)^{n-r}(1+\frac az)^{ r }} 
\\& -\Id_{s>r}\oint _{\Gamma_{0,-a}} \frac{dz}{2\pi \I z} (-z)^{x-y }
\left(\frac{1-az}{1+\frac az}\right)^{ {s-r} }
.\el \ee
 In \cite{JN}, Johansson and Nordenstam show that, for some conjugation by a function $g_n(\tau_1,\eta_1)$, we have: 
 
 
\begin{theorem} (\cite{JN}, Proposition 4.1) For large size $n$ of the Aztec diamond, the point process of interlacing red dots, near the point of tangency of the arctic circle with the boundary of the Aztec diamond, is determinantal for the GUE minor-kernel (\ref{Lminor}); to be precise, it is given by the following limit  (for $a=1$):
$$\bl\lim_{n\to \iy}
\frac{g_n(\tau_1,\eta_1)}{g_n(\tau_2,\eta_2)}  
&\BK^{(1)}\left( 2(n-\tau_1) ,\eta_1; 2(n-\tau_2) ,\eta_2\right)d\eta_2\Bigr|_{\eta_i=n+u_i\sqrt{  n }} 
\\&=
 {\mathbb L}^{\mbox{\tiny GUEminor}}
     (  \tau_1, u_1 ; \tau_2, u_2) du_2 
.\el$$
\end{theorem}
\proof It can be found in \cite{JN}.\qed
The probability (\ref{Bary1'}) can then be deduced from the kernel ${\mathbb L}^{\mbox{\tiny GUEminor}}$.

\bigbreak 
The purpose of this paper is to give an  overview of other tiling models which in the large size limit gives other universal statistics. %
 
 \newpage
 \vspace*{8cm}

\setlength{\unitlength}{0.015in}\hspace*{-2cm}\begin{picture}(0,0)
 
 \put(180,-60){\makebox(0,0){\includegraphics[width=230mm,height=300mm]{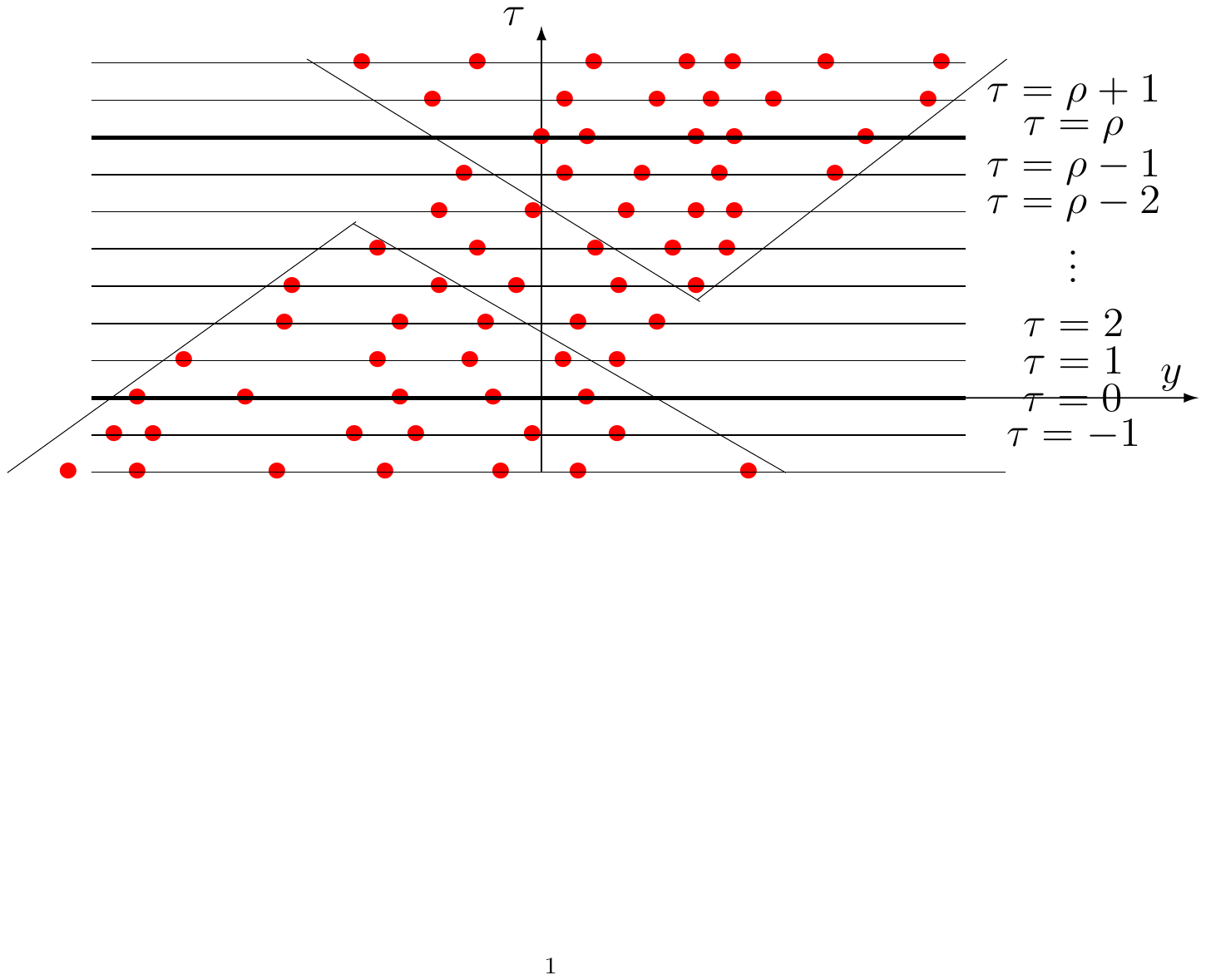}
 }}
   
     
 \end{picture}
 
 \vspace{-4cm}
{\small Figure 2: Doubly interlacing system of points in $\BR$ or $\BZ$ for $\rho>r$, with the two simply interlacing polytopes; here $\rho=7$ and $r=5$.}
 
 \vspace{-2cm}
 
  \setlength{\unitlength}{0.015in}\begin{picture}(0,170)
  \put(100,-200){\makebox(0,0){\includegraphics[width=230mm,height=300mm]{Fig3
  .pdf}}} 

    \end{picture}  

\vspace{6.5cm}
 {\small Figure 3: Doubly interlacing system of points in $\BR$ or $\BZ$ for $\rho=r$, consisting of two singly interlacing sets, subjected to $\rho$ inequalities between the red and black dots connected by the arrows;  here $r=\rho=5$.}

 \newpage

  \section{Double interlacing and the discrete tacnode kernel}

Define levels $-\iy<\tau<\iy$, and numbers $\rho\in\BZ_{\geq 0}$ and $r\in \BZ_{>0}$. Also define the number
\be \label{ntau}\bl
n_\tau:&=r+(\tau-\rho)_{\geq 0} &\mbox{  for  }\tau\geq 0
\\
&=r-\tau&\mbox{  for  }\tau\leq 0,
\el\ee
and an infinite set of vectors ${\bf y}^{(\tau)}:={\bf y}^{(n_\tau) }\in \BR^{n_\tau}$ describing the positions of the $n_\tau$ dots at each level $\tau$.
 They form a {\em doubly interlacing set} if
  (see Fig. 3)
  \be\begin{aligned}
 \dots\succ {\bf y}^{(-2)}\succ {\bf y}^{(-1)} \succ {\bf y}^{(0)}\curlyeqprec   {\bf y}^{(1)} 
\curlyeqprec\ldots \curlyeqprec 
  {\bf y}^{(\rho-1)}
\curlyeqprec {\bf y}^{(\rho)}\prec {\bf y}^{(\rho+1)}\prec{\bf y}^{(\rho+2)}\prec\ldots  
.\end {aligned}\label{xy-interlace}\ee
abbreviated as 
\be\label{yinterlace}\dots \simeq {\bf y}^{(\tau)} \simeq {\bf y}^{(\tau+1)}\simeq\dots\ee
where
$ {\bf y} \curlyeqprec {\bf y}'$ for vectors of equal size ${\bf y},~{\bf y'}\in \BR^r$ means the following: $y_1\leq y_1'\leq y_2\leq y'_2 \leq \ldots \leq y_r\leq y'_r$. So, we have that 
$${\bf y}^{(i)}\in \BR^{(r)}\mbox{ for}~0\leq i\leq \rho\mbox{ and }\left\{\bl &{\bf y}^{(0)}\in \BR^{(r)},~{\bf y}^{(-1)}\in \BR^{(r+1)},~{\bf y}^{(-2)}\in \BR^{(r+2)},\dots
\\&
{\bf y}^{(\rho)}\in \BR^{(r)},~{\bf y}^{(\rho+1)}\in \BR^{(r+1)},~{\bf y}^{(\rho+2)}\in \BR^{(r+2)},\dots\el\right.
. $$
As Fig.2 shows, a doubly interlacing system with $\rho>r$ can be viewed as two single interlacings (within each of the cones), with extra-dots in between, satisfying some inequalities. Fig.3 illustrates a doubly interlacing system for $\rho=r$; this situation reduces to two singly interlacing systems, whose only interactions amount to $\rho$ inequalities, going with each of the arrows in Fig.3.

%

 Consider now the  {\em discrete tacnode kernel} in $ \tau_i\in \BZ$ and $y_i\in \BR $, as introduced in \cite{AJvM1}(formula (14)), and \cite{AJvMskewAzt}(formula (11)), and depending on the geometric integer parameters $\rho,r\in \BZ_{\geq 0}$ and a parameter $\beta\in \BR$,
\be  \label{Ldtac}
 \bl
 {\mathbb L}^{\mbox{\tiny dTac}}_{\rk,\rho,\beta} &  ( \tau_1, y_1 ;\tau_2, y_2)   =- 
{\mathbb H}^{\tau_1-\tau_2} (  y_1-y_2)
~~~~~~~~~~
\\& +
 {\oint_{\!\!\!\Ga_0}\frac{du} {(2\pi\I)^2}\oint_{ \!\!\! L_{0+}}  \frac{ dv}{v-u}\frac{u^{\rho-\tau_1}}{v^{\rho-\tau_2}}
\frac{e^{-\frac{u^2}2  +(\beta+y_1 )  u  }}
{e^{-\frac{v^2}2  +(\beta+y_2 )  v  }}
    }  \frac{ \Theta_r( u, v )} { \Theta_r(0,0)}  ~~~\left\{\bl& { \mbox{$\neq 0$ only} }\\&{\mbox{when $\rho<\tau_1$} }\el\right.
 \\& + 
 { \oint_{\!\!\!\Ga_0}\frac{du} {(2\pi\I)^2}\oint_{\!\!\! L_{0+}}\frac{ dv}{v-u} \frac{u^{  \tau_ 2}}{v^{  \tau_1}}
   \frac{e^{-\frac{u^2}2  +(\beta-y_2 )  u  }} {e^{-\frac{v^2}2  +(\beta-y_1 )  v  }} 
   } 
      \frac{ \Theta_r( u, v )}  
      { \Theta_r(0,0)}  ~~~~~~\left\{\bl& { \mbox{$\neq 0$ only} }\\&{\mbox{when $\tau_2< 0$} }\el\right.
 \\& 
      +  \dis \oint_{  L_{0+}  }    \frac{du} {(2\pi\I)^2} \oint_{  L_{0+}}dv \frac{u^{ -\tau_1}}{v^{\rho-\tau_2}}
\frac{e^{ \frac{u^2}2  -(\beta-y_1 )  u  }}
{e^{-\frac{v^2}2  +(\beta+y_2 )  v  }}
    \frac{ \Theta^+_{r-1}(  u, v )} {  \Theta_r(0,0)} 
    ~~~~~~~\mbox{always $\neq 0$}
 \\
 & -  \oint_{\Ga_0}\!\frac{du} {(2\pi\I)^2} \!\oint_{\Ga_0}\!dv 
 \frac{u^{ \rho-\tau_1}}{v^{ -\tau_2}}
\frac{e^{ -\frac{u^2}2  +(\beta+y_1 )  u  }}
{e^{ \frac{v^2}2  -(\beta-y_2 )  v  }}
    \frac{ \Theta^-_{r+1}( u,v  )} {  \Theta_r(0,0)},   ~~~ \mbox{$\neq 0$  only when}\left\{\bl & \tau_2< 0 \\&\mbox{and} \\&  \rho< \tau_1 \el\right.
\\  =:&    \sum_{k=0}^4  {\mathbb L}_k^{\mbox{\tiny dTac}}
 (\tau_1, y_1;\tau_2, y_2)
\el\ee
where  
 $\Ga_0=$ small circle about $0$  and 
$  L_{0+} =$  upgoing vertical line in $\BC$ to the right of $\Ga_0$. The Heaviside function was defined in (\ref{Heaviside}) and the functions $\Theta_\rk ,~ \Theta^\pm_\rk$ appearing in (\ref{Ldtac}) are as follows:  
%
 \be\bl\Theta_r( u,v)&:=\frac1{r!}\left[     
\prod_{\al=1}^r \oint_{  L_{0+}}\frac{e^{     w_\al^2-2\beta   w_\al}}{    w_\al  ^{\rho }}
 ~\left(\frac{v\!-\!w_\al}{u\!-\!w_\al}\right) \frac{dw_\al}{2\pi \I}\right]\Dt_r^2(w_1,\dots,w_r)
 \\  \Theta^{\pm}_{r\mp1}(  u,v)     &:=\frac{1}{(r\mp1)!}\left[     
\prod_{\al=1}^{ r\mp 1  }\oint_{  L_{0+}}\frac{e^{     w_\al^2-2\beta  w_\al}}{    w_\al  ^{\rho }}
 ~\left( ({u\!-\!w_\al} )\ ({v\!-\!w_\al})\right)^{\pm 1} \frac{dw_\al}{2\pi \I}\right]
 \\
 &~~ ~~\qquad~ ~\qquad \qquad~~~~~~~~~~~~~~~~~~~~~~~~~~\Dt_{r\mp1}^2(w_1,\dots,w_{r\mp1}).
 \label{Theta} 
\el\ee

 \begin{theorem} \cite{AJvM1,AJvM2,AJvMskewAzt} \label{theorem3.1}
 We have that ${\mathbb L}^{\mbox{\tiny dTac}}_{\rk,\rho,\beta}   ( \tau_1, y_1 ;\tau_2, y_2)$ is the correlation kernel for a discrete-continuous determinantal point process on a doubly interlacing set $\dots \simeq {\bf y}^{(\tau)} \simeq {\bf y}^{(\tau+1)} \simeq\dots$ above. 
This kernel is invariant under the involution :
\be \label{involution}
\tau_1\leftrightarrow \rho-\tau_2\mbox{  and  }y_1\leftrightarrow -y_2.\ee
with ${\mathbb L}^{\mbox{\tiny dTac}}_1\leftrightarrow {\mathbb L}^{\mbox{\tiny dTac}}_2$ and with ${\mathbb L}^{\mbox{\tiny dTac}}_i$ being self-involutive for $i=3,4$.
\end{theorem}

\remark Notice ${\mathbb L}^{\mbox{\tiny dTac}}_{\rk,\rho,\beta}   ( \tau_1, y_1 ;\tau_2, y_2)$ contains the GUE-minor kernel, namely
$$\bl {\mathbb L}_0^{\mbox{\tiny dTac}}+{\mathbb L}_1^{\mbox{\tiny dTac}}=&
 {\mathbb L}^{\mbox{\tiny GUEminor}}
     ( \tau_1-\rho, \beta+y_1 ;\tau_2-\rho, \beta+y_2) 
  \\& +  \oint_{\Ga_0}\frac{du} {(2\pi\I)^2}\oint_{  L_{0+}}  \frac{ dv}{v-u}\frac{u^{\rho-\tau_1}}{v^{\rho-\tau_2}}
\frac{e^{-\frac{u^2}2  +(\beta+y_1 )  u  }}
{e^{-\frac{v^2}2  +(\beta+y_2 )  v  }}
      \frac{ \Theta_r( u, v )-\Theta_r(0,0)} { \Theta_r(0,0)}  
\el$$
\be\label{LLGUE}\bl {\mathbb L}_0^{\mbox{\tiny dTac}}+{\mathbb L}_2^{\mbox{\tiny dTac}}=&
 {\mathbb L}^{\mbox{\tiny GUEminor}}
     ( -\tau_2 , \beta-y_2 ;-\tau_1 , \beta-y_1) 
  \\& +  \oint_{\Ga_0}\frac{du} {(2\pi\I)^2}\oint_{  L_{0+}}  \frac{ dv}{v-u}\frac{u^{ \tau_2}}{v^{ \tau_1}}
\frac{e^{-\frac{u^2}2  +(\beta-y_2 )  u  }}
{e^{-\frac{v^2}2  +(\beta-y_1 )  v  }}
      \frac{ \Theta_r( u, v )-\Theta_r(0,0)} { \Theta_r(0,0)}  
\el\ee 
with each of the two  ${\mathbb L}^{\mbox{\tiny GUEminor}}$-kernels corresponding to the two single interlacing polytopes in Fig. 2. 
 Moreover, notice that the kernel heavily depends on the location of the $\tau_i$ vis-\`a-vis the strip $[0,\rho]$, as indicated next to the different terms in (\ref{Ldtac}).

\medbreak

A sketch of the proof of Theorem \ref{theorem3.1} will be given in later sections. Namely, the fact that the ${\mathbb L}^{\mbox{\tiny dTac}}_{\rk,\rho,\beta}$ is the  correlation kernel for a point process corresponding to a doubly interlacing set comes from tiling models. In particular, tilings of rectangular Aztec models \cite{AJvMskewAzt} or hexagon models \cite{AJvM1,AJvM2} with cuts lead to point processes forming doubly interlacing sets on $\BZ^2$, which then in the limit lead to doubly interlacing point processes  on $\BZ\times \BR$ for which the correlation is given by the discrete tacnode kernel ${\mathbb L}^{\mbox{\tiny dTac}}_{\rk,\rho,\beta}$ as in (\ref{Ldtac}). This will be sketched in sections 5 and 6.
%



\vspace*{1cm}
 
\noindent{\em  The discrete tacnode kernel for $\rho=r$. (\cite{ACJvM}  Theorem 1.4)} When $\rho=r$, the kernel $ {\mathbb L}^{\mbox{\tiny dTac}}_{\rk,\rho,\beta}   ( \tau_1, y_1 ;\tau_2, y_2) $, with $ \tau_i,y_i\in \BZ\times \BR$, simplifies to the kernel (\ref{2min0}) below. This is not seen by direct calculation, (i.e., by setting $\rho=r$ in the kernel (\ref{Ldtac})),  but by considering another model, as will be discussed in Subsection 5.2. The kernel is given by  
\footnote{The subscript $\geq -\rho$ refers to the space $\ell^2(-\rho,\ldots,\infty)$.}:
\be
\begin{aligned}
 {\mathbb L}^{\mbox{\tiny dTac}}_{\rho,\rho,\beta}   ( \tau_1, y_1 ;\tau_2, y_2) \Bigr|_{\rho=r}
=&
  \BL^ {\mbox{\tiny GUEminor}}  (\tau_1, -\beta-y_1 ;~\tau_2 , -\beta-y_2 ) \\&~~+ \Bigl\la (\Id - {\cal K}^\beta ( \lambda ,\kappa ))^{-1}_{\geq -\rho} ~{\cal A}^{\beta,y_1+\beta}_{\tau_1 }(\kappa), {\cal B}^{\beta,y_2+\beta}_{\tau_2 }(\lambda)\Bigr\ra
  _{_{ \geq -\rho }}
   \label{2min0}
   \end{aligned}\ee
  where $\BL^ {\mbox{\tiny GUEminor}}$ is defined in (\ref{Lminor}) and where  $\Ga_{\iy}$ and $\Ga_0$ below refer to a large circle about infinity and a small circle about $0$, whereas $L_{0\pm}$ are vertical lines in $\BC$ to the right and to the left of the origin, and where
  %
%
  \be\begin{aligned}
{\cal K}^\beta (\lambda,\kappa)&:=\oint_{\Gamma_0}  \frac{d\zeta} {(2\pi \I)^2 }\int_{ L_{0+}} \frac{d\om}{ \om-\zeta   }
   \frac{e^{- \zeta^2-2 \beta \zeta }} { e^{- \omega^2-2 \beta \omega }}
\frac{\zeta^{\kappa } }{\omega^{\lambda +1}} 
\\
   {\cal A}^{\beta,y_1 }_{\tau_1 }(\kappa)&:=  
 ~
  \oint_{\Gamma_\iy}  \frac{d\zeta}{ (2\pi \I)^2} 
   \!\! \int_{ L_{0+}} \frac{d\om}{\zeta\!-\! \om  }
  \frac{e^{-\frac{\zeta^2}{2}-   y_1  \zeta}} 
  {e^{-\om^2-2 \beta \om}}  
    \frac{\zeta^{-\tau_1 }}{\om^{\kappa +1}} 
\\
{\cal B}^ {\beta,y_2 }_{\tau_2}(\lambda)&:= 
  \oint_{\Gamma_0}  \frac{ d\zeta}{ (2\pi \I)^2}   \int_{ L_{0-}} \frac{d\om}{ \zeta\!-\!\om   }
  ~~\frac{e^{- \zeta^2-2\zeta \beta }} {e^{- \frac{\om^2}{2}- y_2\om       }} 
  \frac{\zeta^{\lambda  }}{\om^{-\tau_2  }}
%
 .\end{aligned}
  \label{defAB}\ee

  
\section{Probability distributions related to the discrete tacnode kernel}

\subsection{Joint probabilities for  the discrete tacnode kernel on doubly interlacing sets ($\rho\geq r$)}
We define the polytope (truncated cone) of doubly interlacing sets for given ${\bf z}= {\bf z}^{(\tau_1 )}$ and $ {\bf z}={\bf z}^{(\tau_2 )}$ (for notation, see (\ref{xy-interlace}) and (\ref{yinterlace}))
 \be\label{cone}
 \CR(\tau_1,{\bf z}^{(\tau_1)};\tau_2,{\bf z}^{(\tau_2)}):=\left\{\begin{array}{llllll}
 {\bf z}^{(\tau_1 )}\simeq {\bf z}^{(\tau_1+1)}\simeq \dots\simeq  {\bf z}^{(\tau_2-1)}\simeq  {\bf z}^{(\tau_2 )}, 
 \\ \mbox{given  } 
 {\bf z}^{(\tau_1 )} \mbox{   and  }{\bf z}^{(\tau_2 )}  
 \end{array}
 \right\}
 \ee
with uniform measure\footnote{$n_\tau $ defined in (\ref{ntau}).} on $\CR(\tau_1,{\bf z}^{(\tau_1)};\tau_2,{\bf z}^{(\tau_2)})$ (Lebesgue measure)
\be \label{unifmeas}d\mu_{{\bf z}^{(\tau_1)}{\bf z}^{(\tau_2)}}( {\bf z}^{(\tau_1+1 )}, \dots,  {\bf z}^{(\tau_2 -1)}))
=\left(\prod_{\tau_1< \tau< \tau_2}
 d{\bf z} ^{(\tau)}\right)
 \Id_{ {\bf z}^{(\tau_1 )}\simeq \dots\simeq{\bf z}^{(\tau_2 )}}
\mbox{   with  } d{\bf z}^{(\tau)}=\prod_{i=1}^{n_\tau} dz_i^{(\tau)}.\ee
The volume of $\CR(\tau_1,{\bf z}^{(\tau_1)};\tau_2,{\bf z}^{(\tau_2)})$ is then given by 
\be\label{vol}
 \mbox{Vol}(\CR(\tau_1,{\bf z}^{(\tau_1)};\tau_2,{\bf z}^{(\tau_2)}))=\int_{\CR(\tau_1,{\bf z}^{(\tau_1)};\tau_2,{\bf z}^{(\tau_2)})}   d\mu_{{\bf z}^{(\tau_1)}{\bf z}^{(\tau_2)}}( {\bf z}^{(\tau_1+1 )}, \dots,  {\bf z}^{(\tau_2 -1)})
 \ee
Define for ${\bf z}^{(\tau)}=(z_1,\dots,z_{n_\tau})$, the determinant of the following matrix of size $n_\tau$,  ($n_\tau$ defined in (\ref{ntau})), 
 :
\be\label{Delta}   \widetilde\Dt_{  }
   (\tau,{\bf z}  )
:=
C^{(r,\rho)}_\tau\det
 \left(\begin{array}{ccc}
  \left.~~~~\begin{array}{cl}z_i^0 
 \\ \vdots
 \\
 z_i^{\tau-\rho-1}\end{array}\right\}
 \\
 \Phi_{\tau-r }( z_i )&
 \\
 \vdots
 \\
 \Phi_{  \max(\tau ,0)
   -1 }( z_i )&
 \end{array}
 \!\!\!\!\!\!\!\!\right)_{1\leq i\leq n_\tau}
   \!\!\!\!\!\!\!\!\!\!\!\!\!\!\!\! \begin{array}{llllll}
 \\&\\&\\& \end{array} 
 \begin{array}{llllll}
 &\hspace*{-1.3cm}\Longleftarrow~\mbox{absent when} ~\tau-\rho\leq 0\\&\\&\\&
 \end{array}
\ee
%
with the understanding that when $\tau-\rho\leq 0$ the pure powers inside the bracket are absent 
 and 
where $\Phi_n (\eta)$ is a Gaussian-type integral along a vertical complex line, which reduces to Hermite polynomials for $k\leq -1$. The integral is given by:
\be \label{Phi}\begin{aligned}
 \Phi_k (\eta) &:=\frac{1}{2\pi \I} \int_{  L_{0+}} \frac{e^{v^2+2\eta v}}{v^{k+1}} dv  
  =\frac{2^{ k}}{ \sqrt{\pi}} 
  \times \left\{\begin{aligned}
  &   
  \int^\infty_0 \frac {\xi^k} {k!} e^{-(\xi -\eta)^2 } d\xi  \quad ,  k\geq 0
  \\ \\ &  e^{-\eta^2}H_{-k  -1}(-\eta)  \quad ,\quad k \leq -1
 \end{aligned}\right. \\ 
  \end{aligned}. \ee
where 
$ H_n=(2x)^n+\dots$ are Hermite polynomials. We also use the two representations for the Hermite polynomials, where $\Ga_0$ is a small circle about $0$ and $L$ a vertical line in $\BC$: 
\be\label{Herm}\bl
  H_j(x) &=
j!\oint _{\Ga_0}e^{-z^2+2xz}\frac{dz}{2\pi \I z^{j+1}} =
 2\sqrt{\pi} 2^{j }  \int_L e^{(w-x)^2 }\frac{w^jdw}{2\pi \I}
   =(2x)^j +\dots
  \el \ee

The following statements are improved versions of Theorem 1.2 and Corollary 1.2 in \cite{AvM2}: 



\begin{theorem}\label{Th:jointP}(Theorem 1.2 in \cite{AvM2})
The joint distribution of the $n_{\tau }$ dots along all levels $\tau$ in between and including $\tau_1$ and $\tau_2$ for  $0\leq \tau_1\leq \tau_2\leq \rho$, 
 for $\rho\leq\tau_1 \leq \tau_2 $ or for $\tau_1\leq   \tau_2\leq 0$ (i.e., the two levels $\tau_i$ belong to the same region, either inside the strip $[0,\rho]$, to the right or to the left of it) is given by
\be\label{multiP}\bl\BP& \left( \bigcap_{\tau=\tau_1}^{\tau_2} \left\{{\bf z} ^{(\tau  )} \in d{\bf z} ^{(\tau )}
\right\} \right)=   c ^{r,\rho}_{\tau_1} c ^{r,\rho}_{\rho- \tau_2}\\&
  \times \widetilde\Dt(\tau_1,\tfrac{{\bf z} ^{(\tau_1)}-\beta}{\sqrt{2}})
  \widetilde\Dt(\rho-\tau_2 ,\tfrac{-{\bf z} ^{(\tau_2)}-\beta}{\sqrt{2}})
 d{\bf z} ^{(\tau_1)}d{\bf z} ^{(\tau_2)}d\mu_{{\bf z} ^{(\tau_1)}{\bf z} ^{(\tau_2)}}
 \left( {\bf z} ^{(\tau_1+1)},\dots,{\bf z} ^{(\tau_2-1)}
 \right),\el\ee
where $ c ^{r,\rho}_{\tau }$ is a $\tau$-dependent constant depending on  the parameters $r$ and $\rho$; for the constant, see \cite{AvM2}, formula (11). 

\end{theorem}

\remark An expression for the probability above (\ref{multiP}) is still unknown for $\tau_1<0$ and $\tau_2>\rho$, except for the case $\rho=r$, where it is known for any $\tau_1\leq \tau_2$; see Theorem \ref{GenTh} in the next section.

\begin{corollary}\label{cor:jointP}In the range of validity of Theorem \ref{Th:jointP}, the following holds:
\be\label{jointP}\bl
\BP&\left(\begin{array}{l} {\bf z}_1^{(\tau_1)}\in d{\bf z}_1   \mbox{   and  }  {\bf z}_2^{(\tau_2)}\in d{\bf z}_2 \end{array}\right)
 \\&~~=    c ^{r,\rho}_{\tau_1} c ^{r,\rho}_{\rho- \tau_2}\widetilde\Dt(\tau_1,\tfrac{{\bf z}_1^{(\tau_1)}-\beta}{\sqrt{2}})
  \widetilde\Dt(\rho-\tau_2 ,\tfrac{-{\bf z}_2^{(\tau_2)}-\beta}{\sqrt{2}})
\mbox{Vol}(\CR(\tau_1,{\bf z}_1;\tau_2,{\bf z}_2))
d{\bf z} _1
d{\bf z}_2,
\el\ee
 while, for any $\tau\in \BZ$,  
\be\label{singleP}\bl
\BP \left(  {\bf z} ^{(\tau )}\in d{\bf z} \right)
=  c ^{r,\rho}_{\tau } c ^{r,\rho}_{\rho- \tau }
 \widetilde\Dt(\tau ,\tfrac{{\bf z} ^{(\tau )}-\beta}{\sqrt{2}})
  \widetilde\Dt(\rho-\tau  ,\tfrac{-{\bf z} ^{(\tau )}-\beta}{\sqrt{2}})d{\bf z}.
 \el\ee
\end{corollary}

\begin{corollary}\label{cor:Pinduct} ({\em marginals})  ~~For all $\tau$, 
 we have that for fixed $  {\bf z} ^{(\tau +1)}$, the integral of $\widetilde\Dt(\tau,.)$
\be\label{Pinduct}\int_{\BR^{n_\tau}}  d{\bf z} ^{(\tau  )}\widetilde\Dt(\tau ,\tfrac{{\bf z} ^{(\tau )}-\beta}{\sqrt{2}})\Id_{ {\bf z} ^{(\tau )}\simeq 
{\bf z} ^{(\tau +1)} }
=\frac{c ^{r,\rho}_{\tau+1}}{  c ^{r,\rho}_{ \tau }}
\widetilde\Dt(\tau +1,\tfrac{{\bf z} ^{(\tau +1)}-\beta}{\sqrt{2}})
\ee
\end{corollary}

\noindent{\em Sketch of Proof of Corollary \ref{cor:jointP} and Theorem \ref{Th:jointP}:} 
One first proves Corollary \ref{cor:jointP} and then Theorem \ref{Th:jointP}. From the table in Fig. 3, we have that for 
$$\bl
\rho\leq \tau_1,\tau_2 &\Longrightarrow 
{\mathbb L}_i^{\mbox{\tiny dTac}}\neq 0 \mbox{ for } i=0,1,3
\\ 0\leq \tau_1\leq \tau_2\leq \rho &\Longrightarrow 
{\mathbb L}_i^{\mbox{\tiny dTac}}\neq 0 \mbox{ for } i=0, 3
\\  \tau_1, \tau_2\leq 0&\Longrightarrow 
{\mathbb L}_i^{\mbox{\tiny dTac}}\neq 0 \mbox{ for } i=0, 2,3
\el$$
Here it is convenient to use representation (\ref{LLGUE}) for that part of the kernel (\ref{Ldtac}).
The integrands of the expressions $\Theta_r(u,v)-\Theta_r(0,0)$ (in (\ref{LLGUE})) and $\Theta_{r-1}^+(u,v)$ contain products of the type 
\be\frac{1}{v-u}\left(\prod_1^r\frac{v-w_\al}{u-w_\al}-1\right)\mbox{ and } 
\prod_1^r(v-w_\al)(u-w_\al)
,\label{prod}\ee
which we expand in $u$ and $v$. One then applies the $u,v$ integrations termwise, enabling one to express ${\mathbb L}_i^{\mbox{\tiny dTac}}$ for $i=1,3$ as bilinear functions in the integrals $\Phi_k$, as in (\ref{Phi}), which depending on the sign of $k$ reduces to Hermite polynomials or integrals. 

  Given the  $n_i:=n_{\tau_i}
$ points at level $\tau_i$ (remember (\ref{ntau})), it is well known that the following probability can be expressed in terms of the discrete tacnode kernel $\widetilde {\mathbb L}^{\mbox{\tiny dTac}}$, given in (\ref{Ldtac}), as a determinant of a matrix of size $n_1+n_2$, which due to the previous considerations can thus be written as 

\be \label{Prob}\bl\BP&\left(\begin{array}{l} y^{\tau_1}_{1,i} \in dy_{1,i}\mbox{, belonging to level $\tau_1$,  for } 1\leq i\leq n_1,\\ y^{\tau_2}_{2,i} \in dy_{2,i}\mbox{, belonging to level $\tau_2$, for } 1\leq j\leq n_2\end{array}\right)
\\& =: p(\tau_1,{\bf y}_1;\tau_2,{\bf y}_2)\prod_{i=1}^{n_1} d y_{1,i}  \prod_{j=1}^{n_2} d y_{2,j}
\el\ee
with density $p( \tau_1,{\bf y}_1;\tau_2,{\bf y}_2)$, expressible as the determinant of a square matrix of size $n_1+n_2$ in terms of the correlation kernel. Using the bilinearity explained after formula (\ref{prod}), this matrix of size $n_1+n_2$ has a very special structure, which for  $0\leq\tau_1<\tau_2$ reads as below; so, we have 
\be \label{22determ} \bl p(&\tau_1,{\bf y}_1;\tau_2,{\bf y}_2)
\\&=
\det \left(
\begin{array}{cccccc}
\left(
  {\mathbb L}^{\mbox{\tiny dTac}}
 (\tau_1 , y_{1,i} ;\tau_1 , y_{1,j})\right)_{1\leq i,j\leq n_1}
&
\left(   {\mathbb L}^{\mbox{\tiny dTac}}
 (\tau_1 , y_{1,i} ;\tau_2 ,y_{2,j})\right)_{{1\leq i \leq n_1}\atop{1\leq  j\leq n_2}}
 \\
 \left(  {\mathbb L}^{\mbox{\tiny dTac}}
 (\tau_2 , y_{2,i} ;\tau_1 , y_{1,j})\right)_{{1\leq i \leq n_2}\atop{1\leq j\leq n_1}}
&
\left(   {\mathbb L}^{\mbox{\tiny dTac}}
 (\tau_2 , y_{2,i} ;\tau_2,y_{2,j}  )\right)_{{1\leq i,j\leq n_2} }
\end{array}\right)
 \\&=2^{n_1+n_2}\det\left(
   \begin{array}{cccc}( A_1^\top(y_{1,i}) B_1(y_{1,j}))_{1\leq i,j\leq n_1} &
   (A_1^\top(y_{1,i}) B_2(y_{2,j}))_{{1\leq i \leq n_1}\atop{1\leq  j\leq n_2}}
 \\
 (A_2^\top(y_{2,i}) B_1(y_{1,j}))_{{1\leq i \leq n_2}\atop{1\leq j\leq n_1}} & ( A_2^\top(y_{2,i} B_2(y_{2,j}))_{1\leq i,j\leq n_2}
 \end{array} \right)
 \\
 &=2^{n_1+n_2}\det \left(\left\la\AR_\al,  {\mathcal B}_\beta\right\ra\right)_{1\leq \al,\beta\leq n_1+n_2}
 =\det\left( \AR \right)
 \det\left( {\mathcal B} \right)
 \el
 \ee
where  $A_i$ and $B_i\in \BC^{n_1+n_2}$ are column-vectors  and where $\AR:=(\AR_\al)_{1\leq \al\leq n_1+n_2}$ is a column of row-vectors and 
${\mathcal B}=({\mathcal B}_\beta)_{1\leq \beta\leq n_1+n_2}$ a row of column-vectors (both ${\mathcal A}$ and ${\mathcal B}$ can be interpreted as square matrices of size $n_1+n_2$):
\be\label{ABscrip}\AR 
 :=\left(\begin{array}{ccc}
 A_1^\top (y_{1,1}) \\ \vdots \\ A_{1}^\top (y_{1,n_1})\\ \\ A_2^\top (y_{2,1})\\ \vdots\\ A_2^\top (y_{2, n_2})
 \end{array}\right)
~\mbox{and}~ {\mathcal B}  
 =\left(B_1(y_{1,1}),\dots,B_1(y_{1,n_1}) ,
 B_2(y_{2,1}),\dots,B_2(y_{2,n_2}) \right).
 \ee
 It turns out that both  ${\mathcal A}$ and  ${\mathcal B}$ are block matrices, with one of the blocks being a zero-matrix $O_{n_1,n_2}$. This enables us to compute these matrices, giving formulas (\ref{jointP}). The one level case involves the upper-left block, and so the Volume-part is not present, thus proving Corollary \ref{cor:jointP}.
  
  Then Theorem \ref{Th:jointP} follows from Corollary \ref{cor:jointP} by the Gibbs property:
  $$\bl
  \BP& \left( \bigcap_{\tau=\tau_1+1}^{\tau_2-1} \left\{{\bf z} ^{(\tau  )}\in d{\bf z} ^{(\tau )}
\right\} \Bigr| {\bf z} ^{(\tau_1)}={\bf z}_1,~{\bf z} ^{(\tau_2)}={\bf z}_2\right)
=\frac{d\mu_{{\bf z}_1{\bf z}_2}( {\bf z}^{(\tau_1+1 )}, \dots,  {\bf z}^{(\tau_2 -1)})}{ \mbox{Vol}(\CR(\tau_1,{\bf z}^{(\tau_1)};\tau_2,{\bf z}^{(\tau_2)}))},\el$$
  which itself follows from it being true for the discrete lozenge model, whose limit leads to the discrete tacnode kernel sketched in Section 6.
   \qed
 
\noindent{\em Proof of Corollary \ref{cor:Pinduct}}:  We apply (\ref{multiP}) to the case $\tau_1=\tau$ and $\tau_2=\tau+1$; notice that then $[\tau_1,\tau_2]$ is either inside the strip or outside the strip, including its boundary, which is the range of validity of Theorem \ref{Th:jointP}. Then we have:
$$\bl
 \BP& \left( {\bf z} ^{(\tau   )}\in A,  {\bf z} ^{(\tau+1  )}\in d{\bf z } ^{(\tau+1 )}
   \right)
 \\&= c ^{r,\rho}_{\tau} c ^{r,\rho}_{\rho- \tau-1}  \int_A  d{\bf z} ^{(\tau  )}\widetilde\Dt(\tau ,\tfrac{{\bf z} ^{(\tau )}-\beta}{\sqrt{2}}) \Id_{ {\bf z} ^{(\tau )}\simeq 
 {\bf z} ^{(\tau +1)} } 
 \widetilde\Dt(\rho-\tau-1 ,\tfrac{-{\bf z} ^{(\tau +1)}-\beta}{\sqrt{2}}) d{\bf z} ^{(\tau +1)}  
 \el$$
For $A=\BR^{n_\tau}$, the left hand side above equals $\BP  \left(    {\bf z} ^{(\tau+1  )}\in d{\bf z} ^{(\tau +1)}
 \right) $, which, using (\ref{singleP}),  equals
$$\bl& \BP  \left(    {\bf z} ^{(\tau+1  )}\in d{\bf z} ^{(\tau +1)}
 \right) 
 =    c ^{r,\rho}_{\tau+1} c ^{r,\rho}_{\rho- \tau-1} \widetilde\Dt(\tau\! +\!1,\tfrac{{\bf z} ^{(\tau +1)}-\beta}{\sqrt{2}}) 
   \widetilde\Dt(\rho\!-\!\tau\!-\!1 ,\tfrac{-{\bf z} ^{(\tau+1)}-\beta}{\sqrt{2}})
 d{\bf z} ^{(\tau +1)} 
\el$$
Comparing the two formulas above for $A=\BR^{n_\tau}$, we have (\ref{Pinduct}), thus ending the proof of Corollary \ref{cor:Pinduct}.\qed

\newpage


\begin{figure}
\hspace*{.1cm}
 \setlength{\unitlength}{0.015in}\begin{picture}(0,170)
  \put(200,-50){\makebox(0,0){\includegraphics[width=210mm,height=275mm]{Fig4
  .pdf}}}    
  
  
    \end{picture}  
\end{figure}
\vspace*{2cm}


{\small Figure 4. Doubly interlacing system of points in $\BR$ for $\rho=r(=5)$, consisting of two singly interlacing sets ${\bf x}^{(n)}$ and ${\bf y}^{(n)}$, with $\rho$ constraints given by the inequalities. This system has a symmetry about an axis running through the middle of the picture from northwest to southeast.}

\subsection{Joint probabilities for  the discrete tacnode kernel on doubly interlacing sets for $\rho=r$ }
We now consider the following doubly interlacing system, as in Fig. 4, or, in short, ``{\it a double cone}\footnote{This is not be confused with the customary definition of a double cone: two cones placed apex to apex; the two cones of the "double cone" as defined in this paper rather have some overlap.}" , given arbitrary ${\bf x}={\bf x}^{(n)}={\bf z}^{(n)}$ and ${\bf y}={\bf y}^{( n)}={\bf z}^{(\rho-n)}\in \BR^n$,
%
 %
   \be
   {\mathcal C}^{(n)}_{{\bf x},{\bf y}} =\left\{\begin{array}{l}  {\bf x}  \succ {\bf x}^{(n-1)} \succ\ldots \succ    {\bf x}^{(1)}\mbox{  for any ${\bf x}^{(i)}\in \BR^i$ }
   \\
   {\bf y} \succ {\bf y}^{(n-1)}\succ \ldots \succ {\bf y}^{(1)} \mbox{  for any ${\bf y}^{(i)}\in \BR^i$}\\
    \mbox{and for $1\leq i\leq \rho$, subjected to}
   \\
   \max  ({\bf y}^{(i)})={  y}^{(i)}_i\leq {  x}^{(\rho-i+1)}_1=\min  (  {\bf x}^{(\rho-i+1)}) \\ 
 \end{array} \right\} \subset \BR^{n(n-1)} ,\label{cone'}\ee
 with independent GUE-minor probabilities on each of the single cones. So we can view the set ${\bf x}  \succ {\bf x}^{(n-1)} \succ\ldots \succ    {\bf x}^{(1)}$ as the interlacing eigenvalues of the consecutive minors of a GUE-matrix $A^{(n)}$ and $ {\bf y} \succ {\bf y}^{(n-1)}\succ \ldots \succ {\bf y}^{(1)}$ as the interlacing set corresponding to the GUE-matrix $B^{(n)}$, where $A$ and $B$ are independent. So, we set
  \be\label{z,x,y}\bl{\bf z}^{(\tau)}&=\left\{\bl &{\bf x}^{(n)} \mbox{ for } \tau=n\geq \rho
 \\ & {\bf x}^{(\rho-i)}\cup 
 {\bf y}^{(i)}\mbox{ for }1\leq\tau=\rho-i\leq \rho-1
 \\&{\bf y}^{(n)}\mbox{ for } \tau=\rho-n\leq 0\el\right. .
 \el\ee





\noindent 
 %

We consider the cone 
$ \CR(\tau_1,{\bf z}^{(\tau_1)};\tau_2,{\bf z}^{(\tau_2)}),$
as in (\ref{cone}), uniform measure 
$d\mu_{{\bf z}^{(\tau_1)}{\bf z}^{(\tau_2)}}( {\bf z}^{(\tau_1+1 )}, \dots,  {\bf z}^{(\tau_2 -1)})
,$
as in (\ref{unifmeas}), and the volume
$ \mbox{Vol}(\CR(\tau_1,{\bf z}^{(\tau_1)};\tau_2,{\bf z}^{(\tau_2)})),$
as in (\ref{vol}). We will show that the probabilities on this interlacing set can be viewed as two independent GUE-minor distributions conditioned by the $\rho$ inequalities in the last line of the cone (\ref{cone'}). Theorem  \ref{GenTh} below is a considerable improvement of Theorem 1.3 in \cite{AvM1}.

\begin{theorem} \label{GenTh}
When $r=\rho\geq 1$, the joint probability of the positions $z^{(\tau)}$ of the dots, as in (\ref{z,x,y}), for arbitrary levels $  \tau_1\leq \tau\leq  \tau_2 $, (i.e., no restriction on the range) is given by expression (\ref{multiP}); to wit 
\be\label{multiP'}\bl\BP& \left( \bigcap_{\tau=\tau_1}^{\tau_2} \left\{{\bf z} ^{(\tau  )}\in d{\bf z} ^{(\tau )}
\right\} \right)=  c ^{\rho,\rho}_{\tau_1} c ^{\rho,\rho}_{\rho- \tau_2}\\&
 \times   \widetilde\Dt(\tau_1,\tfrac{{\bf z} ^{(\tau_1)}-\beta}{\sqrt{2}})
   \widetilde\Dt(\rho-\tau_2 ,\tfrac{-{\bf z} ^{(\tau_2)}-\beta}{\sqrt{2}})
 d{\bf z} ^{(\tau_1)}d{\bf z} ^{(\tau_2)}d\mu_{{\bf z} ^{(\tau_1)}{\bf z} ^{(\tau_2)}}
 \left( {\bf z} ^{(\tau_1+1)},\dots,{\bf z} ^{(\tau_2-1)}
 \right).\el\ee
 For $\tau_1\leq 0\leq \rho\leq \tau_2=\rho-\tau_1$, the constant reads, using the notation (\ref{ntau}) for the number $n_\tau$ of dots belonging to the level $\tau$, and using the kernel ${\mathcal K}^{\beta}$
 in (\ref{defAB}),  \be\label{cc}
 c ^{r,\rho}_{\tau_1} c ^{r,\rho}_{\rho- \tau_2}
 =\frac{\sqrt{2}^{n_{\tau_1}(n_{\tau_1}+1)/2   }
\sqrt{2}^{n_{\tau_2}(n_{\tau_2}+1)/2} }
{\det(\Id- {\mathcal K}^{\beta}
 )_{[-1, -\rho]}},
\ee
with (see notation in Fig. 4 and (\ref{z,x,y}))
\be\label{Kineq}\det(\Id- {\mathcal K}^{\beta} )_{[-1, -\rho]}=\BP^{\mbox{\tiny GUE}} \left(\bigcap _1^{\rho}\{y_i^{(i)}\leq  x_{1}^{(\rho-i+1)} \right). \ee
 \end{theorem}

 \newpage
 
 \vspace*{-.9cm}

 \noindent {\em Sketch of Proof} :
  At first, 
  one takes $\tau_1$ and $\tau_2$ on either side of the strip, i.e. $\tau_1\leq 0$ and  $\tau_2\geq \rho$ and equidistant from the strip; that is $\tau_2=\rho-\tau_1\geq \rho$. From  (\cite{AvM1}, Theorem 1.2), we obtain, upon using in $\stackrel{**}{=}$ the expressions $\widetilde \Dt$ defined in (\ref{Delta}), and using the density $\rho_{\tau }^{ \mbox{\tiny minor}}$ defined in (\ref{GUE-dens}) :
    \be\begin{aligned}
\BP&\Bigl( \bigcap _{\tau=\tau_1}^{\tau_2} 
 {\bf z}^{(\tau)}\in d{\bf z}^{(\tau)}
 \Bigr)  
  \\& \stackrel{*}{=}\frac{\rho_{\rho-\tau_1}^{ \mbox{\tiny minor}}(-{\bf z}^{(\tau_1)}\!+\!\beta) 
   \rho_{\tau_2}^{\mbox{\tiny minor}}( {\bf z}^{(\tau_2)}\!+\!\beta)}
 {\det(\Id- {\mathcal K}^{\beta}
 )_{[-1, -\rho]}
 }
  { d{\bf z}^{(\tau_1)}d{\bf z}^{(\tau_2)} d\mu_ {{\bf z}^{(\tau_1)},{\bf z}^{(\tau_2)}} } 
 \\&\stackrel{**}{=}
 c ^{r,\rho}_{\tau_1} c ^{r,\rho}_{\rho- \tau_2} \det\left(\!\!\!\begin{array}{cccc}&\Phi_{\tau_1-\rho}(\frac{ {\bf z}^{(\tau_1)}_i -\beta}{\sqrt{2}})
\\&\vdots\\&\Phi_{-1}(\frac{ {\bf z}_{ i}^{(\tau_1)}-\beta}{\sqrt{2}})
\end{array}\right)_{1\leq i\leq n_{\tau_1}}\!\!\det \left(\!\!\!\begin{array}{cccc}&\Phi_{-\tau_2 }(\tfrac{-{\bf z}_{ i}^{(\tau_2)}-\beta}{\sqrt{2}})
\\&\vdots\\&\Phi_{ -1}(\tfrac{-{\bf z}_{ i}^{(\tau_2)}-\beta}{\sqrt{2}})
\end{array}\right)_{1\leq i\leq n_{\tau_2}}  
\\&~~~~~~~~~~~~~\times { d{\bf z}^{(\tau_1)}d{\bf z}^{(\tau_2)} d\mu_ {{\bf z}^{(\tau_1)},{\bf z}^{(\tau_2)}} } 
\\&\stackrel{***}{=}
 c ^{r,\rho}_{\tau_1} c ^{r,\rho}_{\rho- \tau_2}\widetilde\Dt(\tau_1,\tfrac{{\bf z} ^{(\tau_1)}-\beta}{\sqrt{2}})
  \widetilde\Dt(\rho-\tau_2 ,\tfrac{-{\bf z} ^{(\tau_2)}-\beta}{\sqrt{2}})
 d{\bf z} ^{(\tau_1)}d{\bf z} ^{(\tau_2)}d\mu_{{\bf z} ^{(\tau_1)}{\bf z} ^{(\tau_2)}}
 .\end{aligned}
 \label{15}\ee
The equality $\stackrel{*}{=}$ above follows from the fact that for $\tau_1\leq 0\leq \rho\leq \tau_2$, the probability (\ref{15}) can be computed using the explicit form of the kernel (\ref{2min0}) for $\rho=r$. That same probability can also be computed by noticing that for $\rho=r$, a doubly interlacing set of points reduces to two singly interlacing systems, subjected to $\rho$ inequalities indicated by the arrows in Fig. 4; therefore the probability on the left hand side of (\ref{15}) can be expressed as the product of two GUE-minors probabilities (\ref{Bary1'}), conditioned by the inequalities $ y_i^{(i)}\leq  x_{1}^{(\rho-i+1)}$. Therefore by comparing the two expressions, we have established (\ref{Kineq}).
  
The equality $\stackrel{**}{=}$ follows from the observation that, 
since the indices $k=\tau_1-\rho<0$ and $k=-\tau_2<0$, all the $\Phi_{-k}$'s, as in (\ref{Phi}), reduce to Hermite functions: 
$$\Phi_{-k}(\frac{\eta}{\sqrt2})=
2^{-k}\frac{e^{-\frac{\eta^2}{2}}}{\sqrt{2\pi}}\sqrt 2 H_{k-1}(-\frac{\eta}{\sqrt 2})=(\sqrt 2)^{-k}\frac{e^{-\frac{\eta^2}{2}}}{\sqrt{2\pi}}((-  \eta)^{k-1}+\dots).
$$
So, elementary row operations show that these determinants involving the $\Phi$'s are the same as the GUE-minor densities $ \rho_{k}^{\mbox{\tiny minor}} $ as in  (\ref{Bary1'}), up to a multiplicative constant as in (\ref{cc}). The determinants in $\stackrel{**}{=}
$ of (\ref{15}) are then recognized to be the 
$\widetilde \Dt$'s defined in (\ref{Delta})
for the range $\tau_1\leq 0\leq \rho\leq \tau_2$, leading to the expression $\stackrel{***}{=}
$ in (\ref{15}).

Finally, using (\ref{Pinduct}) of Corollary \ref{cor:Pinduct}, we go from $\tau_1\to \tau_1+1$, by integrating out $z^{(\tau_1)}$ (i.e., taking the marginal) and doing this over and over again, thus shrinking the interval $[\tau_1,\tau_2]$. Once $\tau_1$ enters the strip $[0,\rho]$, the constant (\ref{cc}) will become more complicated. This establishes expression  (\ref{multiP'}) in the range $-\iy<\tau_1\leq \tau_2<\iy$, with $\rho\leq \tau_2$. Using the involution (\ref{involution}), the formula (\ref{multiP'}) will hold for any $\tau_1\leq \tau_2$, establishing Theorem \ref{GenTh}. 
\qed

\section{Domino tilings of skew-Aztec rectangles and double interlacing}
 
The purpose of sections 5 and 6 is to discuss two different tiling models, both leading to a system of doubly interlacing dots and the discrete tacnode kernel. This section deals with domino tilings of a so-called {\em skew-Aztec rectangle} or {\em Aztec rectangle} for short. To explain the model, a standard Aztec diamond, as in Fig. 5(a), has 
 the property that the two upper-most adjacent squares and the lower-most adjacent squares have {\em opposite orientation}, and the same for the two left-most squares and the two right-most squares. We now consider a rectangular model $\DR$, as in Fig. 5(b), where the top two and bottom two adjacent squares of the rectangle have the {\em same orientation} (blue-white) and the two most
 left and two most right as well ($\mbox{\footnotesize blue}\atop\mbox{\footnotesize white}$). This forces the model to have two indentation (``cuts'', for short) on each of the two sides; they can be viewed as nonconvexities. 
 See Fig. 5(a) and 5(b) for a regular Aztec diamond versus a skew-rectangular Aztec diamond.  

 Let $\pl \DR_U,\pl \DR_R,\pl \DR_L,\pl \DR_D$  denote the {\bf U}pper-left, upper-{\bf R}ight, lower-{\bf L}eft and lower-right ({\bf D}own) sides of the rectangle; see Fig. 5(b). The rectangle has width $n=\#\{\mbox{blue squares}\}>0$ along $\pl D_U$ and length $ m+M\geq 1$ along $\pl \DR_R$, with $m=\#\{\mbox{white squares}\}$ and  $M=\#\{\mbox{blue squares}\}$ along the side $\pl \DR_R$, separated by two ``cuts", which are taken symmetric about the middle of the rectangle along the sides $\pl \DR_R$ and $\pl \DR_L$.  By symmetry, $m$ is also the number of blue squares and $M$ the number of white squares along the $\pl \DR_L$-side. Assume $m\geq 0$ and $M\geq 1$.

 We need two different systems of coordinates on the skew-Aztec rectangle (with different origins): $(\xi,\eta)$ and $(s,u) $, as in Fig. 5(c), related by 
  \be(\xi,\eta)\to (s,u)=\left( 
    \eta\! +\!1 ,~\tfrac 12 (\eta\!-\!\xi \!+
   \!1)\right), \mbox{with} ~-1\leq\xi\leq 2(m\!+\!M),~ -2\leq \eta\leq 2n+1\label{changevar}\ee

\newpage

 \newpage







\newpage
\vspace*{2cm}\hspace*{0cm}
\setlength{\unitlength}{0.015in}\begin{picture}(0,170)

\put(160,-25)
 {\makebox(0,0){\includegraphics[width=190mm,height=254mm]{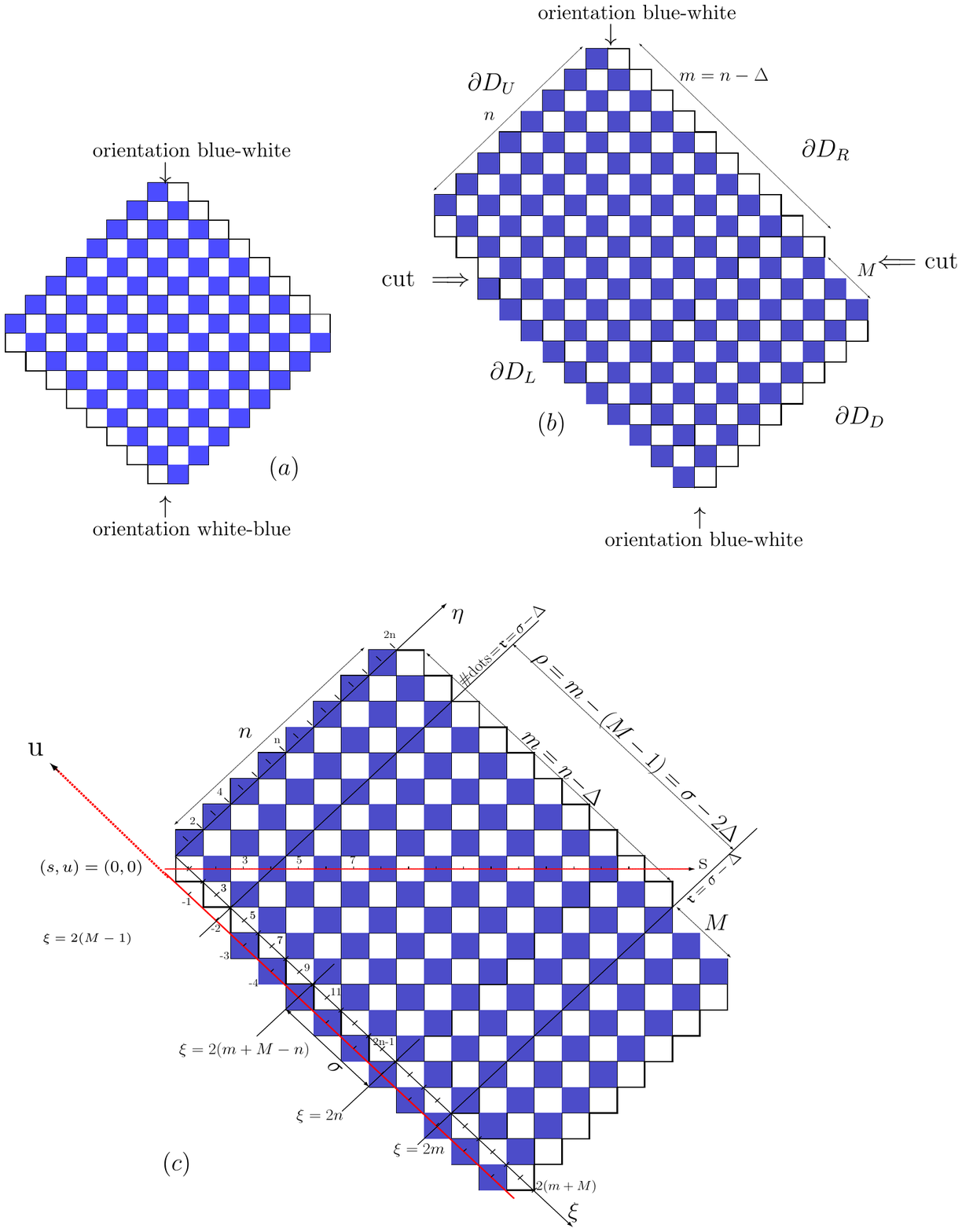}}}
 \end{picture}

\vspace*{8cm}
{\small Figure 5: A standard Aztec diamond (a) , with opposite orientation top and bottom (also left and right); a skew-Aztec rectangle (b) with equal orientation top and bottom (also left and right). Here $n=8,~m=10, ~M=3,~\Dt=-2,~\rho=8,\sg=4,~r=6$, with $(\xi,\eta)$- and $(s,u)$-coordinate-systems and strip $\{\rho\}$ in (c).}

\newpage
 
 %

\newpage
\newpage

The middle of the most left boundary of the diamond serves as origin of the $(\xi, \eta)$ system, while the origin of the $(s,u)$-system is at  $(\xi, \eta)=(0,-1)$.  The middle of the blue squares are at $(\xi,\eta)\in 2\BZ\times (2\BZ+1)$, with $0\leq \xi\leq  2(m+M-1) $ and $-1\leq \eta\leq 2n-1$. In $(s,u)$ coordinates, they are at $(s,u)\in 2\BZ\times \BZ $, with $0\leq s\leq 2n$ and $-(m+M-1)\leq u\leq n $. One defines two crucial integers $\Delta=-\ka$,
\be\Dt:=-\ka:=n-m\in \BZ, \label{Dt}\ee 
 and $\sigma$ (the second formula in (\ref{sg}) follows at once from (\ref{Dt})) 
 \be \label{sg}
 \sg:=n-(m+M-n)+1=n-M+\Dt+1\in \BZ.
  \ee
 
The integer $\Dt$ measures how much the Aztec rectangle differs from two overlapping regular Aztec diamonds. The integer $\sg$ measures the ``amount of overlap", if one were to ignore the cuts and view the Aztec rectangle as two overlapping diamonds.

{\em The strip $\{\rho\}$} is defined by 
the the $\rho +1$ parallel lines $\xi \in 2\BZ$ (i.e., through the black squares) between the lines $\xi=2(M-1)$ and $\xi=2m$; so, the strip is bounded by the two lines passing through the cuts and has width (see Fig.5(c))
\be \label{rho}
  \rho:=|m-(M-1)|.
  \ee  

  {\em Domino tilings, Height functions, level curves, point processes and double interlacing}. 
    Consider a covering of the skew-Aztec rectangle by dominos, horizontal ones with the blue square to the left or to the right $(H_L,~H_R)$ and vertical ones with the blue square on the upper-side or on the lower-side $( V_U, ~V_D)$, as in Fig. 1(b). 
      Let each domino carry an appropriately chosen height function $h$, as indicated in Fig. 1(b),  
       such that the red lines are its level curves of height $h+\tfrac 12$. 

 Therefore any such domino covering 
    produces a set of $n+m$ red  (nonintersecting) level curves of successive heights $\tfrac12, \tfrac32, \tfrac52,...$,  
     extending all the way from the left edge of the blue boundary squares along $\pl \DR_U$ and $\pl \DR_L$ to the right edge of the white boundary squares along $\pl \DR_R$ and $\pl\DR_D$, with a {\em specified boundary condition}. 
 
    Indeed, the heights along the boundary of the rectangle (indicated by the integers in Fig. 6.) are completely specified, since 
    the heights of the left-outer side of each blue square or the right-outer side of each white square, as in Fig. 1(b), always increases by $1$ from top to bottom, 
     whereas the heights of the left-outer side of each white square or the right-outer side of each blue square, as in Fig. 1(b), always remains constant.

\newpage

\vspace*{1.6cm}
\hspace*{ 0cm}
\setlength{\unitlength}{0.015in}\begin{picture}(0,170)

 \put(160,30){\makebox(0,0){\includegraphics[width=180mm,height=214mm]{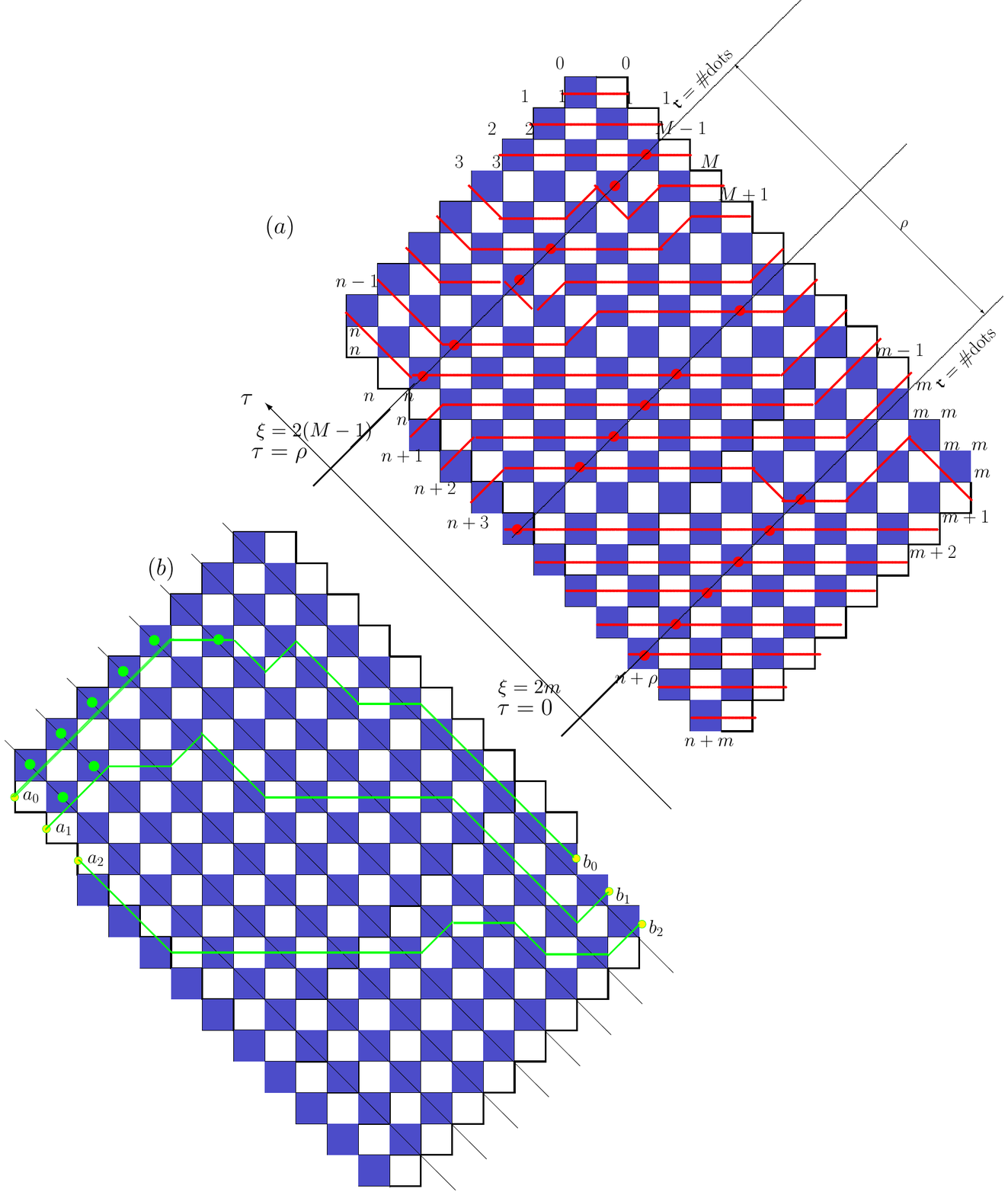}}}
 
\end{picture}

\vspace*{5cm}

 {\small Figure~6. (a): Tiling of the Aztec rectangle Fig.5(c), together with the associated red level curves, yielding  the point process $\PR_{\tiny{red}}$ of red dots along the lines $\xi\in 2\BZ$. The integers along the boundary give the heights. The lines $\xi =2(M-1), ~\xi = 2m$ (or $\tau=\rho$ and $\tau=0$) are the boundaries of the strip $\{\rho\}$. The tiling in (b) leads to the point process $\PR_{\tiny{green}}$, given by the intersection of the green level curves (departing from $M$ contiguous points) with the lines $\eta\in \BZ_{\tiny{odd}}$.}

\newpage

 
 \newpage

\newpage

   \vspace*{1.3cm}
 
\setlength{\unitlength}{0.015in} \begin{picture}(0,0)

\put(144,-200) 
 {\makebox(0,0){\includegraphics[width=219mm,height=278mm]{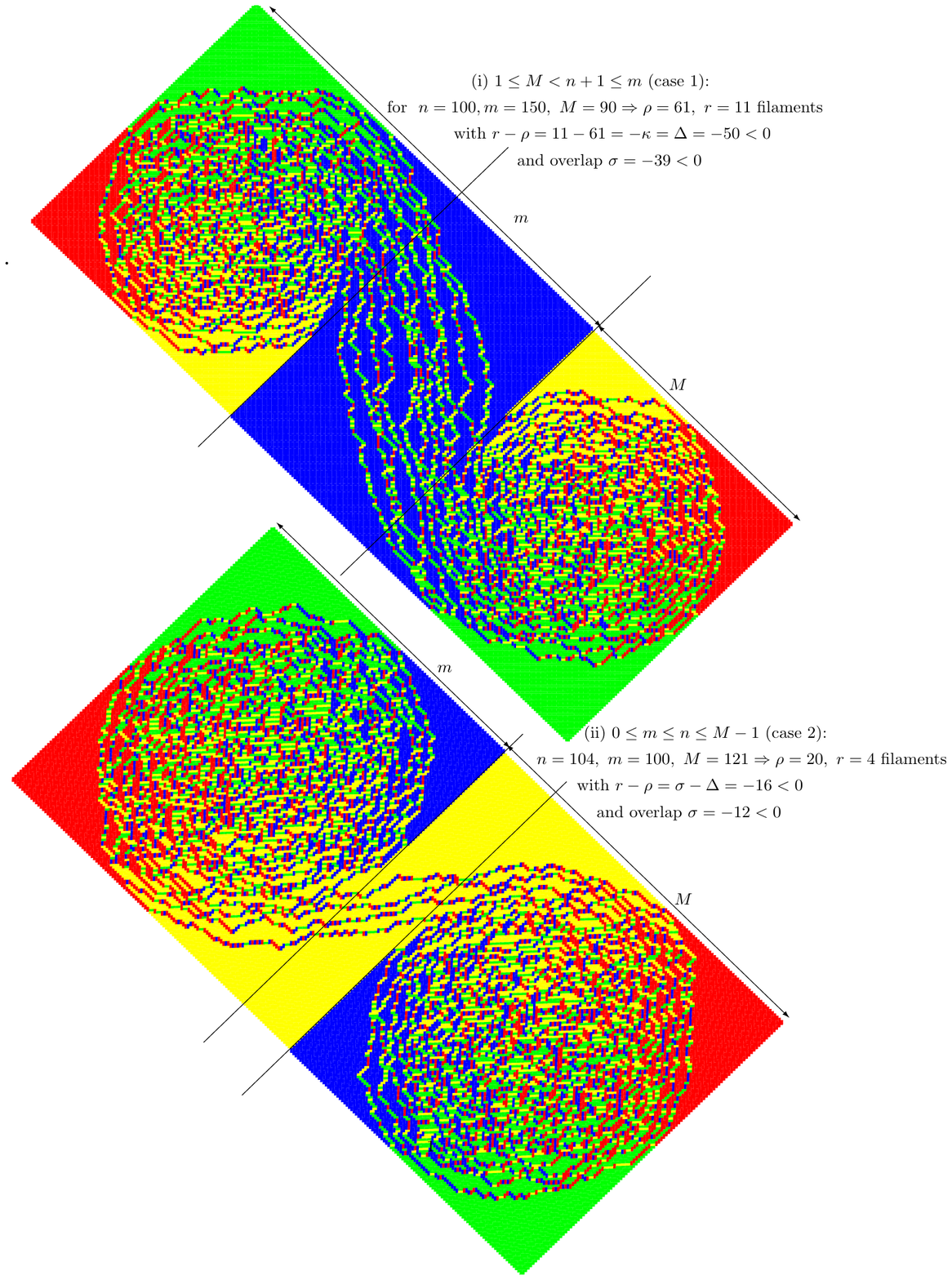}}}  

\end{picture}

 \vspace*{15cm}
  
 {\footnotesize  Figure 7 . Simulation of random tilings of an Aztec rectangle for $1\leq M< \min(m,n+1) $  
   and for $1\leq m+1< \min(M,n+1) $  (cases 1 and 2 of Theorem \ref{tilability}), with two different values of $n,~m,~M$ with $r$ number of filaments and strip $\{\rho\}$ of size $\rho$. (Courtesy of Sunil Chhita)}
\newpage


 
 

 

%


 
 

 

%
 \newpage

       The {\em point process  $\PR_{{\mathcal R}ed}$} is defined  by the set of successive lines $\xi\in 2\BZ$ equipped with the {\em red dots} assigned to the middle of the blue square each time the line  $\xi\in 2\BZ$ intersects the level curves in that square; see the dominos on top of Fig. 8 (III) for individual dominos and Fig. 6(a) for an example of a tiling of an Aztec rectangle.


   One also shows that the difference of the heights, between the extremities, and measured along the parallel lines $\xi\in 2\BZ$  (i.e., through the blue squares) within the strip $\{\rho\}$, 
  always equals a fixed number $r$ to be specified. In contrast, outside the strip, 
  that difference of heights (between the extremities) goes up by $1$ each time one moves one step to the left of the strip ($\xi\to\xi-2$) or one step to the right ($\xi\to\xi+2$) of the strip $\{\rho\}$. So, the number of intersection points of the parallel lines (within the strip) with the consecutive level lines  (whose heights increase each time by $1$) is given by that same number $r$ and so
\be \label{rk}r:=\mbox{$\#\{$red dots along the lines $\{\xi\in 2\BZ$ $\}$ $\overbrace{\mbox{between $\xi=2(M\!-\!1)$ and $\xi=2m$} }^{^{\mbox{\tiny strip}~\{\rho\}}}$  $ \}$}.
\ee 
How these numbers $r$ and $\rho$ will be expressed in terms of the geometrical data of  the model will be discussed in Theorem \ref{tilability}.
In the situation of Fig. 5(c) one is dealing with $r=n-M+1$ red dots along the $\rho+1=m-M+2$ parallel lines $\xi \in 2\BZ$  within the strip $\{\rho\}$, bounded  by $2(M-1)\leq\xi\leq 2m$.

The integer $r$ in (\ref{rk}) can be expressed in terms of the geometric data $n,~m,~M$, in the same way as formula (\ref{rho}) for $\rho$; however the formula for $r$ will depend on the situation, as will be seen in Theorem \ref{tilability} below.

Finally, the point process, called $\PR_{{\mathcal R}ed}$, formed by the system of successive lines $\xi\in 2\BZ$ equipped with the red dots is a {\em doubly interlacing system}, as defined in (\ref{xy-interlace}) with $\rho$ and $r$ as in section 3, by letting $n,m,M$ grow large, keeping $n-M$ and $m-M$ finite.

\bigbreak
     
   {\em Tilability}.  Such Aztec rectangles may not always be tilable; they will be in the cases below. As mentioned before, we shall express  $\rho, ~r$ defined in (\ref{rho}) and (\ref{rk}), in terms of the basic geometrical data $n, ~m,~M$. 
      
      \medbreak

    \begin{theorem} (\cite{AJvMskewAzt}, Theorem 2.1)\label{tilability}
    The skew-Aztec rectangle  is tilable if and only if  
$$ \bl \mbox{ Case 1:} &~~~  1\leq ~~~M ~~~ \leq ~~~\min (m    ,n+1   )   
\\ 
\mbox{or  Case 2:}&~~~1\leq ~ m+1~\leq ~~\min (M ,n+1 ).
\el$$
The associated point process $\PR_{\mathcal R ed}$ consists of $r$ (red) dots on each of the $\rho+1$ lines $\xi \in 2\BZ$ within the strip $ \{\rho\}$ of width $\rho$, with the number of dots per line increasing one-by-one on either side of the strip, up to $n$, where 
 \be \label{rho,r} 
 \bl \rho&= |m-(M-1)| \\&= |\sg-2\Dt |\el
~~~ \mbox{and}~~~
 \bl r
  &=\max (n-(M-1), ~n-m)\\&=\max(\sg-\Dt,\Dt).\el 
   \ee
   
   \end{theorem}
   
To be more specific, we have
 \newline Case 1. $ \rho=m-(M-1)=\sg\!-\!2\Dt\geq 1$ ~and ~ $r=n\!-\!(M\!-\!1)=\sg\!-\!\Dt\geq 0$, \newline \hspace*{1cm}with $ r-\rho=n-m = \Dt$ 
  \newline  Case 2.  $\rho=(M-1)-m=2\Dt- \sg \geq  0$ and ~ $r=n-m=\Dt\geq 0$, \newline \hspace*{1cm}with $r-\rho= n-M+1 =\sg-\Dt$
 \vspace*{-.4cm} \be\label{rho,r'}\ee 
 
 {\em Sketch of Proof of Theorem \ref{tilability}}: It relies on the nonintersecting path description in the $(s,u)$ coordinates on the rectangle, as in Figs. 5(c). An equivalent path description appears in Fig. 8. The details of the proof can be found in \cite{AJvMskewAzt}, Theorem 2.1.\qed
 
 \subsection{Aztec rectangles with $1\leq M\leq  n+1\leq m$ and $\Dt=r-\rho\leq 0$  }
 The present paper deals with {\em Case 1} and $\Dt=r-\rho\leq 0$. For the notational convenience, we will often set $\ka:=-\Dt$.

 Putting the customary probability
 \be\label{Prob}\footnotesize {\mathbb P}(\mbox{domino tiling $T$})=
 \frac{a^{\#\mbox{vertical domino's in $T$}}}{\sum_{\mbox{all possible tilings in $T$}}a^{\#
  \mbox{vertical domino's in $T$}}}\ee
  on the tilings by giving the weight $0<a\leq 1$ on vertical dominos and weight $1$ on horizontal dominos, 
one obtains a point processes  
 $\PR_{{\mathcal R}ed}$ of 
  red dots, depending on $\xi\in 2\BZ$; an example is in Fig. 6. 
  The purpose of this paper will be to show that $\PR_{{\mathcal R}ed}$ is a determinantal point process, with kernel 
$\BK_{ n,r,\rho}^{^{\textrm{\tiny red}}}
 (\xi_1,\eta_1;\xi_2,\eta_2)
 $, which will be given in rescaled form in (\ref{Lquadr}). 
An appropriate  asymptotic limit of the kernel $\BK_{ n,r,\rho}^{^{\textrm{\tiny red}}}(\xi_1,\eta_1;\xi_2,\eta_2)$, after rescaling, will be the discrete tacnode kernel ${\mathbb L}^{\mbox{\tiny dTac}}_{r,\rho,\beta}( \tau_1, y_1 ;\tau_2, y_2) $ as in (\ref{Ldtac}).  

Fig. 7 contains two different {\em simulations} of domino tilings of a skew-rectangle (for $a=1$), one for case 1 and another for case 2.

\medbreak

We now consider the following {\em discrete-continuous} rescaling $(\xi,\eta)\in \BZ^2\to (\tau,y)\in (\BZ,\BR)$, 
  together with a rescaling of the parameter $a$, appearing in the probability (\ref{Prob}). We will let  $n,m,M\to \infty$, keeping $r=n-M+1$ and $\rho=m-M+1$ and an extra-parameter $\beta$ fixed. So we have a map from the $(\xi,\eta)$-variables to the new  rescaled ones
   $(\xi,\eta)\in \BZ_{\mbox{\tiny even}} \times \BZ_{\mbox{\tiny odd}}  \to (x,y)\in  \BZ\times \BR $, or, in view of the change of variable (\ref{changevar}), $(s,u)\in \BZ_{\mbox{\tiny even}} \times \BZ  \to (x,y)\in  \BZ\times \BR $;  namely
   for $t\to 0$, we have\footnote{Since $s_i\in \BZ_{\mbox{\tiny even}}$, it is convenient to introduce $s_i'=s_i/2$. Also, since the middle of the blue squares are given by $\eta\in \BZ_{\mbox{\tiny even}}$, we automatically have $\Dt\eta=2$ } 
%
\be\label{scaling}
n=\frac{1}{t^{ 2}},~a=1+\frac{\beta}{ \sqrt n }  ,~  \xi_i :=  2(n-x_i),~~~
\eta_i :=  n+  y_i\sqrt{2n}   -1,~\frac {\Dt \eta}2= dy \sqrt{\frac{n}{ 2}} .
\ee
\be \label{sr}\bl
   s_i=2s'_i &= \eta_i +1 =  \frac 1{t^2}+\tfrac{y_i\sqrt{2}}{t}  ,~~u_i=\tfrac 12
 (\eta_i-\xi_i+1)=\tfrac 12 (-\frac 1{t^2}+\frac{y_i\sqrt{2}}{t}+2x_i)
 \el \ee
 At a much later stage, we will further replace\footnote{Expressed in terms of $\tau_i$, we have $\xi_i=2(m- \tau_i)$ in (\ref{scaling}).} (remember $\ka=m-n$)
 \be\label{scaling'}x_i\to  \tau_i=x_i+\ka,~~y_i\to y_i'=y_i\sqrt2 
 .\ee
 The latter amounts to rescaling about the point $(\xi,\eta)=(2m,n)$, which is the middle of the right hand boundary of the strip $\{\rho\}$; see Fig. 6(a).

 %
 We now define the following functions of $u$ in the rescaled variables $(x_i,y_i)$, with $s_i$ given by (\ref{sr}); all of them have nice limits, upon letting $t\to 0$:
 \be\bl\label{fncts}
 h(u)&:= (1+a^2tu )^{n}
(1-tu)^{n+1 }=e^{-u^2+2\beta u}(1+O(t)),
\\ \FR(u ) &:=
 u^{ -x_1}(1+a^2tu )^{s'_1-1}
(1-tu )^{n-s'_1} = u ^{-x_1}e^{-\frac 12 u ^2+(\beta+y_1\sqrt{2})u }(1+O(t)) 
 \\
\GR_{ }(u ) &:=
 u ^{-x_2-\ka}(1+a^2tu )^{s'_2}
(1-tu)^{n+1-s'_2}=u ^{-x_2-\ka}e^{-\frac 12 u^2+(\beta+y_2\sqrt{2})u }(1+O(t))
\\\Phi(v)&:=\oint_{\ga_{\sg_2^+}}\frac{h(u)du}{2\pi \I \GR(u)(v-u)}
 \mbox{   , for $|v|<\sg_2^+$.  (radius $\sg_2^+$ slightly larger than $\sg_2 $)}
\el\ee
 Also define, in terms of the Vandermonde $\Dt_r$, an integral about the boundary of the annulus $\ga_{\sg_2}-\Ga_0$, (see the radii, just after (\ref{Lsca}))
 \be\label{Omega}\bl   \Om (u,v  ) &:= \left(
\prod_{\ell=0}^{r-1}\oint_{\ga_{\sg_2}-\Ga_0}\frac{du_\ell  }{2\pi \I u_\ell^{\ka+\ell+1}  h(u_\ell) }\frac{ u_\ell-v}{u_\ell-u}
\right)\Dt_r(u_0,\dots,u_{r-1})
.\el\ee

%
%

\newpage

\newpage

\vspace*{1.8cm}

\setlength{\unitlength}{0.015in}\begin{picture}(0,0)

 \put(170,-120){\makebox(0,0){\includegraphics[width=180mm,height=266mm]{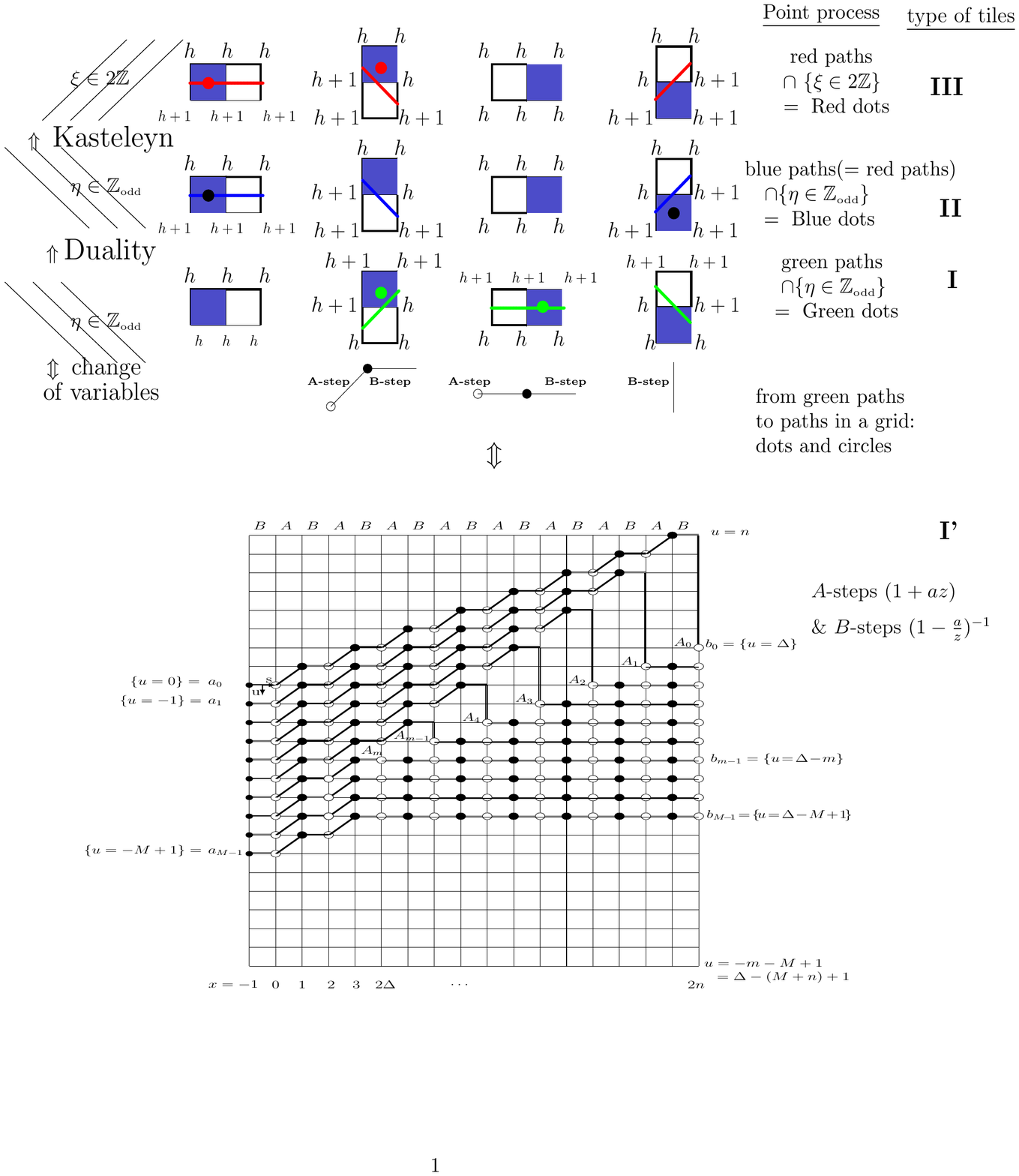}}}

\end{picture}

\vspace{13.8cm}

{\small Figure 8. From bottom to top, a tiling by dominos with the green level curves; at their intersection with the lines $\eta\in \BZ_{\tiny\mbox{odd}}$, we put  green dots  (${\bf  I}$); they are equivalent to the zig-zag paths on the graph in $(s,u)$ coordinates (${\bf  I'}$). Then by duality, one obtains the tiles with blue level curves, intersected by the same $\eta\in \BZ_{\tiny\mbox{odd}}$, giving blue dots (${\bf  II}$). Upon intersection of the same level curves by the lines $\xi\in \BZ_{\tiny\mbox{even}}$, one obtains the point process of red dots (${\bf  III}$). }

 \newpage


\newpage

\newpage

\begin{theorem}(\label{theo:Kred}\cite{AJvMskewAzt}, Proposition 5.1) The point process of red dots $\PR_{\mathcal R ed}$ is determinantal with a kernel $\BK_{ n,r,\rho}^{^{\textrm{
 red}}}
  (\xi_1,\eta_1 ;\xi_2,\eta_2)$ given by  $ {\mathbb L}^{^{\textrm{\tiny red}}}_{n,r,\rho}
(x_1,y_1;x_2,y_2)$ in the rescaled variables and  after conjugation
  \be\label{Lquadr}\bl
 \BK_{n,r,\rho}^{^{  \textrm{\tiny red}}}  &(\xi_1,\eta_1 ; \xi_2,\eta_2) 
\frac{\Dt \eta_2}{2}
  = (-1)^{s'_2-s'_1 } a^{u_2-u_1}t^{x_2-x_1 }
  {\mathbb L}^{^{\textrm{\tiny red}}}_{n,r,\rho}
(x_1,y_1;x_2,y_2)
\sqrt{2}dy_2\tfrac{ 1+a^2 }{  2}
,\el\ee
where  (the $s'_i,u_i$ appearing on the right hand side of (\ref{Lquadr}) must be thought of as given by (\ref{sr}); i.e., expressed in terms of the $(y,x)$)
\be\bl  \label{Lsca}
 {\mathbb L}^{^{\textrm{\tiny red}}}_{n,r,\rho}
&:=-  {\mathbb L}_0 
+(-1)^\ka   ( {\mathbb L}  _1+{\mathbb L}  _2)
\\&:= \Id_{x_1>x_2 }\oint_{\ga_{\rho_1}}\frac{\FR(u) du}{2\pi \I u^\ka \GR(u)}
\\
&~~~  -\!\oint_{\ga_{\rho_2}}\!\! \frac{u^r \FR(u)du}{2\pi \I}\!\! \oint_{\ga_{\sg_2}-\Ga_0} \frac{ v^{-\rho}\Phi(v)dv }{2\pi \I (u\!-\!v)h(v) }  \frac{\Om (u,v)}{\Om (0,0) }
 -\oint_{\ga_{\rho_2}} \frac{\FR(u)du}{2\pi \I  u^{\ka}\GR(u)}
 \el\ee
with kernels in terms of the functions (\ref{fncts}) and (\ref{Omega}) and in terms of circles $\ga$ about $0$ of radii $a< \rho_1<\sg_1 <\sg_2
 <\rho_2
  <a^{-1}$, and a circle $\Ga_0$ within $\ga_{\sg_2}$.

\end{theorem}

\noindent{\em The proof of Theorem \ref{theo:Kred}}, appearing in Section 4 of \cite{AJvMskewAzt} consists of several steps, which will be sketched below: 

{\bf Three different point processes $\PR{\mbox{\tiny red}}, ~\PR{\mbox{\tiny blue}}, \PR{\mbox{\tiny green}}$}. At first, Fig. 8 contains three different dominos equipped with different height functions. The dominos of type III are the same as in Fig. 1(b) and lead to the desired point process $\PR{\mbox{\tiny red}}$; i.e., the point process of lines $\xi\in 2\BZ$ and red dots given by the intersections of the lines $\xi\in 2\BZ$ with the red level curves (see Fig. 6(a)). The dominos of type II are the same as before (same dominos, same height function), except that the point process $\PR_{\mbox{\tiny blue}}$ of blue dots along $\eta \in \BZ_{\mbox{\tiny odd}}$ is given by the intersection of the same level curves, as before, with the lines $\eta \in \BZ_{\mbox{\tiny odd}}$. Type I describes a dual set of dominos;   {\em duality} amounts to the simultaneous interchanges of the colours of the squares and the height functions: (as depicted in the tiles going from Step II to Step I in Fig. 8) 
$$\mbox{blue square}\leftrightarrow\mbox{white square}
,~~~~~h\leftrightarrow h+1.$$
This duality also amounts to rotating the dominos, height function and level curves of type II by $180^o$ giving the configuration of type I in Fig. 8. Intersecting the green level curves with the lines $\eta\in \BZ_{\mbox{\tiny odd}}$ gives us the point process $\PR_{\mbox{\tiny green}}$. Close inspection would show that to a green level curve for the point process $\PR_{\mbox{\tiny green}}$, there is a blue level curve filling up the tiles which were not visited by $\PR_{\mbox{\tiny blue}}$. 
 
 The main difficulty of the red path in Fig. 6(a) is that  they depart from the left edge of squares along different  sides of the domain $\DR$, namely $\pl \DR_U$ and $\pl \DR_L$; the red paths end up at different sides as well: $\pl \DR_R$ and $\pl \DR_D$. In other terms, they are not synchronized!
 
 To remedy this difficulty, we consider   dominos of type {\bf I} (Fig. 8) instead. From their level curves, it is easy to see that the green paths of type {\bf I} in Fig. 8 can only begin at the left-edge of the $M$ white squares along $\pl \DR_L$ and can only end up at the the right edge of the $M$ blue squares along $\pl \DR_R$. {\em So, the main point of this duality is that the green level curves start and arrive at a contiguous set of squares along $\DR_L$ and end up at a contiguous set of blue squares along $\DR_R$ as depicted in Fig. 6(b)}!

Remembering the change of variables (\ref{changevar}) $(\xi,\eta)\to (s,u) $, we now express the green level curves of type I in $(s,u)$-coordinates and replace them  by $A$-steps and $B$-steps with white circles and black circles, corresponding respectively to a white square and a black square, as is illustrated by type I' in Fig. 8. 

Notice that between the contiguous set of starting and the contiguous set of end points there is a shift $u=\Dt=-\ka=n-m=r-\rho$ (see (\ref{Dt})), as is clearly seen in Fig.8({\bf I'}). In the formulas later, we will express the A-steps by $(1+az)$ and $B$-steps by $(1-\frac az)^{-1}$; this will reflect itself  in the transition functions $\psi_{s_1,s_2}(z)$ given in (\ref{transitionf}).

\vspace{1cm}

\noindent {\bf The point process $\PR_{\mbox{\tiny green}}$ is determinantal with kernel ${\mathbb K}^{\textrm{green}}$}. 
 Denote by $D_\al(f)$ the Toeplitz determinant of an analytic function $f$; namely,
\be\label{Toeplitz}D_\al(f)=\det\left(\hat f_{i-j}\right)_{1\leq i,j\leq \al},
~~\mbox{with}~~~  \hat f_k=\oint_{S^1}\frac{dz}{2\pi \I z}\frac{f(z)}{z^k}.\ee
From an analytical point of view, the starting point for the construction is the following proposition: 

  \begin{proposition}~~(Johansson \cite{Jo16}) The paths in the $(s,u)$-grid in Fig. 8 or in other terms the  point-process $\PR_{green}$ of green dots is determinantal with kernel:
 \be\label{Kgreen}\bl
{\mathbb K}^{\textrm{green}}_{ } (s_1 ,y_1;s_2,y_2) 
&= - \Id_{s_1 < s_2}  p_{s_1;s_2} (u_1,u_2) + \widetilde {\mathbb K}^{\textrm{green}}_{ } (s_1,u_1;s_2,u_2)
\el\ee
where (with radii $a<\rho_1< \sg_1<\sg_2<
 \rho_2<a^{-1}$) 
\be\label{Kgreen'}
\bl
p_{s_1, s _2} (x,y)  &: =
  \int_{\ga_{\rho_1}} \zeta^{x-y} \vp_{s_1,s_2}(\zeta){d\zeta\over  2\pi \I \zeta}
\\ \widetilde {\mathbb K}^{\textrm{green}} (s_1 ,u_1;s_2,u_2) 
    &:=
    \int_{\ga_{\rho_1}} {z^{u_1}dz \over  2\pi \I z}
 \int_{\ga_{\rho_2}} \frac{w^{-u_2}dw}{  2\pi \I w} 
  \psi_{s_1,2n+1} (z) \psi_{0,s_2} (w) 
\\&~~~~~~~~~~~~~ \times {D_{M-1} \left[\psi_{0,2n+1} (\zeta)\left(1-{\zeta \over w}\right) \left(1-{z\over \zeta}\right) \right]   \over D_{M} \left[\psi_{0,2n+1}(\zeta)\right]},
 \el
\ee
 in terms of the transition functions (with $\psi_{s_1,s_2}$ involving the shift $-\Dt= \ka=m-n= \rho-r$)
  \be\label{transitionf}
\bl
\varphi_{s_1,s_2} (\zeta)&= 
{(1+a\zeta)^{[\tfrac {s_2}2]-[\tfrac {s_1}2]}  \over  (1-\frac a\zeta)^{s_2-[\tfrac {s_2}2]-
s_1+[\tfrac {s_1}2] }}  
~\mbox{and}~
\psi_{s_1,s_2}(\zeta)    
=\left\{ \bl  &\varphi_{s_1,s_2}(\zeta)\mbox{  ~~       for  }
s_2<2n+1
\\
&
\zeta^{-\Dt } ~\varphi_{s_1,s_2}(\zeta)\mbox{   for  } s_2=2n+1\el\right.,
\el\ee
thus leading in (\ref{Kgreen'}) to Toeplitz determinants $D_{M-1}$ and $D_M$ for a singular symbol.
  
\end{proposition}


\noindent {\em Sketch of Proof}:  The Lindstr\"om-Gessel-Viennot Theorem and a Eynard-Mehta-type argument for nonintersecting paths enables one to write down a kernel for the $M$ paths in  Fig. 8({\bf I'}); namely for $0<s_1,s_2<2n+1$, we have:
\be\bl\label{LGV} {K}^{ {green}} &(s_1  ,u_1;s_2,u_2)
\\=& -\Id_{s_1<s_2}\widehat \varphi_{s_1,s_2}(u_1,u_2) \sum_{i,j=1}^M \widehat \varphi_{s_1,2n+1}(u_1,1\! -\!  i+\Dt)(A^{-1})_{ij}\widehat \varphi_{0,s_2}(1\! -\! j,u_2)
,\el\ee
where $\widehat\varphi_{s_1,s_2}(u_1,u_2)$ is a Fourier transform 
and $A$ a matrix of size $M$:

 \be\bl \label{transFourier}
  \Id_{s_1<s_2}
  \widehat\varphi_{s_1,s_2}(u_1,u_2)  &  = \Id_{s_1<s_2}\oint_{\ga_{\rho_1}} \zeta^{u_1-u_2} \varphi_{s_1,s_2}(\zeta){d\zeta\over  2\pi \I \zeta}  
  \\ A=(A_{ij})_{ 1\leq i,j\leq M}~&\mbox{with} ~A_{ij} :=\widehat\varphi_{0,2n+1}(0,i-j+\Dt) 
  \el \ee
  Since the expression   $\varphi_{s_1,s_2}$   is a product of functions admitting a Wiener-Hopf factorization, the kernel (\ref{LGV}) can be written as a double integral involving a ratio of Toeplitz determinants as explained in \cite{Jo16}.\qed

 So, formula (\ref{Kgreen}) for ${\mathbb K}^{\textrm{green}}$ involves a {\em Toeplitz determinant of a singular symbol} (\ref{transitionf}), with a singularity $\zeta^{-\Dt}$ (special case of Fisher-Hartwig singularity). This can be dealt with using the following {\em key Lemma}, involving the resolution of the singularity $\zeta^\ka$: (remember $\ka=-\Dt=m-n$)
\be\label{plambda}
 \zeta^\ka~~\Longrightarrow ~~\prod_{i=1}^\ka (1- \frac{\zeta}{\lb_i} )=: p_{\bm\lb}(\zeta) ,~~\mbox{with}~\lb_i\neq 0.  \ee


\begin{lemma} \label{lambda-integration}The following holds:
%
%
$$\large
\mbox{$\dis D_n[\zeta^{\pm \kappa}f(\zeta)]
=(-1)^{\kappa n}    \oint_{(\Ga_0)^\kappa}
\prod_1^\kappa\frac{\lb_j^{n }d\lb_j}{2\pi \I \lb_j}
D_n \left[   {p_{ \lb }(\zeta^{\pm 1} })
 f(\zeta)\right].$}$$
 \end{lemma}
   \proof This is given in Lemma 4.2 of \cite{AJvMskewAzt}.\qed

\medbreak


 Then by the Case-Geronimo-Borodin-Okounkov formula (CGBO), we express the Toeplitz determinant in terms of a Fredholm determinant of another kernel; namely 
\be\label{Toepl}\bl  D_{M-1} & \left[\zeta^\ka\varphi_{0,2n+1} (\zeta)\left(1-{\zeta \over w}\right) \left(1-{z\over \zeta}\right) \right] =\frac{(-1)^{\ka (M-1)} (1-\frac aw)^{n+1}}{(1-\frac zw)(1+az)^n }(1+a^2)^{n(n+1)}
\\
&\times\oint_{(\Ga_0)^\ka}
\left[\! \prod_1^\kappa\frac{\lb_j^{M -2}d\lb_j}{2\pi \I  }\!\right] \frac{(p_{\lb}(a))^{n+1}}{p_{\lb}(z)}\det (\Id\!\!-\!\! {{\mathcal K}^{(\lb)}_{k,\ell}(w^{-1},z)}
   )_{\geq M-1} ,
\el\ee
where ${\mathcal K}^{(\lb)}_{k,\ell} (w^{-1},z)$ is a CGBO-type kernel ($\ga_{\sg_i}$ are circles of radius $\sg_i$ with $\sg_1<\sg_2$)
\be\label{KBO}\bl
{\mathcal K}^{(\lb)}_{k,\ell} (w^{-1},z) &:= \oint_{\ga_{\sg_1}}\frac{(-1)^{k+\ell}du}{(2\pi \I)^2} \oint_{ \ga_{\sg_2}  }\frac{dv}{(v-u)} \frac{u^{\ell}}{v^{k+1}} 
 \frac{(1-\frac uw)(1-\frac zv)}{(1-\frac vw)(1-\frac zu)} \frac{\rho_{\lb }( u)}{\rho_{\lb }( v)}
 \\&=
  {\mathcal K}^{(\lb)}_{k+1,\ell+1}(0,0)-(z-w)h_{k+1}^{(1)}(w^{-1},\lb)
  \frac 1z h_{\ell+1}^{(2)}(z,\lb),  \el\ee
 where  ($p_\lb(u)$ as in (\ref{plambda})):
\be \label{rholambda}\rho_{\lb }( u) :=\rho(u)p_\lb(u):=(1+au)^n(1-\frac au)^{n+1}
p_\lb(u).\ee
The expression ${\mathcal K}^{(\lb)}_{k,\ell} (w^{-1},z) $ in (\ref{KBO}) can then be written as 
 as a rank one perturbation of the kernel ${\mathcal K}^{(\lb)}_{k+1,\ell+1} (0,0) $ involving two new functions $h_{k+1}^{(i)}$ for $i=1,2$, yielding the second expression (\ref{KBO}); see Lemma 4.3 in \cite{AJvMskewAzt}. 
Substituting formula (\ref{Toepl}) in the expression (\ref{Kgreen'}) leads to an expression for the kernel  $  {\mathbb K}^{\textrm{green}}$  
  of the determinantal process of green dots; its precise formula is given in Proposition 4.5 of \cite{AJvMskewAzt}.

 \bigbreak

 \noindent \noindent {\bf The point process $\PR_{\mbox{\tiny blue}}$ is determinantal with kernel ${\mathbb K}^{\textrm{blue}}$, using duality between   $\PR_{\mbox{\tiny green}}$ ({\bf I}) and $\PR_{\mbox{\tiny blue}}$ ({\bf II}) in Fig. 8:} As mentioned before, the domino configurations in steps I and II in Fig. 8 are dual to each other. 
  It is known that the kernel for the dual process of blue dots is then given by 
\be\label{Kblue}
\BK^{ {blue}} (n,u_1 ; n, u_2) = \Id_{\{u_1=u_2\}} - \BK^{ {green}}(n,u_1;n,u_2).
\ee
for a fixed level $s_1=s_2=n$, which by appropriate conjugation, using a semi-group property, can be extended to any levels 
$s_1,s_2$, yielding a kernel $\BK^{ {blue}}  ( s_1  ,u_1; s_2  ,u_2)$; its expression can be found in \cite{AJvMskewAzt}, Lemma 4.6.

\bigbreak 

\noindent {\bf Going from $\PR_{\mbox{\tiny blue}}$ ({\bf II}) to $\PR_{\mbox{\tiny red}}$ ({\bf III}):} : 
 As already pointed out, the height functions and thus the (blue and red) level curves are the same. The only difference is that in case II, the point process $\PR_{blue}$ is generated by the intersection points of the level curves with the lines $\eta\in \BZ_{\mbox{\tiny odd}}$, whereas in case I the point process $\PR_{red}$ is generated by the intersection points of the level curves with the lines $\xi\in  \BZ_{\mbox{\tiny even}} $.
 %
 
  So, going from {\bf II} to {\bf III}, can be dealt with by using a {\em fortunate fact} that  already was observed in \cite{ACJvM}. Namely representing the tiling by a dimer model, where each white (blue) square gets replaced by a white (blue) dot of coordinates $(w_1,w_2)$ ($(b_1,b_2)$) and a link each time a domino covers a white and blue square. Then the Kasteleyn matrix is the adjacency matrix for that dimer model, which in this instance is nearest neighbour (see \cite{Kast,Kast1}).  Then we state the following Proposition: 
  
  \begin{proposition} (\cite{AJvMskewAzt}, Proposition 4.8 and Lemma 4.9) The inverse Kasteleyn matrix coincides with the blue kernel $\BK^{blue}$:
  \begin{equation}
\bl
  K_{Kast}^{-1}& \Bigl((w_1,w_2),(b_1,b_2)\Bigr)  = -(-1)^{(w_1-w_2+b_1-b_2+2)/4}
\\
&  \times\BK^{blue}
   \left( {b_2+1   } ,  {b_2-b_1+1  \over   2} ,  {w_2 + 1  },   {w_2-w_1+1  \over   2} \right)
.\el
\end{equation}
This then leads to (remember $s=2s'$ from (\ref{sr}))
\be\label{Kred}
 \bl \BK_{ n,r,\rho}^{^{\textrm{
 red}}}
  (\xi_1,\eta_1&;\xi_2,\eta_2)
    \\& = \BK^{blue}_{ } \ \left( 2s'_2,u_2 ; 2s'_1\!+\!1,u_1  \right)
\!  -a ~ \BK^{blue}_{} \ \left(2s'_2,u_2 ; 2s'_1\!-\!1,u_1\!-\!1\right)
.\el   
\ee
To wit, if $(2s'_1,u_1)$ (or $(\xi_1,\eta_1)$ in $(\xi,\eta)$ coordinates) is the middle of a blue square in Fig. 5(c), then $(2s'_1,u_1)+(1,0)$ and $(2s'_1,u_1)-(1,1)$ are the middle of the two adjacent white squares, the one to the right and the one just below. 
\end{proposition} 

\noindent{\em Sketch of Proof}: Since $K_{Kast}$ is a nearest-neighbor matrix, it suffices to show that $\BK^{blue}$ satisfies a difference equation, given by $K_{Kast}\BK^{blue}\sim \Id$; first in the bulk and then along the boundary; see section 2.2 in \cite{ACJvM} or Proposition 4.8 in \cite{AJvMskewAzt}.

  Kenyon's Theorem \cite{Ke,B } enables us then to express the correlation kernel $\BK^{red}$ in terms of the Kasteleyn kernel and its inverse $K_{\mbox{\tiny Kast}}^{-1}$, which involves the blue kernel evaluated at the two adjacent white squares to the right and below the black square.  \qed

 \vspace{1cm}
\noindent{\em Proof of Theorem \ref{theo:Kred}}:
 To summarize, in order to obtain the kernel $\BK_{ n,r,\rho}^{^{\textrm{
 red}}}
  (\xi_1,\eta_1 ;\xi_2,\eta_2)$, use the kernel $\BK^{green}_{}$ in {\bf I} (Fig.8), which for {\bf II}(Fig.8) leads to $\BK^{blue}_{}$ and then further to $ \BK_{ n,r,\rho}^{^{\textrm{
  red}}}$ for the tiling {\bf III}(Fig.8). 

  So, one uses expression (\ref{Kgreen}) for $\BK^{green}_{}$, which contains a Toeplitz determinant, given by expression (\ref{Toepl}). Then one uses the extended version (\ref{Kblue})  of $\BK^{blue}_{}$, with  the expression of $\BK^{green}_{}$ substituted. Finally, $\BK^{red}_{}$ expressed in terms of $\BK^{blue}_{}$, as in (\ref{Kred}) of Proposition 5.5, gives us, after some further work, the following expression:
  %
  %
 %
  \be 
\bl
\BK_{ n,r,\rho}^{^{\textrm{\tiny red}}}
 (\xi_1,\eta_1;\xi_2,\eta_2) 
  & = \frac{\widetilde\LR}{\widetilde\LR(1)} (-{\mathbb K}_0^{(\lb)}+(-1)^\ka {\mathbb K}_1^{(\lb)})(\xi_1,\eta_1 ; \xi_2,\eta_2)
,\el\label{Lkernel}
\ee
where $ {\mathbb K}_1^{(\lb)}$ is a perturbation of a kernel $S_1^\lb$ with an inner-product between vector-functions, $a_u^{s}(k)$ and $b_u^{s}(k)$; one of them being acted upon with a resolvent of the kernel ${\mathcal K}^{(\lb)}_{k,\ell}$, as in (\ref{KBO}),
 \be \bl
  {\mathbb K}_0^{(\lb)}&=(1+a^2)  \Id_{\xi_1 < \xi_2} \int_{\Ga_{0,a}} {dz  \over  2\pi \I}  z^{(\xi_1-\xi_2)/2}   
{(1+az)^{ \frac{\eta_1-\eta_2}2-1 }  \over   (z-a)^{ \frac{\eta_1-\eta_2}2+1 }}
\\  
 {\mathbb K}_1^{(\lb)}&=  
 \left[
  \bl &S_1^\lb(2s'_2,u_2 ; 2s'_1\!-\!1,u_1\!-\!1) 
  \\&+ \left\langle \left(I- {\mathcal K} ^{(\lb)}(0,0)\right)^{-1}_{\geq M} a_{u_1-1,1}^{(2s_1-1)} , b^{(2s_2)}_{u_2} \right\rangle_{\geq M}
  \el \right],\label{Li}
  \el\ee
  
  \noindent The operator $\widetilde \LR$ in (\ref{Lkernel}), again containing the same kernel (\ref{KBO}), reads as follows \footnote{$g_a=(1+a^2)^{n(n+1)}$}:  \be \bl\label{Ltilde}\widetilde \LR (f):=
 g_a \int_{\ga_R^\ka} 
 \prod^\ka_{j=1} \frac{d\lb_j (\lb_j\!-\!a)^{n+1}}{2\pi \I \lb_j^{r+1}}  \det (I\!-\!{\mathcal K}_{k,\ell}^{(\lb)} {(0,0)})_{\geq M } f(\lb). 
%
\el\ee

   Finally, one inserts the scaling (\ref{scaling}) for the $(\xi,\eta)$'s and the equivalent scaling (\ref{sr}) for the $(u,s)$'s into the expression (\ref{Lkernel}) for $\BK_{ n,r,\rho}^{^{\textrm{\tiny red}}}$. Next, pulling through the integrations coming from the $\widetilde \LR$-operator (\ref{Ltilde}), leads to expression (\ref{Lquadr}) for $\BK_{n,r,\rho}^{^{  \!\textrm{\tiny red}}}$, with (\ref{Lsca})  for $\BL_{n,r,\rho}^{^{  \!\textrm{\tiny red}}}$. This ends the sketchy proof of Theorem \ref{theo:Kred}.\qed

 So, the two scalings (\ref{scaling}) and (\ref{scaling'}) combined   
  \be \label{scaling''}
   \xi_i  = 2m -2\tau_i,~
 \eta_i +1 =  n +  y_i\sqrt{  n} ,~a=1+\frac{\beta}{\sqrt{n}} ,~\frac {\Dt \eta}2=\frac {dy} 2\sqrt{n}
, \ee
 will now be used in the final Theorem:

%


\newpage

\vspace*{.5cm}

\setlength{\unitlength}{0.015in}\begin{picture}(0,0)

\put(150,-190){\makebox(0,0){\includegraphics[width=220
 mm,height=300
  mm]{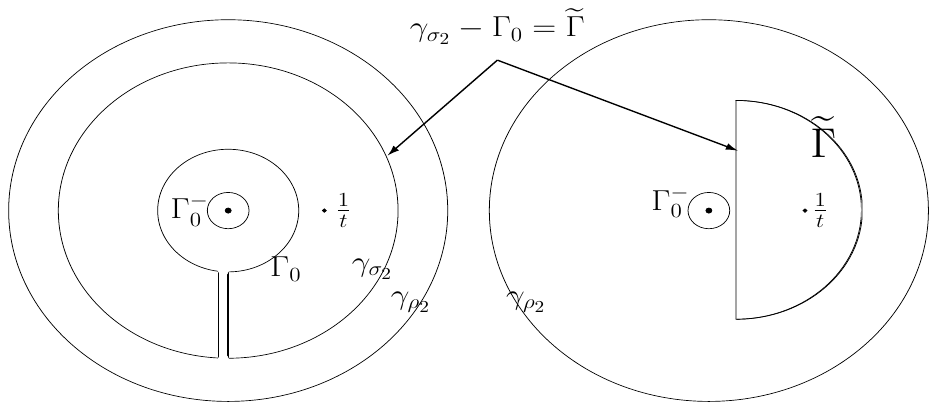}}}

\end{picture}
 
\vspace{1cm}
{\small Figure 9. Turning the annulus $\ga_{\sg_2}-\Ga_0$ into a semi-circle $\widetilde \Ga$, of radius $1/t^2$ and containing $1/t>0$, to the right of $\Ga_0^-$.}

\vspace*{.7cm}

 
\begin{theorem}\label{theo:asympt}(Adler, Johansson, van Moerbeke \cite{AJvMskewAzt}, Theorem 2.2):  Consider an Aztec rectangle, such that
 $$ 1\leq M< n+1\leq m \mbox{ and so }
 \rho-r=m-n=\ka>0.$$
   We let $n\to \infty$, while keeping $r=n-M+1$,~ $\rho=m-M+1$ and $\beta$ fixed. The following scaling-limit  ($\tau_i  \in \BZ,~~y_i\in \BR$) holds, with $ {\mathbb L}^{\mbox{\tiny dTac}}_{r,\rho,\beta}$ as in (\ref{Ldtac}):
   \be\label{limitRectangle} \begin{aligned}
   \lim_{{n\to \iy}\atop{\rho,~r ,~\beta~\mbox{\footnotesize fixed}}} \frac{2 (-a)^{\frac{\eta_1-\eta_2}{2}}}{1+a^2} (\tfrac{a}{\sqrt{n}})^{\frac{\xi_2-\xi_1}{2}} {\mathbb K}_{ n,r,\rho}^{red}   (\xi_1,&\eta_1 ;  \xi_2,\eta_2) 
\frac{\Dt \eta_2}{2} \Bigr|_{\footnotesize\mbox{scaling (\ref{scaling''})}} 
\\& =
   {\mathbb L}^{\mbox{\tiny dTac}}_{r,\rho,\beta}  \left( {\tau_1}, y'_1 ;  {\tau_2}, y'_2\right) dy'_2. 
 \\  &= \mbox{ \em ``Discrete Tacnode Kernel''}.    \end{aligned}
       \ee 
       The discrete tacnode kernel ${\mathbb L}^{\mbox{\tiny dTac}}_{r,\rho,\beta} $ has its support on a doubly interlacing set of points, as defined in (\ref{xy-interlace});  $\tau=0$ and $\tau=\rho$ are the boundary lines of the strip $\{\rho\}$, as depicted in Fig. 6(a).
   \end{theorem}


\noindent {\em Sketch of the Proof of Theorem \ref{theo:asympt}}: The starting point will be the rescaled kernel ${\mathbb K}_{ n,r,\rho}^{red}$ for the red dots point process,  in particular ${\mathbb L}_{ n,r,\rho}^{red}$ as in (\ref{Lquadr}) and (\ref{Lsca}), which in its present form is not amenable to taking a limit for $t\to 0$. Indeed the functions appearing in (\ref{Lsca}), and in particular in $\Phi(v)$ and $\Om(u,v) $(\ref{Omega}) all  
have poles at $u=0$ and at $u=\frac 1t$ only; so, one must express the kernel ${\mathbb L}^{^{\textrm{\tiny red}}}_{n,r,\rho}$ as in (\ref{Lsca}) in terms of contours about these two poles. From (\ref{Lsca}), we have that 
\be \label{K1'}
\bl
 (-1)^\ka({\mathbb L} _1\!+\!{\mathbb L} _2)
=& -\oint_{\ga_{\rho_2}\simeq   \Ga^-_0+\widetilde \Ga ^+} \frac{u^r \FR(u)du}{2\pi \I} \oint_{\ga_{\sg_2}-\Ga_0\simeq \widetilde \Ga}  \frac{ v^{-\rho}\Phi(v)dv }{2\pi \I (u\!-\!v)h(v) }  \frac{\Om  (u,v)}{\Om  (0,0) }
\\&  -\oint_{\ga_{\rho_2}} \frac{\FR(u)du}{2\pi \I u^\ka \GR(u)},
\el \ee
where the annulus $\ga_{\sg_2}-\Gamma_0$, containing the pole $1/t$,  can be deformed into a half circle to the right of the origin, which we choose to be of radius $1/t^2$, as in Fig. 9; so $\ga_{\sg_2}-\Gamma_0\simeq\widetilde \Gamma$, appearing on the right hand side of Fig. 9. Then $\ga_{\!\rho_2}$ for $\rho_2>\sg_2$ can be deformed into $\ga_{\rho_2}\simeq   \Ga^+_0+\widetilde \Ga ^+$, with $\widetilde\Gamma^+,~\Ga_0^+$ slightly larger than $\widetilde \Gamma,~ \Ga_0$. Interchanging the $u$ and $v$-integrations gives us for ${\mathbb L}  _1$ in (\ref{K1'}) the following expression (using $\ka=\rho-r$, as in (\ref{rho,r'})):
\be\label{K1''}
\oint_{\widetilde \Ga}\frac{A(v)dv}{2\pi \I} \oint_{\Ga_0^-} \frac{B(u)du}{2\pi \I(u-v)} +
\oint_{\widetilde\Ga}\frac{dvA(v)}{2\pi\I}\oint_{\widetilde \Ga^+\simeq ~\underbrace{\tiny\mbox{$\Ga_{\!\!\{v\}}+\sum_{\ell=0}^{r-1} \Ga_{\!\{u_{\ell}\}}$
  } }_{**}}\frac{duB(u)}{2\pi\I(u-v)}
\ee
where $\Ga_{\{v\}}$ and $\Ga_{\{u_\ell\}}$ are small circles about the corresponding points and where $A(v)$ and $B(u)$ are given by (see (\ref{fncts})):
\be\label{K1'''}A(v):=\frac{\Phi(v)}{v^\rho h(v)}=\frac{1}{v^\rho h(v)}\int_{\ga_{\sg_2^+}=\underbrace{\tiny \mbox{$\Ga_0^++\widetilde \Ga^+$}}_{*}}\frac{dw~h(w)}{2\pi \I \GR(w)(v\!-\!w)}
,~~B(u):= {u^r\FR(u)\Om (u,v)} ,
\ee
where the $u$-integrand (containing $B(u)$) in the second term of (\ref{K1''}) has poles $u=v$ and $u=u_j$ for $0\leq j\leq r-1$, coming from the poles of $\Om (u,v)$. So, in the end, the expression (\ref{K1''}) splits into 4 terms, after taking into account ${\mathbb L}_2$ and after some cancellations, two terms from (*) in (\ref{K1'''}) and two other terms from (**) in (\ref{K1''}). In order to take the limit of these integrals, one uses the asymptotics for the functions $\FR(u),~\GR(u),~h(u)$ in (\ref{Lsca}) and (\ref{Omega}); one notices that the integrals along the half-circle-part of $\widetilde \Ga$ (radius $1/t^2$; see Fig. 9) tends to $0$ for $t\to 0$, and that the integral along the vertical segment of $\widetilde \Ga$ tends to an integral over the vertical imaginary line $L_{0+}$ for $t\to 0$. So, using the minor change of variables (\ref{scaling'}), namely $x_i\to  \tau_i=x_i+\ka,~~y_i\to y_i'=y_i\sqrt2$,  one finds the $4$ terms $ {\mathbb L}_i^{\mbox{\tiny dTac}}$ with $1\leq i\leq 4$ in the discrete tacnode kernel (\ref{Ldtac}); the ${\mathbb L}_0$-part of the kernel in (\ref{Lsca}) tends to the Heaviside part $ {\mathbb L}_0^{\mbox{\tiny dTac}}$. Omitting the primes in $y_i'$'s, this is precisely formula (\ref{Ldtac}).

This ends the very sketchy proof of the main Theorem of this section. Further details can be found in sections 10 and 11 of \cite{AJvMskewAzt}.\qed


 \subsection{ Aztec rectangles formed by two overlapping Aztec diamonds ($\rho=r$)}

 We now overlap two identical Aztec diamonds (as described in Fig. 14), with the second one being rotated by $90^o$. This leads to a special instance of a skew-Aztec rectangle, as  considered in section 5, Fig. 5(b), but where $m=n$, implying $\Dt=0$ by (\ref{Dt}). Overlapping means that $1\leq M\leq n$,  \footnote{The extreme case where the two Aztec diamonds are set side by side, we would have  $M=n+1$ and so $\sg=\Dt=0$.} which happens to satisfy  {\em Case 1} of Theorem \ref{tilability}; so, this implies {\em tilability}. Then identities (\ref{Dt}) and (\ref{rho,r'}) imply $r-\rho=0=\Dt$ and so the strip $\{\rho\}$ has width: 
  \be\label{rho=r}\rho=r=n-M+1=\sg,~  ~~ n-M=\rho-1\mbox{ and } \Dt=0
   .\ee
  
  The arguments follow essentially the same steps as in Fig. 8. One first computes the kernel for the point process of green dots (type I in Fig. 8), whose level curves have contiguous start and end points. This is simpler in this case due to the fact that $\Dt=n-m=r-\rho=0$, which removes the singularity in the Toeplitz determinants; see (\ref{Kgreen'}) and (\ref{transitionf}); so the $\lb_j$ integration in Lemma \ref{lambda-integration} is unnecessary. 
  According to formula (\ref{Kgreen}), the double $(z,w)-$integration will act on the Fredholm determinant $\det(\Id-{\mathcal K}_{k,\ell}(w^{-1},z))$; this kernel equals the second expression in (\ref{KBO}). Then, at first $(z-w)$ can be taken out, up to some error; secondly the $(z,w)$- integration acts on $\det(\Id-{\mathcal K}_{k,\ell}(w^{-1},z))$ by integrating each of the functions $h_{k+1}^{(i)}$ in (\ref{KBO}), again up to an error. These two facts are possible due to the statement: 
  
  \begin{lemma}\label{LA1} ((Johansson, \cite{Jo13}) 
Given a trace-class operator $K$ and a rank one operator $a\otimes b$, given two continuous functions $F(z)$ and $G(w)$ and vectors $a^z,~b^w$ depending continuously on $z$ and $w$ and given an arbitrary constant $c$, the following two identities hold :%
\begin{multline}
\det (I-K+c~a\otimes b)=(1-c)\det(I-K)+c\det(I-K+a\otimes b).
 \end{multline}
and 
 \begin{multline}
\int_{|z|={r_1}}\frac{dz}{z}\int_{|w|={r_2}}\frac{dw}{w}F(z)G(w)\det (I-K +a^z\otimes b^w)\\
= 
 \det\left( I-K +\left( \int_{|z|={r_1}}F(z)a^z\frac{dz}{z}\right)\otimes\left(\int_{|w|={r_2}}G(w)b^w\frac{dw}w \right)  \right)  \\
 +\left[ \left(\int_{|z|={r_1}}F(z)\frac{dz}{z}\right)    \left(  \int_{|w|={r_2}} G(w)\frac{dw}{w}\right)-1\right]\det(I-K ).
 \label{A1}\end{multline}

\end{lemma}

    Next one proceeds by duality to the process of blue dots generated by a dual height function. To obtain the kernel for the point process of red dots, one again uses a Kasteleyn type of argument.  
  So, one obtains a kernel in $\xi,\eta$ for the process of red dots which has a somewhat similar structure as (\ref{2min0}). Then we have the following statement (see \cite{ACJvM}, Theorem 1.4 and some background in \cite{AJvM0}):


\begin{theorem}\label{theo:asymptOverlap}(Adler, Chhita, Johansson, van Moerbeke  \cite{ACJvM}):  Consider an Aztec rectangle formed by two overlapping Aztec diamonds, i.e., such that $m=n$ and $M=n-(\rho-1)$, implying $\rho=r$. 
 %
  For $n\to \infty$, while keeping $r=\rho=n-M+1$ and $\beta$ fixed, the following scaling-limit  ($\tau_i  \in \BZ,~~y_i\in \BR$) holds, with $ {\mathbb L}^{\mbox{\tiny dTac}}_{\rho,\rho,\beta}$ as in (\ref{2min0}):
   \be\label{limitRectangleSunil} \begin{aligned}
   \lim_{{n\to \iy}\atop{\rho,~\beta~\mbox{\footnotesize fixed}}} &\frac{2 (-a)^{\frac{\eta_1-\eta_2}{2}}}{1+a^2} (\tfrac{a}{\sqrt{n}})^{\frac{\xi_2-\xi_1}{2}} {\mathbb K}_{ n,r,\rho}^{red}   (\xi_1,\eta_1 ;  \xi_2,\eta_2) 
\frac{\Dt \eta_2}{2} \Bigr|_{\footnotesize\mbox{scaling (\ref{scaling''})}} 
\\=&  \BL^ {\mbox{\tiny GUEminor}}  (\tau_1,\! -\beta\!-\!y_1 ; \tau_2 , \!-\beta\!-\!y_2 ) 
\\&~+ \Bigl\la (\Id - {\cal K}^\beta ( \lambda ,\kappa ))^{-1}_{\geq -\rho} ~{\cal A}^{\beta,y_1+\beta}_{\tau_1 }(\kappa), {\cal B}^{\beta,y_2+\beta}_{\tau_2 }(\lambda)\Bigr\ra
  _{_{ \geq -\rho }} 
 .\end{aligned}
       \ee  
\end{theorem}

 \section{Lozenge tilings of nonconvex hexagons and double interlacing}

In this section we present results on random lozenge tilings of non-convex polygonal regions.
As already mentioned in the introduction, non-convex figures are particularly interesting due to the appearance of new statistics for the tiling fluctuations, caused by the non-convexities themselves or by the interaction between these non-convexities. What is the asymptotics of the tiling statistical fluctuations in the neighborhood of these non-convexities, when the polygons tend to an appropriate scaling limit?



\newpage

\vspace*{1.3cm}

\setlength{\unitlength}{0.015in}\begin{picture}(0,0)

     {\makebox(310,-290) {\rotatebox{0} {\includegraphics[width=200mm,height=295mm] {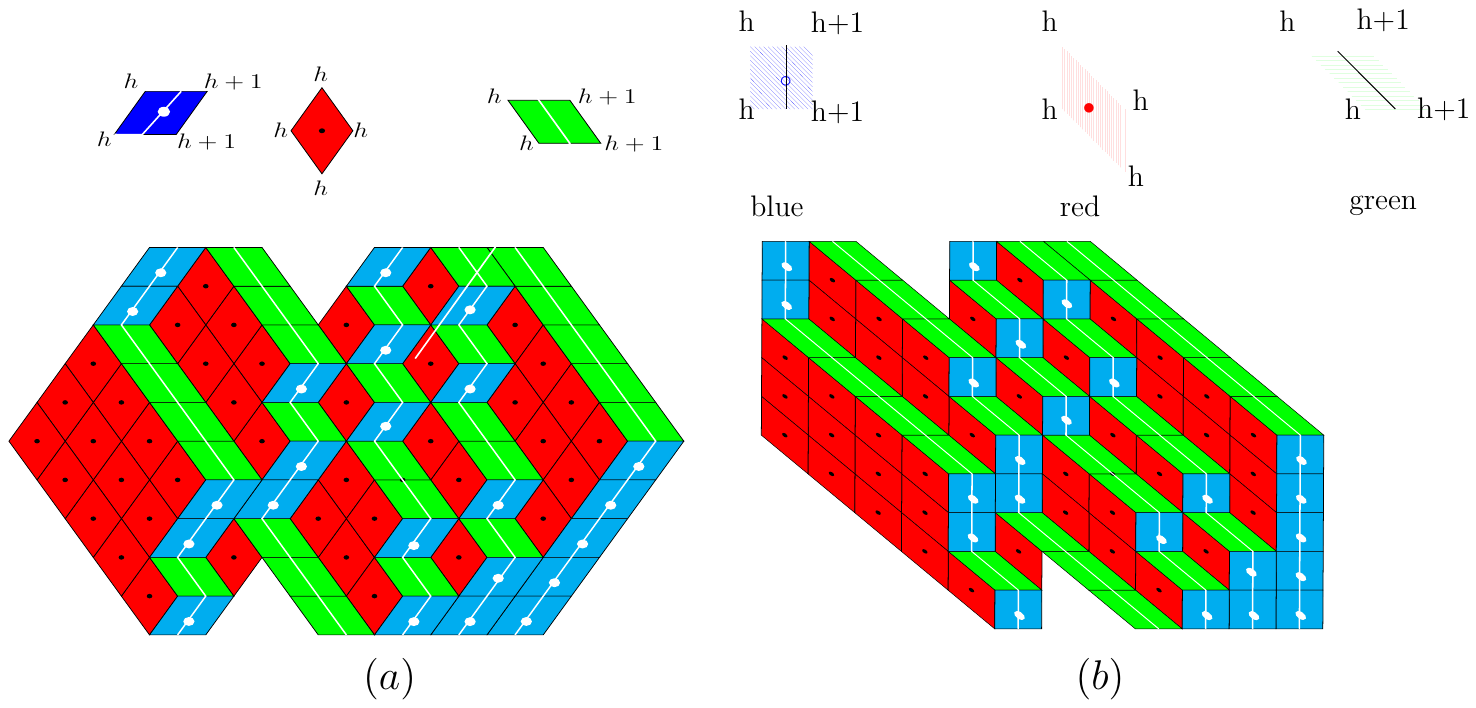}    }}}
    
\end{picture}

    \vspace*{2.5cm}
      
  {\small Figure 10: Lozenges of three different types (blue, red, green) covering a hexagon with cuts on opposite sides $(a)$. In $(b)$ appears the affine transformation of the lozenges and the hexagon. The white curves on the tiles above are level curves for the height $h$ given on the tiles; the $n_1+n_2$ white curves (see (\ref{tilab})) on the hexagon with cuts are the level curves corresponding to the tiling in the figure. }

\vspace*{.7cm}

Tilings of non-convex domains were investigated by Okounkov-Reshetikhin \cite{OR2} and Kenyon-Okounkov \cite{KO} from a macroscopic point of view. Further important phenomena for nonconvex domains appear in the work of Borodin and Duits \cite{BD}, Borodin and Ferrari \cite{BF}, Borodin, Gorin and Rains \cite{BGR}, Duits \cite{D}, Defosseux \cite{Defos}, Metcalfe \cite{Metc}, Petrov \cite{Petrov1,Petrov2}, Gorin \cite{Gorin1}, Novak \cite{Nov}, Bufetov and Knizel \cite{Buf}, Duse and Metcalfe \cite{Duse2,Duse1}, and Duse, Johansson and Metcalfe \cite{DJM}; see also the papers by V. Gorin and L. Petrov \cite{GorinPetrov} and Betea, Bouttier, Nejjar and Vuletic \cite{BBNV}.

In this section, we consider a hexagon ${\bf P}$ with cuts (non-convexities) along two parallel sides, say, along the top and the bottom sides, as depicted in the left-top figure in Fig. 10. 
  The configuration proposed here is different from the ones considered before. The motivation is that this new feature, cuts on opposite sides, will lead to new statistics in some appropriate scaling limit, going beyond Petrov's work \cite{Petrov2} on hexagons with one or several cuts on one side only, and yet inspired by some of his techniques. 

\medbreak 

{\em The geometry of non-convex hexagons, two systems of coordinates and the strip $\{\rho\}$}: We now tile the non-convex hexagon ${\bf P}$ (with cuts) as in Fig. 10, with lozenges of three different shapes, characterized by the red, blue or green color.  
 From the point of view of coordinates, it will be convenient to affine transform the hexagon to the one on the right hand side of Fig. 10. It is clear from the top of Fig. 10 how this transform affects the lozenges. Like for the dominos, we have a height function on the lozenges, and level curves of height $h+1/2$ for both, the lozenges and the affined transformed ones, as appears in Fig. 10. We now need to introduced some integers characterizing the non-convex hexagon ${\bf P}$. Fig. 11 gives an example of a hexagon with two opposite cuts, each of same size $d$; the lower cut is at a distance $m_1$ and $m_2$ from the lower-left and lower-right corner of the hexagon and the upper cut is at a distance $n_1$ and $n_2$ from the upper-left and upper-right corner of the hexagon. The two remaining parallel edges have sizes $b $ and $c\sqrt{2}$, with $N:=b+c$. As we shall see later, \be \label{tilab} m_1+m_2=n_1+n_2\ee will be required for tilability.  So this is a hexagon $(b,c\sqrt{2},n_1+n_2+d,b, c\sqrt{2},m_1+m_2+d)$ with two cuts of size $d$. All this information is contained in Fig. 11.

   We now have {\em two systems of coordinates} $(m,x)$ and $(\eta, \xi)$ as in Fig. 11(a), related by
    \be\eta=m+x+\tfrac 12,~~~\xi=m-x-\tfrac 12 ~~~~~ \Leftrightarrow ~~~m =\tfrac 12(\eta+\xi),~~x=\tfrac 12 (\eta-\xi-1) .\label{Lcoord}\ee
    where $1\leq m\leq N$ are the horizontal (dotted) lines in Fig. 11(a) and $x$ the running integer variable along those lines. The origin $x=0$ of the $x$-coordinate is such that the left-most and right-most corners of the grid are given by $x=-d-b-c-1/2$ and $x=m_1+m_2-1/2$. 
         In Fig. 11, the origin $(m,x)=(0,0)$ is given by the black dot on the lowerline ${m=0}$ of ${\bf P}$, whereas the origin $(\xi,\eta)=(0,0)$ is the circle lying $1/2$ to the left of $(m,x)=(0,0)$.

      To this configuration Fig. 11, we also associate {\em a strip $\{\rho\}$}, determined by  the extension of the two oblique lines $\eta=m_1$ and $\eta=  n_1 +b-d$ belonging to the cuts; thus, the strip $\{\rho\}$  has width $\rho$ which is given by
    \be\label{rho'} \rho=  n_1-m_1+b-d  =m_2-n_2+b-d.\ee

 The integers along the bottom line $\notin {\bf P}$ are labeled by $ y_{d+N}<\dots<y_2<y_1$, with \be\label{y's} y_1=m_1-1,\dots,~y_d=m_1-d,~y_{d+1}=-d-1,\dots, ~y_{d+N}=-d-N.\ee The integers along the top line $\notin {\bf P}$ are labeled by $ x_{d+N}<\dots<x_2<x_1$, with 
     \be\bl \label{x's} x_{d+N}&=-d-N,\dots,
     x_{c+d+1}=-c-d-1,~x_{c+d}=n_1-c-d,~\dots,\\&x_{c+1}=n_1-c-1,~x_c=m_1+m_2-c,\dots,x_1=m_1+m_2-1.
   \el\ee 
 So much for the geometry of such a non-convex hexagons!

  \newpage
   
    \vspace*{.2cm}

\setlength{\unitlength}{0.017in}\begin{picture}(0,0)

 \put(140,-150)
{\makebox(0,0) 
 {\includegraphics[width=200mm,
 height=300mm]
 {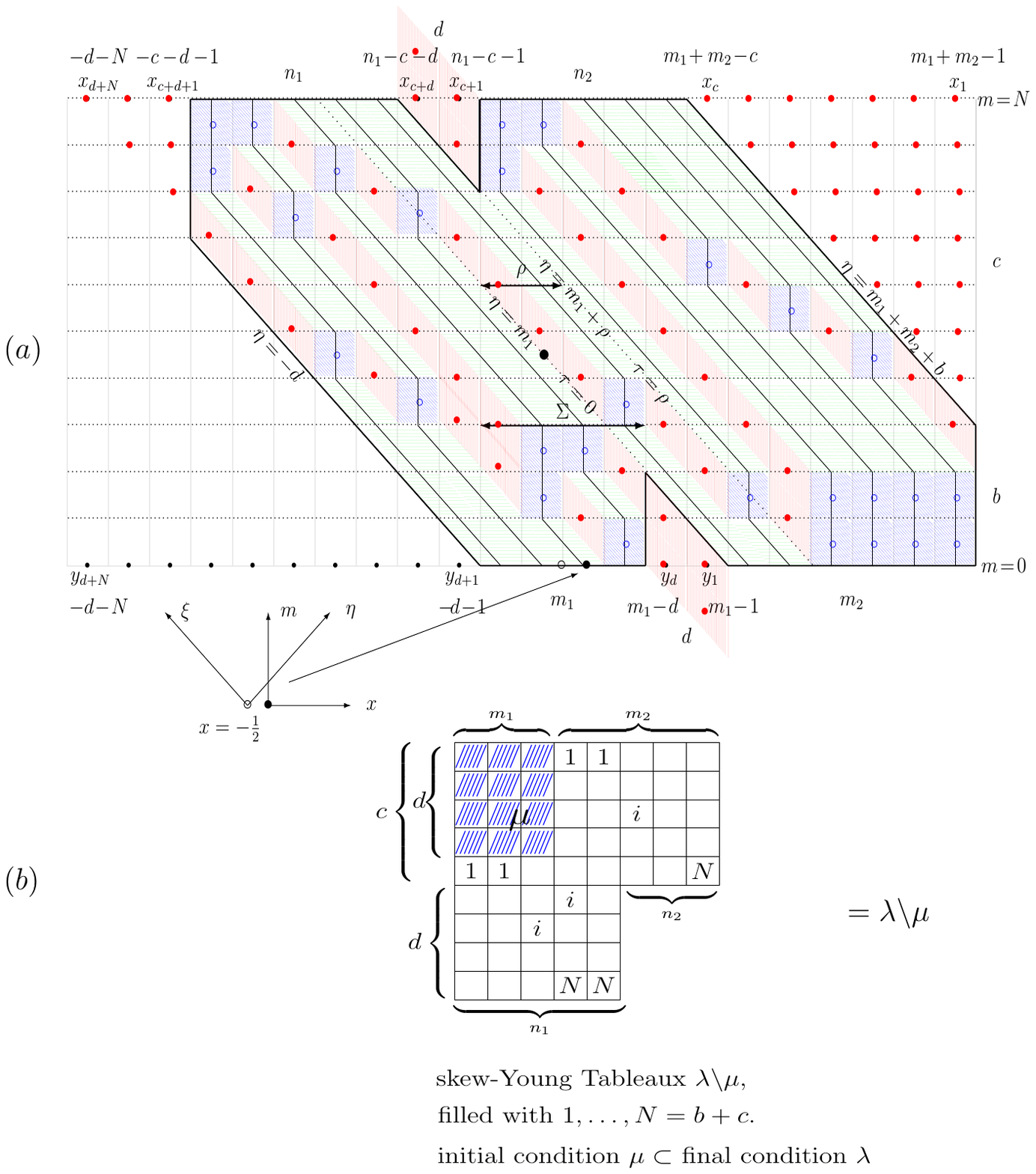} }}
 
\end{picture}

\vspace*{13.6cm}

{\small Figure 11. (a) Tiling of a hexagon with two opposite cuts of equal size, with red, blue and green tiles. The blue (resp. red) dots are in the middle of the blue (resp. red) tiles. The $(m,x )$-coordinates have their origin at the black dot on the bottom-axis $m=0$ and the $(\eta,\xi )$-coordinates at the circle given by $(m,x )=(0,-\tfrac 12 )$. The strip $\{\rho\}$ is bounded by $\eta=m_1$ and $\eta=m_1+\rho$; alternatively, setting $\tau=\eta-\rho$, $\{\rho\}$ is bounded by $0\leq \tau\leq \rho$.  Here $d=2$, $n_1=n_2=5,~m_1=4,~m_2=6,~ b=3,~c=7,~N=10$, and thus $~ r=1,~ \rho=2.$
\newline (b) This is an equivalent description of the red dots in terms of skew Young tableaux  $\lb\backslash\mu$, filled with numbers $1,\dots, N$, where $N=b+c$.}

   \newpage



\newpage

 \vspace*{-.3cm}

 \setlength{\unitlength}{0.017in}\begin{picture}(0,0)
\put(150,-155){\makebox(0,0) {\includegraphics[width=220mm,height=258mm]
 {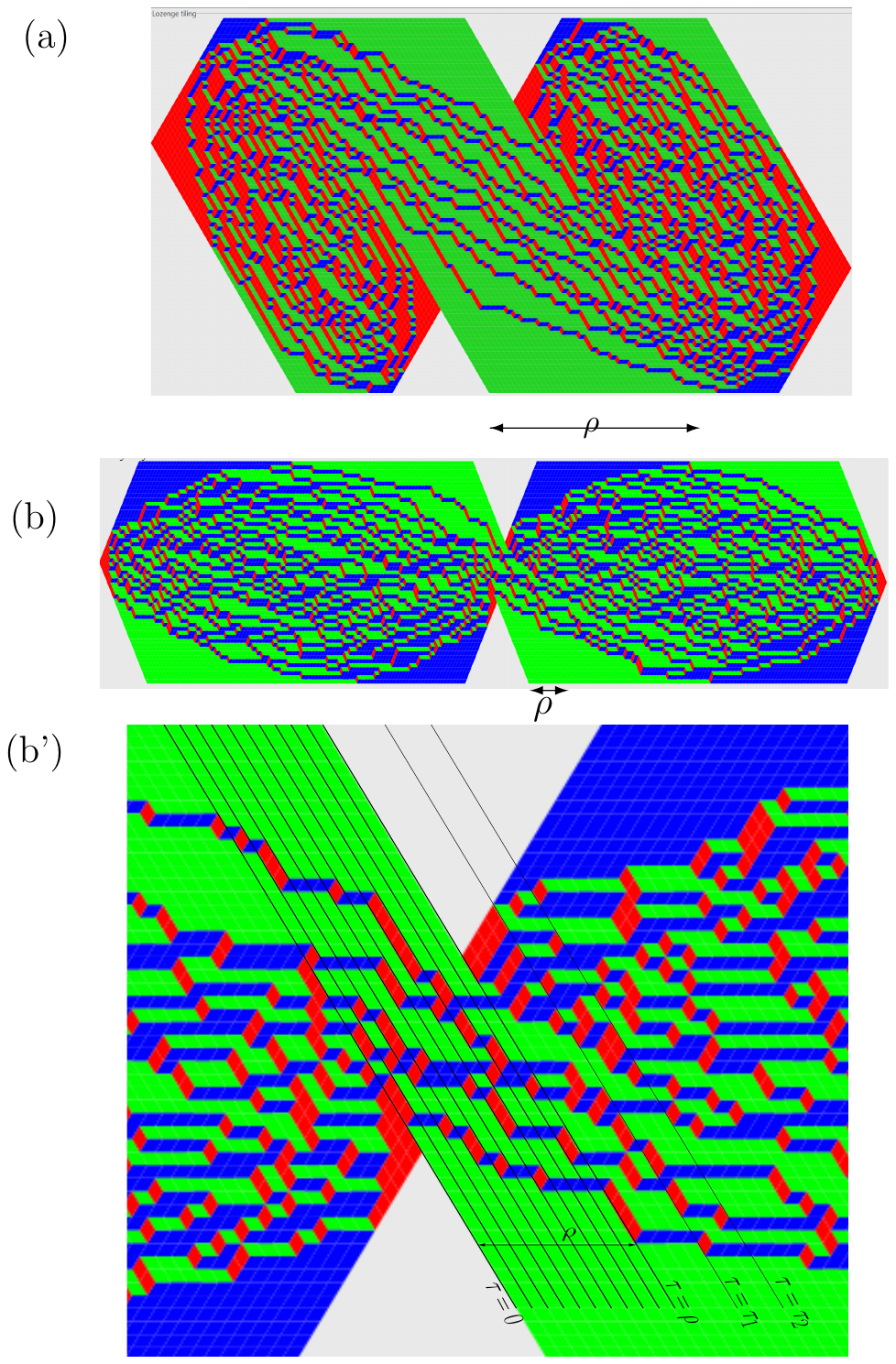}}}  
  \end{picture}

\vspace{12.5cm}

 {\small Fig. 12. Two computer simulations with different parameters (referring to the parameters in Fig. 11):  
\newline$\bullet$ Panel(a):~$n_1 = 50, n_2 = 30, m_1 = 20, m_2 = 60, b = 30, c = 60, d = 20$, so $\rho=40$, $r=10$  
 \newline$\bullet$  Panel(b):~$n_1 =  105, ~n_2=95,~m_1=m_2=100,~b=25,~c=30,~d=20$, so $\rho=10 $, $r=5$ 
 \newline$\bullet$ Zooming Panel(b) about the region between the cuts gives Panel(b'), where $\rho+1 =11$ is the number of oblique lines in the strip $\{\rho\}$  with $0\leq \tau \leq \rho$; for $\tau=\eta-m_1 $, see Fig. 11(a). Each such line carries $r=5$ blue tiles each. The lines $\tau =\tau_\al\geq \rho$   each carry $\tau_\al-\rho+r$ blue tiles. The ${\mathbb L}^{\mbox{\tiny dTac}}$-kernel describes the asymptotic statistics of the blue tiles along the oblique lines within and in the neighborhood of the strip $\{\rho\}$.(Courtesy of Antoine Doeraene)
 }
 
   \newpage

\newpage


    \medbreak
    {\em Lozenge tilings, level curves and the point processes $\PR^{\mbox{\tiny blue}}$ of blue dots}: The domain $\bf P$ is randomly tiled with green, red and blue lozenges (as in Fig. 10 or Fig. 11), except that the cuts are forced-filled with red lozenges (or equivalently with red dots), as is done in Fig. 11(a). The two left and right upper-triangles should be filled with red tiles as well; however for the sake of clarity,  we only put red dots. More generally, if one were to have many more cuts, they would all be filled with red lozenges.
    %
  Fig. 12 contains two {\em different simulations}  (a) and (b) for different values of $m_i,n_i,b,c,d$. Fig. 12 (b') zooms about the middle of (b), showing more clearly what occurs in the neighborhood of the strip.
  
     We consider the discrete-time point process $\PR^{\mbox{\tiny blue}}$ of blue dots along the oblique lines $-d+1\leq \eta\leq m_1+m_2+b-1$, $\eta \in \BZ$; more specifically, the blue dots (or blue tiles) belonging to the intersection of the parallel oblique lines $x+m=k-\tfrac 12$  with the horizontal lines $m=\ell-\tfrac 12$ for $k,\ell \in \BZ$  (see Fig. 11). So, the blue dots are parametrized by $(\eta,\xi  )=(k,2\ell-k-1 )\in \BZ^2$, with $ k,\ell\in \BZ$. 
         It follows that the $(\eta,\xi )$-coordinates of the blue dots satisfy $\xi+\eta=1,3,\dots,2N-1$. 
         
         Tiling the hexagon with the tiles equipped with the level curve (as on top of Fig. 10) automatically induces level curves on the hexagon ${\bf P}$. It is easily seen that the  $n_1+n_2=m_1+m_2$ level curves in ${\bf P}$, depart from the integer points at the bottom of ${\bf P}$, and end up at the integer points at the top of ${\bf P}$, excluding the integer points of both cuts; in Fig. 10 the level curves are indicated by the white curves and in Fig. 11 by the black curves. As was noticed in the context of Aztec rectangles, the height is uniquely specified along the boundary of the domain $\bf P$. One picks the height $h=0$ along the left-side of the domain          as in Fig. 10(b), $h=n_1+n_2$ along the right-side of the domain and grows one-by-one from $h=0$ up to $h=n_1+n_2$ along the upper- and lower-boundary.

         The point process $\PR^{\mbox{\tiny blue}}$ lives on  oblique lines, parallel to the strip $\{\rho\}$, departing  from the points $m=0,~x\in \BZ $ in the $(m,x)$-coordinates. So, the number of blue dots along oblique lines $\eta=k\in \BZ$ equals the number of times it will intersect level curves and also equals the difference of heights measured along $\eta=k$ between its intersection points with the horizontal lines $m=0$ and $m=N$. 
               So, these level curves contain the same information as the blue tiles (or the blue dots).
         
  It is easily shown that the number $r$ of blue dots along each of the $\rho+1$ parallel lines $\eta=k$ within the strip $\{\rho\}$ (i.e., $m_1\leq \eta\leq m+\rho$) is always the same and equals:
  \be \label{r'}r:=b-d.\ee 
  Moving away from the strip to the left or to the right has the effect of having the number of dots per line $\eta=k$  first increase one by one and then decrease one by one after a while. Of course, the plan is to show that the point process  $\PR^{\mbox{\tiny blue}}$ of these blue dots is determinantal, with correlation kernel ${\mathbb K}^{\mbox{\tiny blue}}(\xi_1,\eta_1;\xi_2,\eta_2)$. This is the kernel which in the scaling limit should lead to the discrete tacnode kernel (\ref{Ldtac}).
  
  {\em The point process  $\PR^{\mbox{\tiny red}} $ of red dots} will also play an important role. We put red dots in the middle of each red lozenge. They  belong to the intersections of the vertical lines $x\in \BZ$ and the horizontal lines $m=0, \ldots , N$. 
  The initial condition at the bottom $m=0$ is given by the $d$ {\em fixed red dots} at integer locations in the lower-cuts, whereas the final condition at the top $m=N$ is given by the  $d+N$ {\em fixed red dots} in the upper-cuts, including the red dots to the left and to the right of the figure, all at integer locations. Notice that the process of red dots on $\widetilde {\bf P}$ form an interlacing set of integers starting from $d$ fixed dots (contiguous for this two-cut model) and growing one by one to end up with a set of $d+N$ (non-contiguous) fixed dots. This can be viewed as a ``truncated" Gel'fand-Zetlin cone!

  \medbreak
   
   {\em The dual height function, its level curves and filaments between arctic circles}:  The height function and level curves on the tiles (as in Fig. 13) lead to a dual set of $b+d$ level curves, whose levels also increase one by one, going from left to right. Referring to Fig. 11(a), these level curves start, either at the left vertical edge of the domain $\bf P$ (of length $b$), or at the vertical edge (of length $d$) of  the upper-cut and end up at the vertical edge of  the lower-cut or at the right vertical edge of the domain $\bf P$. Similarly, the blue dots are at the intersection of these level curves with the same lines $\eta\in \BZ$. As is apparent from Fig. 11, among the $b+d$ level curves, one sees that  only $b+d-2d=r$ traverse the strip $\{\rho\}$. These $r$ level curves depart from the left-vertical edge of $\bf P$ and end up at its right-vertical edge.   They are the $r$ ones which in the limit will become the filaments, seen in the simulation of Fig. 12. The filaments are given by a succession of red and blue tiles as is seen in the zoomed version Fig. 12 (b') of Fig. 12 (b).

  \newpage
  
   \setlength{\unitlength}{0.017in}\begin{picture}(0,0) 
      \put( 150, -120){\makebox(0,0) {\includegraphics[width=224mm,height=285mm] {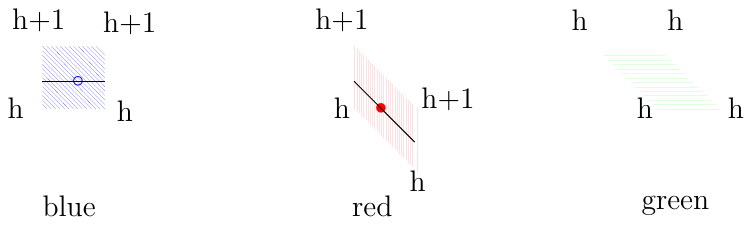} }}  
\end{picture}  

\vspace*{ 1cm}

{\small Fig. 13 : Lozenges with level curves dual to the level curves given in Fig. 10(b).}

 \vspace{.8cm}

   
\noindent {\em The scaling of the geometry and the running variables}:    
%
 \be\mbox{Fixing} ~\left\{ \begin{aligned}
  \rho&=  n_1-m_1+b-d  =m_2-n_2+b-d 
  =\{\mbox{width of the strip $\{\rho\}$}\} 
\\
 r&=b-d=\#\{\mbox{blue dots on the parallel lines in the strip $\{\rho\}$}\} ,
\end{aligned}\right.
\label{r,rho}\ee
we let the following data go to infinity together, according to the following scaling:
   \newline
    {\em (i) Scaling of geometric data:} Let the size of the cuts $=d\to \iy$, and let the parameters $m_1,m_2,n_1,n_2,b,c,d\to \iy$, where we allow additional fixed parameters $\ka, \beta_i,\ga_i$; the  new scale is as follows: 
\be\label{scalinggeom}\begin{array}{lllll}
b=d+r&& c=\kappa d
\\
m_1=\tfrac{\kappa+1}{\kappa-1}
 ( d+\sqrt{\frac{\kappa}{\kappa-1}}  \beta_1\sqrt{d}+ \ga_1) && n_1 = m_1+(\rho-r)   
\\  
m_2=\tfrac{\kappa+1}{\kappa-1}
 ( d+\sqrt{\frac{\kappa}{\kappa-1}}  \beta_2\sqrt{d}+ \ga_2) &&  n_2 = m_2-(\rho-r)  ,  
\\ \\
\beta= \frac{\beta_1 +\beta_2}{2\sqrt{2}},~~&&~ 1<\kappa<3 ,~\beta_i,~\gamma_i=\mbox{free parameters.}
 \end{array} \ee
%
%
\newline
{\em (ii) Scaling of running variables:}   $(\eta_1,\xi_i)\in \BZ^2\to (\tau_i,y_i)\in (\BZ \times\BR ) ,$  about the point  $(\eta_0,\xi_0) $  given by the black dot in Fig. 11(a), (halfway point along the left boundary of the strip $\{\rho\}$ shifted by $(-\tfrac 12, \tfrac 12)$), upon setting $a=2\sqrt{\frac{\kappa}{\kappa-1}}$,
\be\begin{aligned}
 (\eta_i,\xi_i)  &= (\eta_0,\xi_0)+\left(\tau_i,~-\tfrac {  (\kappa+1)}a \bigl( { y_i+\frac{\beta_1-\beta_2}{2\sqrt{2}}  }  
  \bigr)\sqrt{2d} \right)
  \\&\mbox{    with   }
  (\eta_0,\xi_0)  =(m_1,N-m_1 -1).
   \end{aligned}
\label{eta-xi}\ee
 We show the following asymptotic result :
\begin{theorem} \label{Th1}(Adler, Johansson, van Moerbeke (\cite{AJvM1}, Theorem 1.3 and \cite{AJvM2}, Theorem 1.2) Given the scaling (\ref{scalinggeom}) and (\ref{eta-xi}) above, the correlation kernel ${\mathbb K}^{\mbox{\tiny blue}}$ of blue dots tends to the discrete tacnode kernel:
\be\begin{aligned}
 \lim_{d\to \infty} (-1)^{\tfrac 12 
   (\eta_1+\xi_1-\eta_2-\xi_2)}
&\left(\sqrt{d}\frac{\kappa + 1}{2a}\right)^{\eta_2-\eta_1 
 }
{\mathbb K}^{\mbox{\tiny blue}}( \eta_1 ,\xi_1;\eta_2,\xi_2)\frac 12 \Dt\xi_2
\\
&= -\sqrt{2}~{\mathbb L}^{\mbox{\tiny dTac}}_{r,\rho,\beta}   \left( \tau_1, y_1 ;\tau_2, y_2\right)  
dy_2 ,\end{aligned}
 \label{limit}\ee 
\end{theorem}

%
\proof 
{\bf Step I. Skew-Young Tableaux and skew-Schur polynomials}. The point process $\PR^{\mbox{\tiny red}} $ of {\em red dots} on horizontal lines satisfy the following interlacing condition:
 \be x_{i+1}^{(\ell)}<
x_{i}^{(\ell-1)}\leq x_{i}^{(\ell)}\label{interl}\ee
in terms of the levels $m=\ell$ in the $(m,x)$ coordinates. So, this forms a truncated Gelfand-Zetlin cone with prescribed top and bottom. The prescribed set at the bottom are the red dots in the lower cut $y_1=m_1-1,  y_2=m_1-2 ,  y_d=m_1-d$ (see (\ref{y's})), whereas the red dots at the top are those to the right of ${\bf P}$, those in the cut and those to the left of ${\bf P}$; i.e., the set (\ref{x's}).

Moreover, remembering $N=b+c$ and setting for each level $0\leq \ell\leq N$, 
$$\nu_i^{(\ell)} =x_i^{(\ell)}+i ,
$$
leads to a sequence of partitions, due to the interlacing (\ref{interl}),
\begin{equation}
 \nu^{(0)} \subset \nu^{(1)} \ldots \subset \nu^{(N )}  ,
\label{1}
\end{equation}
with prescribed {\bf initial and final condition},
$$ \mu:=  \nu^{(0)}   ~~~~~  \mbox{    and    }   ~~~~~ \lb:=  \nu^{(N)}  
$$
and 
such that each skew diagram $\nu^{(i)} \backslash \nu^{(i-1)}$ is a horizontal strip\footnote{\label{ft3}An horizontal strip is a Young diagram where each row has at most one box in each column. }. 
Note the condition $x_i\geq y_i$ for all $1\leq i\leq d+N$ along the cuts, according to (\ref{x's}) and (\ref{y's}), guarantees that $\mu\subset \lb$. In fact the consecutive diagrams $\nu^{(i)} \backslash \nu^{(i-1)}$ are horizontal strips, if and only if the precise inequalities $x_{i+1}^{(m)}<
x_{i}^{(m-1)}\leq x_{i}^{(m)}$ hold.

 Putting the integer $i$'s in each skew-diagram $\nu^{(i)} \backslash \nu^{(i-1)}$, for $1\leq i\leq N$, is equivalent to a skew-Young tableau filled with exactly $N$ numbers $1,\dots,N$ for $N=b+c$.  An argument, using the height function, which is similar to the one in \cite{ACJvM}, shows that all configurations are equally likely. So we have {\bf uniform distribution} on the set of configurations. 
 So,we have:
   \be\label{config}\begin{aligned}
   &\{\mbox{All configuration of red dots}\}   
   \\& \Longleftrightarrow  \{\mbox{skew-Young Tableaux of shape $\lb\backslash \mu$ filled with numbers $1$ to $N $ }\}        
 \\
  &\Longleftrightarrow  \left\{\begin{aligned}
& 
  \mbox{$\mu  = \nu^{(0)} \subset \nu^{(1)} \ldots \subset \nu^{(N )} = \lambda  $, with $\mu\subset \lb$ fixed,}
  \\ 
   &\mbox{with $\nu^{(m)}\backslash \nu^{(m-1)}=\mbox{horizontal strip} 
 $},
\end{aligned}\right\}
\end{aligned}
\ee
with equal probability for each configuration, the number being expressed in terms of skew-Schur polynomials
 \be\begin{aligned}
   &\#\{\mbox{configuration of red dots}\}   
        = s_{\lb\backslash \mu} (\underbrace{1,\dots,1}_N,0,0\ldots)  ,   
 \end{aligned}
 \label{number}\ee. 
where
\begin{equation}
s_{\lambda\backslash \mu} (u) := \det (h_{\lambda_i -i -\mu_j  +j} (u))_{1\leq i, j \leq n} =\det(h_{x_i -y_j   } (u))_{1\leq i, j \leq n} ,~~\ \textrm{for} \ n \geq \ell(\lambda),
\label{2}
\end{equation}
where $h_r(u)$ is defined for $r\geq 0$ as
\be\label{seriesh}
\prod_{i\geq 1} (1-u_i z)^{-1} = \sum_{r\geq 0} h_r (u) z^r,
\ee 
and $h_r(u)=0$ for $r < 0$.

\medbreak

\bigbreak

 \noindent {\bf Step II.  A $q$-dependent probability and its $q$-kernel}. Computing the uniform distribution on the configurations (\ref{config}), and then deducing its correlation kernel turns out to be quite singular. Therefore, we first $q$-deform this distribution, deduce its corresponding $q$-kernel and then let $q\to 1$. Indeed, 
 %
  a $q$-dependent probability measure $\BP_q$, for $0<q\leq 1$, on the set of red dot-configurations or, what is the same, the space of all $N+1$-uples of partitions, is given as follows  
%
\be\begin{aligned}
\BP_q(&\nu^{(0)}, \nu^{(1)} , \ldots \, \nu^{(N )},~~\mbox{such that}~~\mu=\nu^{(0)}\subset \nu^{(1)} \subset \ldots \subset \nu^{(N )}=\lb)
\\
&=
\frac{(q^{-1})^{ \sum^{N-1}_{i=1}|\nu^{(i )}|- (N-1)|\nu^{(0)}|} }
{s_{\lambda\backslash \mu} (q^{1-N}, q^{2-N},\ldots, q^{-1}, q^0) }
\Id_{\nu^{(0)}\subset \nu^{(1)} \subset \ldots \subset \nu^{(N )}}
\Id_{\nu^{(0)}=\mu}  \Id_{\nu^{(N )}=\lb},
\end{aligned}\label{qprob}
\ee
which obviously for $q\to 1$ leads to the uniform probability. 

Since the $\PR^{\mbox{\tiny red}} $-process of red dots has a nonintersecting paths description, one expects to have a Karlin-McGregor formula for the $q$-dependent probability measure $\BP_q$, as in  (\ref{qprob}), and we obtain a $q$-dependent kernel ${\mathbb K}^{(q)}(m,x;n,y)$ by adapting the Eynard-Mehta-type arguments in Borodin-Ferrari-Pr\"ahofer \cite{BFP} and in Borodin-Rains \cite{BR} to this new circumstance of an initial condition $\mu$ and a final condition $\lb$; i.e., a``{\em two boundary}"-problem; see section 4 and, in particular, Proposition 4.2 in  \cite{AJvM1}. Before proceeding, we need to take the limit $q\to 1$ to obtain the ${\mathbb K}^{\mbox{\tiny red}}$-process; see (\cite{AJvM1}, Theorem 7.1). Using inclusion-exclusion, one is able to maximally reduce the number of integrations to at most $r+3$, leading to the following proposition (see Proposition 8.1 in \cite{AJvM1}):

\begin{proposition}\cite{AJvM1} For $(m,x)$ and $(n,y)\in {\bf P}$, the determinantal process of red dots is given by the kernel, involving at most $r+3$-fold integrals, with $r=b-d$, as in (\ref{r,rho}):
   \be \begin{aligned}
 \lim_{q\to 1}q^{(d+m )(x-y)}  {\mathbb K}^{(q)}( m,x;n,y)=  {\mathbb K}^{\mbox{\tiny red}} &(m,x;n,y) =: \BK_0+\tfrac{(N-n)!}{(N-m-1)!}(\BK_1+ \tfrac {1}{r+1}\BK_2) ,
  \end{aligned}\label{Kernlim4}\ee
where
$$\bl
\BK_0:=&-\frac{(y-x+1)^{(n-m-1)}}{(n-m-1)!}\Id_{n>m}\Id_{y\geq x}
\\
\BK_1:=& \oint_{ {\Gamma( {x +{\mathbb N}}) }}\frac{dvR_1(v)}{2\pi \I}
\oint_{\Gamma_{\infty}} \frac{dz}{2\pi \I(z\!-\!v)R_2(z)}   
 \frac{  \Om^{+ }_r (v,z)}{  \Om^{ +}_r (0,0)} 
\\
 \BK_2:=& \oint_{ {\Gamma( {x +{\mathbb N}}) }}\frac{dvR_1(v)}{2\pi \I}
 \oint_{ \Ga_{  \tau   
  }}  
   \frac{dz   }{2\pi \I R_2(z)h(z) }   
 \frac{ \Om^{-}_{r+1} (v,z)}{  \Om^{+}_r (0,0)} , 
\el$$
 and  where $R_1(z),~R_2(z), h(z)$ are ratios of monic polynomials with roots depending on certain subsets of $(x_1,\dots,x_{d+N})$ at the top of  ${\bf P}$ and $(y_1,\dots, y_{d+N})$ at the bottom of  ${\bf P}$. Also
\be\begin{aligned}
\Gamma(x+{\mathbb N})&:=\mbox{contour containing the set}~  x+{\mathbb N}=\{x,x+1,\ldots\}
\\
\Gamma_{\infty}&:=\mbox{very large contour containing all the poles of the $z$-integrand}
\\
\Ga_{\tau }&:=\Ga(y\!+\!n\!-\!N,\dots,\min(y_1 \!-\!N,y) )\Id_{ \tau<0},\mbox{    with  }\tau:=(y+n)-(y_1+1)\in\BZ.
\end{aligned}
\label{cont0}\ee
and where
 \be\begin{aligned} 
 \Om^\pm_k (v,z)&:=\left(\prod_{\al=1}^k  \oint_{\Gamma(\LR)}
\frac{du_\al  h(u_\al) (z-u_\al)^{\pm 1} }{2\pi \I ~(v-u_\al)  }
    \right)
  \Dt^2_k (u).
 \end{aligned}\label{Omr}\ee
\end{proposition}

\remark When the domain has several cuts, \cite{AJvM1} contains a formula similar to (\ref{Kernlim4}), but with additional terms. When the domain $\bf P$ is a hexagon $a,b,c$ (i.e., without cuts), formula (\ref{Kernlim4}) reduces to Johansson's \cite{Jo05b} formula for ${\mathbb K}^{\mbox{\tiny red}} $. With one or several cuts on one side only, one recovers Petrov's formula \cite{Petrov2}.

\medbreak

{\bf Step III. Going from the ${\mathbb K}^{\mbox{\tiny red}}$-kernel to the ${\mathbb K}^{\mbox{\tiny blue}}$-kernel}.  The integrands contain various polynomials involving the geometric data of the model.   

 As mentioned, we actually need the kernel ${\mathbb K} ^{\mbox{\tiny blue}}$ of blue dots along the oblique lines. To do so, we need the Kasteleyn matrix for the corresponding dimer model, which in this case is a honeycomb lattice; the arguments are similar to the ones used in the transition from step II to step III in section 5. 
 One then proves the following statement, the details of which can be found in (\cite{AJvM2}, Lemma 4.1):

\begin{proposition}\label{Theo:L-kernel}\cite{AJvM2} 
The ${\mathbb K} ^{\mbox{\tiny blue}}$-process of blue dots and the $ {\mathbb K}^{\mbox{\tiny red}}$-process of red dots have kernels related as follows:
\be\begin{aligned} {\mathbb K}^{\mbox{\tiny blue}}(\eta &,\xi ;\eta',\xi') 
 =-{\mathbb K}^{\mbox{\tiny red}}\left(m -\tfrac 12,x ;
m' +\tfrac 12,x' \right),
\end{aligned}\label{L-kernel'}\ee
where $(m,x)$ and $(m',x')$ are the same geometric points as $(\eta,\xi)$ and $(\eta',\xi')$, expressed in the new coordinates (\ref{Lcoord}).
\end{proposition}


Putting the ${\mathbb K}^{\mbox{\tiny red}}$-kernel given by (\ref{Kernlim4}) into the formula (\ref{L-kernel'}) gives the formula for the ${\mathbb K}^{\mbox{\tiny blue}}$-kernel.

Concatenating the two expressions (\ref{Kernlim4}) and (\ref{L-kernel'}) for the kernels ${\mathbb K}^{\mbox{\tiny red}}$ and ${\mathbb L}^{\mbox{\tiny blue}}$ gives the expression for ${\mathbb K}^{\mbox{\tiny blue}}$, of which one takes the scaling limit for $d\to \iy$. After a considerable amount of contour changes and rewriting \cite{AJvM2}, one obtains in the limit the desired discrete tacnode kernel, ending this very sketchy proof of Theorem \ref{Th1}.\qed

\section{A soft tacnode universality class }
A different probing will lead to yet another universality class, as explained in \cite{AJvM0}. Namely, we overlap two identical Aztec diamonds, with the second  one being rotated by $90^o$, as is done in Subsection 5.2; see Fig 14. As pointed out, this leads to a special instance of a skew-Aztec rectangle, as  considered in section 5, but where $m=n$, implying $\Dt=0$.  So we have identities (\ref{rho=r}). Again, 
for convenience, we will pick $n$ to be even! In Fig. 16 the size $n$ happens to be odd ($n=7$); our apologies!

For this model we define the probability of the domino tiling as in (\ref{Prob}), but keeping $0<a<1$. We will be considering a different scaling and so a different limiting statement, which we know explain. 

From Jockush, Propp, Shor \cite{JPS} a
nd Johansson \cite{Jo05c} we know that the tiling of a single Aztec diamond leads in the limit for $n\to \iy$ to a disordered region bounded by an Arctic ellipse (\ref{ellips}) of excentricity $\sqrt{1-\frac qp}=\sqrt{1-a^2}$, centered at the center of the diamond, with a frozen region outside the ellipse; see Fig. 15. {\em Sliding the two Aztec diamonds close enough such that the two ellipses touch} will lead to two tangent identical ellipses within the skew-Aztec rectangle, centered at the point of tangency, which coincides with the center of the configuration. The ellipses  satisfying the equations:
  \be \label{ellips}  \frac{(x \pm\tfrac r2)^2}{p}+\frac{(y \pm\tfrac r2)^2}{q}=1,~~ \mbox{with} ~q=\frac{a }{a+a^{-1}}\mbox{,  }p=1-q=\frac{a^{-1} }{a+a^{-1}},~r=\frac{2}{a+a^{-1}}
.\ee
This represents the picture if one would  lets $n\to \iy$, with tiles having size $1/n$, after dividing the size of the picture by $n$. So, here we will be probing in the neighborhood of that point of tangency.

To continue on the geometry, we will henceforth pick $M$ to be odd, merely for convenience! As before (see Fig. 5(c)) we consider the $(\xi,\eta)$-coordinates and  $(z,x)$-coordinates; the latter are the same as the $(s,u)$-coordinates in Fig. 5(c), but shifted by $ \left[\tfrac M2\right]$, so as to have the $z$-axis pass through the center of the configuration, where the tangency of the two arctic ellipses will take place; so, expressed in $\xi,\eta$-coordinates, we have (see Fig. 16(a))
$$z=\eta+1,~~~x=\tfrac12 (\eta-\xi )+ \tfrac M2 .$$

\newpage

\vspace*{1cm}

\begin{picture}(0,170)

\put(160,20)
 {\makebox(0,0){\includegraphics[width=200mm,height=260mm]{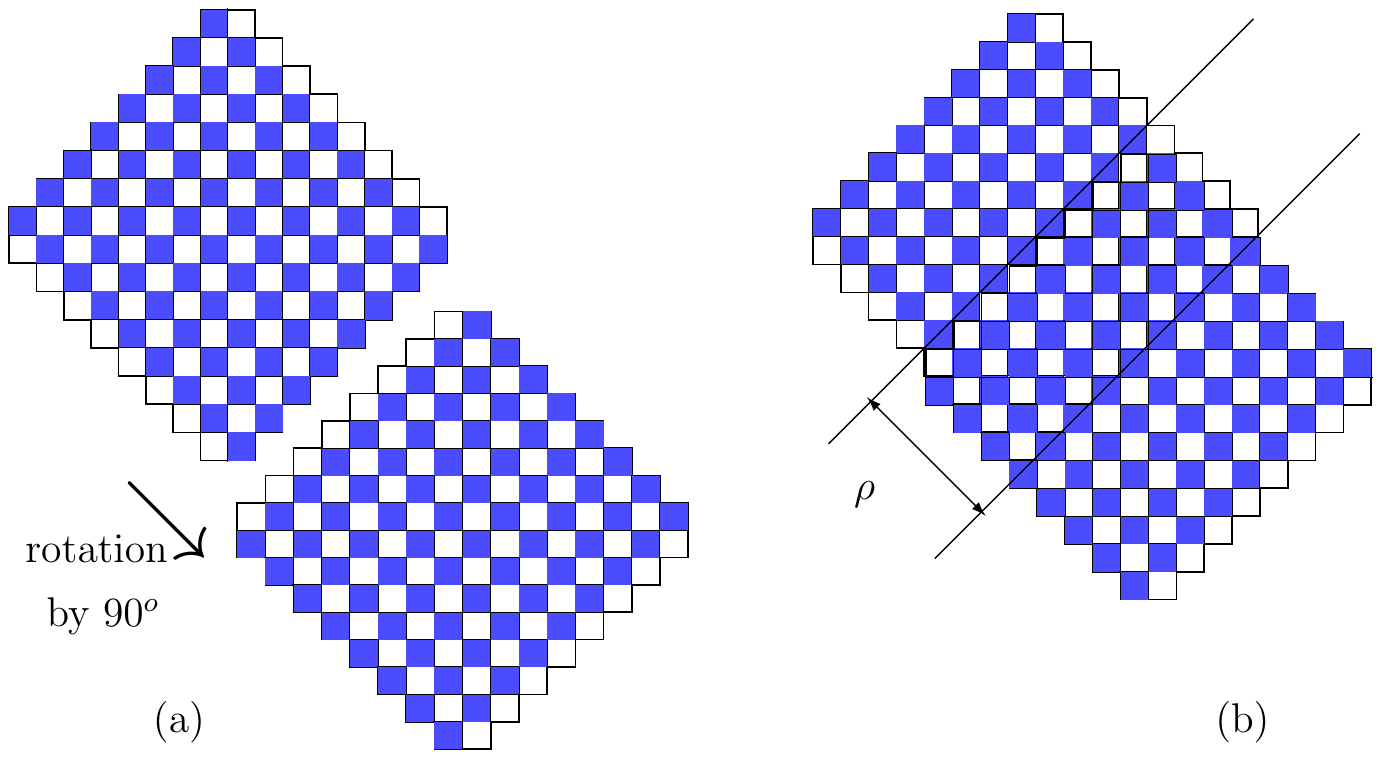}}}
  \end{picture}
  
  \vspace*{-3.7cm}
  {\small Figure 14.  An Aztec rectangle formed by sliding two Aztec diamonds with opposite ``orientations" over each other; here $n=m=8,~M=5$ and the strip $\{\rho\}$ has width $\rho=r=n-M+1=\sg=4$ and $\Dt=0$.}

 \begin{picture}(0,0)

  \put(160,-190)
 {\makebox(0,0){\includegraphics[width=192mm,height=264mm]{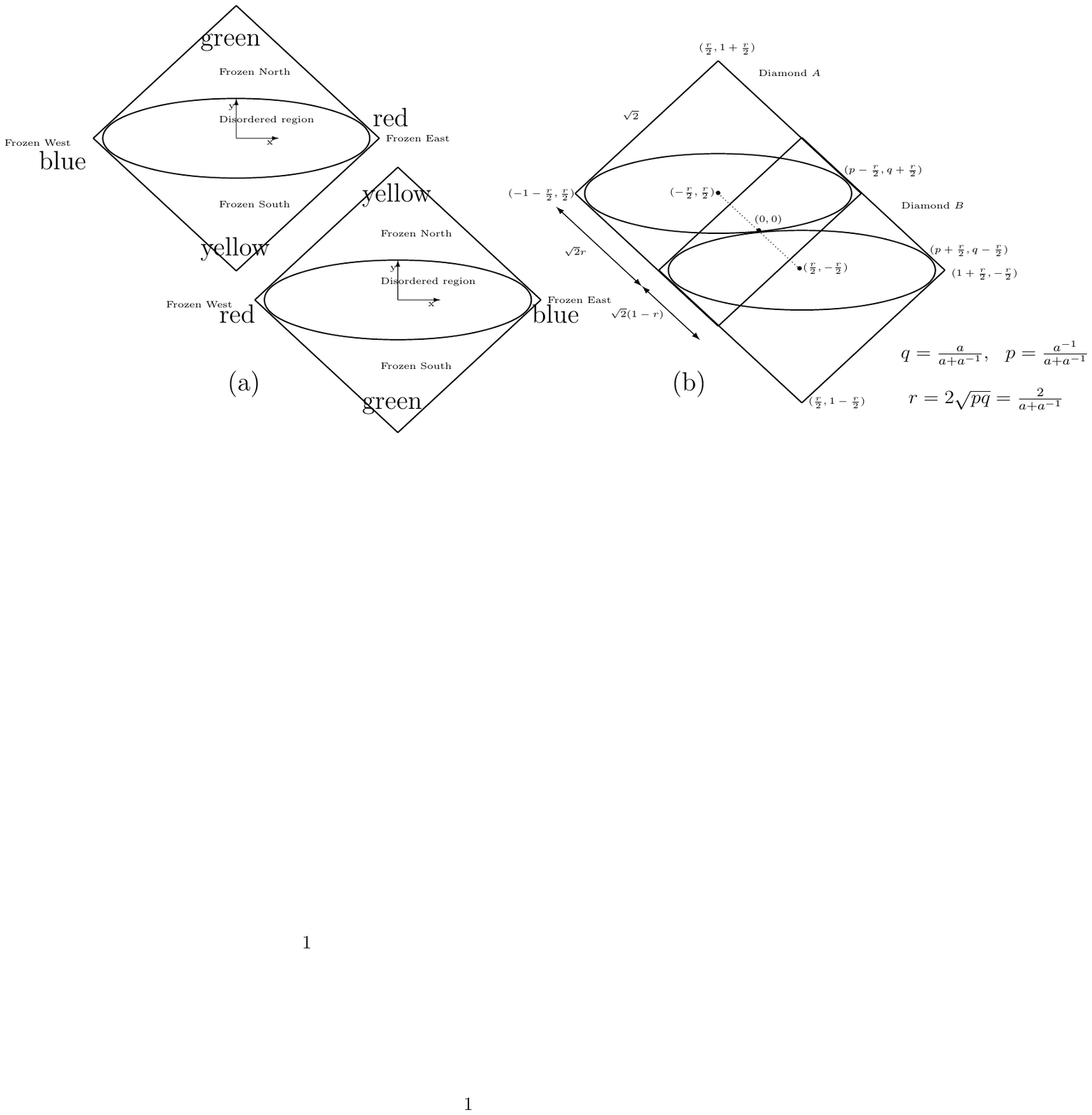}}}
\end{picture}  
 
 \vspace*{7.5cm}
 
 {\small Figure 15. Given $0<a<1$, figure (a) represents two Aztec diamonds for very large $n$, each with an inscribed arctic curve (ellipse as given by equation (\ref{ellips})), separating the solid and liquid regions. The solid parts in each of the diamonds are covered by tiles of the same colours as the tiles in Fig. 16(c) and  the ones in the simulation of Fig. 17. In Figure (b), we slide the two Aztec diamonds over each other in such a way that the two arctic ellipses merely touch, producing 7 different frozen regions. The smaller the parameter $a$, the larger is the amount of overlap.} 
  \newpage


\newpage

\begin{picture}(0,0)
   
  \put(145,0) 
 {\makebox(0,-200){\includegraphics[width=170mm,height=220mm]{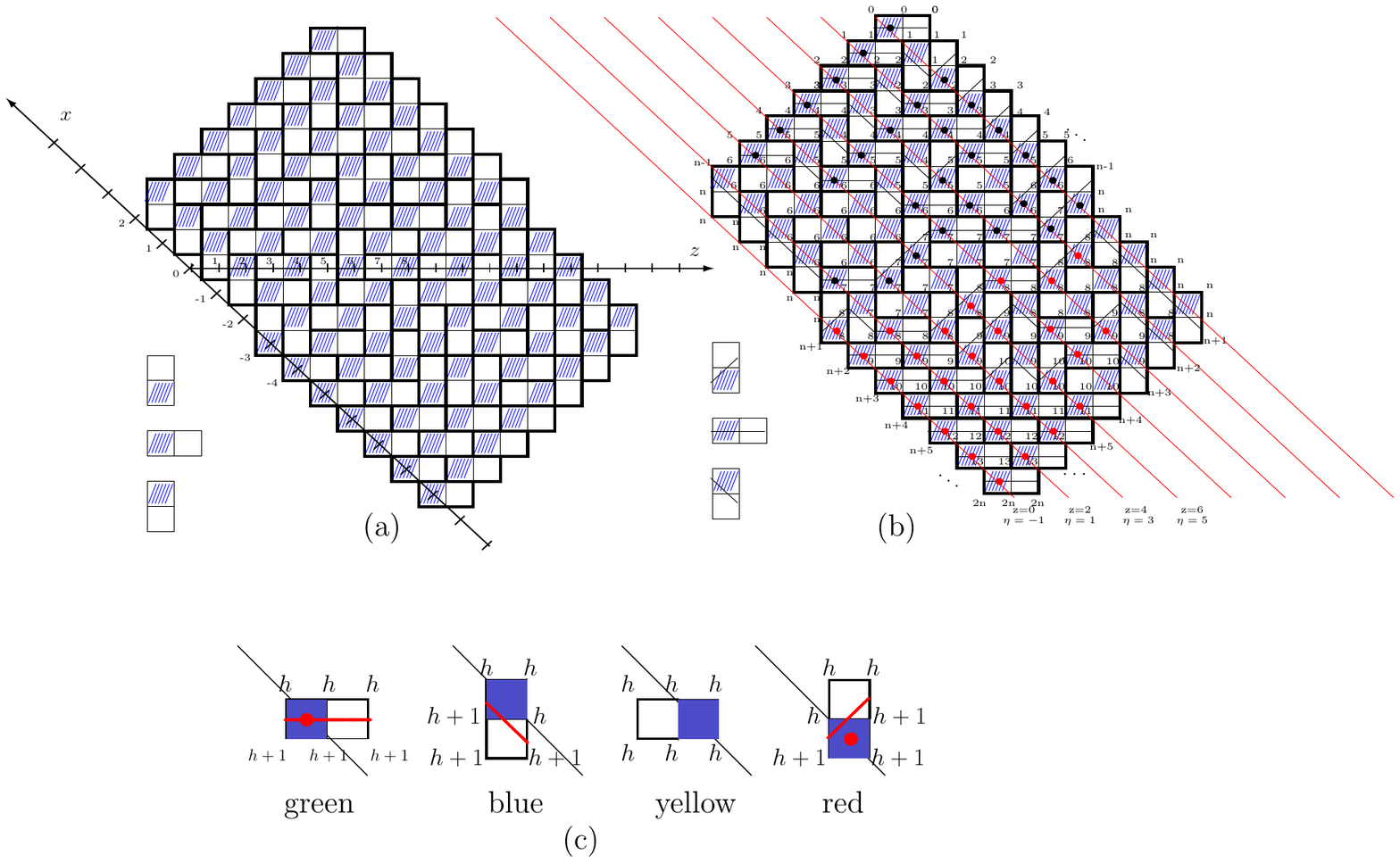}} }
 \end{picture}

 \vspace{6cm}

{\small Figure 16.  A domino tiling of an Aztec rectangle with $m=n$ and $\Dt=0$, with axes $(z,x)$ such that the coordinates of the middle of the blue squares are $(z,x)\in  \BZ_{\tiny \mbox{even}}\times \BZ$; see (a). The tiling of this rectangle with dominos (c), equipped with a height function and a level line, as in (c), defines a height function on the whole of the rectangle with prescribed height along the boundary, given by the numbers appearing in fig. (b). The colours of the tiles correspond to a simulation in Fig. 17. This defines a point process of black and red dots  $x=k$ on parallel lines $z=2\ell$, with $(z,x)\in  \BZ_{\tiny \mbox{even}}\times \BZ$ by recording the intersection of the lines $z=2k$ with the level lines for the height function, as in (b).} 

\vspace{1cm}


  We now cover this configuration with tiles (as in Fig. 16(c)), equipped with a {\em height function}, {\em a level curve} and a {\em red dot} each time the oblique lines $z=2k$ through the blue squares intersect the level curve.  So, for this model (as in Fig. 16(b)), we consider the {\em point process} of intersection points of the oblique lines $z=2k$ for $0\leq k\leq n$ (through the blue squares) with the level lines for the height function, as drawn in Fig. 16(b). Notice this process is very different from the point process considered in section 5, where the set of lines are perpendicular to the lines in this configuration. We distinguish two groups of level lines, the ones departing from the left-upper side of the rectangle and the ones departing from the right-lower side, which explains the presence of red and black dots in Fig. 16(b).

The dot-
 particles $x\in \BZ$ on the successive lines $\{z=2k \}$ for $1\leq k\leq n$, as in Fig. 16(b) form a determinantal point process, under the tiling probability (\ref{Prob}), with correlation kernel $ {\mathbb K}^{\mbox{\tiny \rm twoAzt}}_{n,\left[ M/2 \right]}$ 
 given by a perturbation of the one-Aztec diamond  kernel ${\mathbb K}_{n+1}^{\mbox{\tiny \rm OneAzt}}$,
  \be\begin{aligned}
  {{\mathbb K}_{n }^{\mbox{\tiny OneAzt}}
 (z_1,x_1;z_2,x_2)}=&\frac {(-1)^{x_1\!-\!x_2}}{(2\pi \I)^2}  \oint_{\Gamma_{0}}du 
  \oint_{\Gamma_{0,u,a}}\frac{dv}{v\!-\!u} 
 \frac{v^{-x_1}}{u^{1-x_2}} \frac{(1+au)^{n -\tfrac{z_2}{2}}(1-\tfrac au)^{\tfrac{z_2}{2}}} 
 {(1+av)^{n -\tfrac{z_1}{2}}(1-\tfrac av)^{\tfrac{z_1}{2}}}
 \\&  -\Id_{z_2>z_1}\oint _{\Gamma_{0,a}} \frac{du}{2\pi \I u} u^{x_1-x_2}
\left(\frac{1+au}{1-\frac au}\right)^{\tfrac{z_2-z_1}2}
 , \end{aligned} \label{OneAztec}\ee
 Up to a conjugation, this kernel is the same as the kernel (\ref{2.7}), but with the change of variables $(w,z)\to (-v,-u)$. We now have the following statement, which has appeared in (\cite{AJvM0}, Theorem 1.1); namely, 
   \begin{theorem} For $(z_i,x_i)\in  \BZ_{\tiny \mbox{even}}\times \BZ$, the point process of dots (given by $x\in \BZ $) along the lines $z=2k$ for $1\leq k\leq n$, described above, is determinantal, with correlation kernel given by a perturbation of the one-Aztec diamond  kernel ${\mathbb K}_{n+1}^{\mbox{\tiny \rm OneAzt}}$ above (\ref{OneAztec}):
%
 %
 \be
  \begin{aligned}
\lefteqn{ \hspace*{-.3cm}{(-1)^{x_1-x_2}   
 {\mathbb K}^{\mbox{\tiny \rm twoAzt}}_{n,\left[\!\tfrac M2\!\right]}  
       (z_1,x_1;z_2,x_2) }}\\
&= {\mathbb K}_{n+1}^{\mbox{\tiny \rm OneAzt}}
  \bigl(2(n+1)-z_1 ,\left[\!\tfrac M2\!\right]-x_1+1;2(n+1)-z_2 ,\left[\!\tfrac M2\!\right]-x_2+1\bigr) %
 \\  
 &~~~~+\left\la(\Id-K )_{\geq M}^{-1}a_{-x_2,\tfrac{z_2}{2}}(k),b_{-x_1,\tfrac{z_1}{2}}(k) \right\ra _{\geq M }    .\end{aligned}  %
 \label{TwoAztec}\ee
 with an inner-product $\la~,~\ra$ of two functions,
 involving the resolvent of a kernel $K
 $ and functions $a_{-x,z/2}(k)$ and $b_{-x,z/2}(k)$, all defined in (\cite{AJvM0}, formula (14)) \footnote{The subscript ${}_{\geq M}$ refers to taking the resolvent and the inner-product $\la~,~\ra$ over the space of integers $[M,\infty)$.}.

 \end{theorem}
%

We now define fixed quantities $v_0,~A,~\lb ,~\theta$, all in terms of the fixed parameter $0<a<1$ figuring in the probability (\ref{Prob}):
\be\begin{aligned}
v_0&:= -\frac{1\!-\!a}{1\!+\!a} <0,~~A^3:=\frac{a(1+a)^5}{(1\!-\!a)(1\!+\!a^2)}, 
~\lb :=-Av_0 >0,~~ \theta :=\sqrt{\lb(a+a^{-1})} >0.
\end{aligned}\label{4}\ee
In terms of the parameters (\ref{4}) and some extra ``{\em pressure parameter}" $\sg$, we finally introduce the scaling of the geometric parameters $n,M$ and the scaling of the coordinates ``levels and positions" $ (z_i,x_i)\in  \BZ_{\tiny \mbox{even}}\times \BZ \to (\tau_i,\xi_i)\in \BR^2$:
\be\label{scalingTac}
\begin{aligned}
n&=m=2t , ~~~\hspace*{3cm}~~M 
 =\frac{4t}{a+a^{-1}}+  2\sigma  \lb
  t^{1/3}+1,
\\
z_i &  =2t+2(1+a^2)\theta\tau_i t^{2/3},~~~~~~~~~~~x_i =2a^2\theta \tau_i t^{2/3}+\xi_i \lb t^{1/3}  
\end{aligned}
 \ee
In other words, the geometric parameters $ M,~\rho$ tend to $\iy$, together with $n=2t\to \iy$. Moreover,  the $(z_i,x_i)$-scaling (\ref{scalingTac}) is Airy process-like: namely {\em $(2/3)$-in the tangential} and {\em $(1/3)$-in the transversal} direction with regard to the arctic ellipse at the middle point $(z,x)=(n,0)$; i.e., ($q$ as in (\ref{ellips}))  
$$M\sim \frac{2n}{a+a^{-1}}
\mbox{ and } \rho=r\sim n\left(1-\frac{2}{a+a^{-1}}\right)
$$
$$z_i-n\sim \tau_i t^{2/3}\mbox{ and } 
x_i-q(z_i-n)\sim \xi_it^{1/3},\mbox{ for } n\to \iy.$$
We now define the {\em soft tacnode kernel} 
 (or continuous tacnode kernel) for $(\tau_i,\xi_i)\in \BR^2$, as follows, with $q_\sg(\tau,\xi):=e^{\tau(\sg-\xi)+\frac{2\tau^3}{3}}$:
\be\begin{aligned}\label{Kctac}
  {\mathbb K}^{\rm cTac} (\tau_1, \xi_1;\tau_2,\xi_2 )  
&=\frac{q_\sigma (\tau_1,\xi_1)}{q_\sigma (\tau_2,\xi_2)}{\mathbb K}^{\mbox{\tiny \rm AiryProcess}}(\tau_2,\sigma-\xi_2+\tau_2^2;\tau_1,\sigma-\xi_1+\tau_1^2) 
\\
&~~~+
 2^{1/3} \int_{2^{2/3}\sg}^{\iy}\left((I- K_{\mbox{Ai}}
 )^{-1}_{_{\geq 2^{2/3}\sg} }
 \AR_{\xi_1-\sg}^{ \tau_1}\right)(\lb)
\AR_{\xi_2-\sg}^{ -\tau_2 }(\lb)d\lb
,\el\ee
which is a "{\em perturbation}" of the usual Airy process kernel, as defined by Johansson (\cite{Jo05c} and \cite{Jo16}):
\be\bl
 {\mathbb K} ^{\mbox{\tiny AiryProcess}}(&\tau_1,\xi_1; \tau_2,\xi_2)  =\int_0^{\infty}e^{\lambda (\tau_2-\tau_1)}
 \Ai (\xi_1+\lambda)  \Ai (\xi_2+\lambda)d\lambda\\
&-
\tfrac{\Id_{\tau_2>\tau_1}}{\sqrt{4\pi (\tau_2\!-\!\tau_1)}}\mbox{exp} \left(-\tfrac{(\xi_1-\xi_2)^2}{4(\tau_2-\tau_1)}+\tau_1(\xi_1\!+\!\sigma)-\tau_2(\xi_2\!+\!\sigma)+\tfrac 23 (\tau_1^3-\tau_2^3)\right)
.\end{aligned}\label{AiryP}
\ee
The perturbation term in ${\mathbb K}^{\tiny\mbox{cTac}}$ (\ref{Kctac}) is an inner-product (integral) of two continuous functions, one of them being acted upon by the resolvent of the Airy kernel, much as the discrete perturbation in ${\mathbb K}^{\mbox{\tiny \rm twoAzt}}_{n,\left[  M/2 \right]} $ (see (\ref{TwoAztec})) before taking the double scaling limit.

\newpage

\newpage

\setlength{\unitlength}{0.015in}\begin{picture}(0,170)

 \put(160,-10) {\makebox(0,0){\includegraphics[width=240mm,height=280mm]{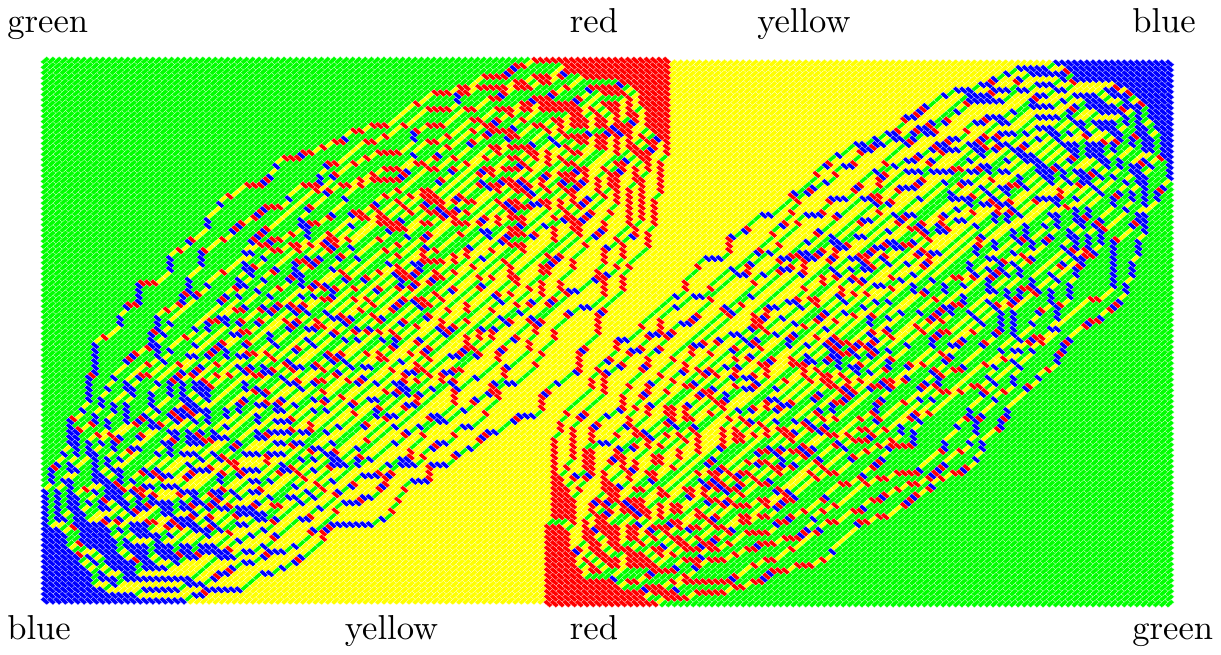}}}
 
 \end{picture}
 
 \vspace*{-1.8cm}
   
 {\small Figure 17.  Simulation of the ${\mathbb K}^{\mbox{\tiny \rm twoAzt}}_{n,[M/2]}$-process for large $n$, thus resembling  a continuous tacnode process $ {\mathbb K}^{\rm cTac}$ in the neighborhood of the tacnode, with the frozen regions being the same on either side of it. This figure rotated  clockwise by $45^o$ corresponds to Fig. 16, with the colours above corresponding to the colours of the tiles in Fig. 16(c). Here  $n=100,~M=81,~\mbox{overlap}=\rho=r=n(1-\frac 2{a+a^{-1}})=20,~a=1/2$.  (courtesy of Eric Nordenstam)}
 
 \vspace*{1cm}

The kernels (\ref{Kctac}) and (\ref{AiryP}) above contain the Airy function, the Airy kernel $K_{ \Ai }$ and extensions $\Ai^{(s)} (x)$ and ${\cal A}_\xi^\tau$, defined by
\be\bl
 \Ai^{(s)}(x) &:=\frac{1}{2\pi i}\int _{\nearrow \atop \nwarrow }  dz~e^{ z^3/3+z^2s- zx }=e^{ sx+\tfrac 2 3s^3 }\mbox{Ai}(x+s^2) ,~~\Ai (x)=\Ai^{(0)}(x)
 \\
 K_{ \Ai }^{ }(x,y) &:=\!\int_0^{\infty}\!\!
\Ai^{ }(x\!+\!u)  \Ai^{ }(y\!+\!u)du,~
\\
{\cal A}^{\tau}_\xi(\kappa) 
 &:= \Ai^{(\tau)} ( \xi+2^{1/3} \kappa)- \int_0^\infty \Ai^{(\tau)} (-\xi+2^{1/3} \beta )
\Ai (\kappa+ \beta )d\beta  
.\el\ee

 \noindent We now have the following statement (\cite{AJvM0}, Theorem 1.2):

 \begin{theorem}  Given the scaling (\ref{scalingTac}), the scaling limit of the kernel (\ref{TwoAztec}) reads as follows, for  $(\tau_i,\xi_i)\in \BR^2$,
\be\label{tacKernel}\bl
\lim_{t\to \infty} 
		 (\tfrac{1-a}{1+a})^{x_2-x_1+\tfrac{z_1-z_2}2}(-1)^{x_2-x_1}& {\mathbb K}^{\mbox{\tiny \rm twoAzt}}_{n,[M/2]}( z_1,x_1 ; z_2,x_2)dx_2 \Bigr|_{\mbox{\tiny scaling (\ref{scalingTac})}} \\& ~~~~~~~~~~~~~~~~~~={\mathbb K}^{\rm cTac} (\tau_1, \xi_1;\tau_2,\xi_2 )d\xi_2
		.\el\ee
		
\end{theorem}

A simulation for $n=100$ and $M=81$ is given in Fig. 17, showing the presence of only one (yellow) phase near the tacnode. There has been a extensive literature on the subject. This continuous tacnode kernel has been obtained for other models, like two groups of nonintersecting random walks \cite{AFvM} or Brownian motions (Johansson \cite{Jo13}), starting from two different points and forced to end up at two different points. Tuning appropriately the starting and end points turns the model into two groups of random processes merely touching, producing a tacnode; the fluctuations near the tacnode are governed by the comtinuous tacnode process. Riemann-Hilbert aspects and interesting variations on the tacnode theme have been discussed by  (\cite{DelvKuiZh},  \cite{DelvVeto},\cite{Delv},\cite{FerrariVetBr}, \cite{FerrariVet}) and others.

 %
  
 

 
 


\begin{thebibliography}{10}

  \bibitem{ACvM} Mark Adler, Mattia Cafasso and Pierre van Moerbeke : {\em From the Pearcey to the Airy process}. Electron. J. Probab. 16 (2011), no. 36, 1048-1064.  
 
\bibitem{ACJvM} Mark Adler,  Sunil Chhita, Kurt Johansson and
Pierre van Moerbeke: {\em Tacnode GUE-minor processes and double Aztec diamonds}, Probab. Theory Related Fields {\bf 162}, no. 1-2, 275-325 (2015)  

\bibitem{AFvM} Mark Adler, Patrik Ferrari and P. van Moerbeke : {\em Nonintersecting random walks in the neighborhood of a symmetric tacnode}.  Ann. Probab. {\bf 41} (2013), no. 4, 2599-2647.

\bibitem{AJvM0}  Mark Adler, Kurt Johansson and Pierre van Moerbeke:  {\em Double Aztec diamonds and the tacnode process}. Adv. Math. {\bf 252} (2014), 518-571.


 
  \bibitem{AJvM1}  Mark Adler, Kurt Johansson and Pierre van Moerbeke:  {\em Tilings of non-convex Polygons, skew-Young Tableaux and determinantal Processes}. Comm. Math. Phys. {\bf 364}, 287-342 (2018) (arXiv:1609.06995)

\bibitem{AJvM2}  Mark Adler, Kurt Johansson and Pierre van Moerbeke:  {\em  Lozenge tilings of hexagons with cuts and asymptotic fluctuations: a new universality class},  Math. Phys. and Geom. {\bf 21}: 9 (53pp)  (2018)
(arXiv:1706.01055) 

\bibitem{AJvMskewAzt}Mark Adler, Kurt Johansson \& P. van Moerbeke: {\em A singular Toeplitz determinant and the discrete tacnode kernel for skew-Aztec rectangles}, The Annals of Applied probability. {\bf 32}, 1234-1294 (2022). (arXiv:1912.02511)



 \bibitem{AvM1}  Mark Adler and Pierre van Moerbeke, {\em Coupled GUE-minor Processes}, Intern. Math. Research Notices, 21, 10987-11044 (2015) . (arXiv:1312.3859)

\bibitem{AvM2}  Mark Adler and Pierre van Moerbeke: {\em Probability distributions related to tilings of non-convex polygons}, Journal of Math. Phys 59, 091418 (2018) (Special volume in memory of  Ludvig Faddeev) (arXiv:1810.04692)


\bibitem{Ag} Amol Aggarwal: {\em Universality for lozenge tiling local statistics}(arXiv:1907.09991)

\bibitem{Ag-Go} Amol Aggarwal, Vadim Gorin: {\em Gaussian Unitary Ensemble in random lozenge tilings}. Probab. Theory Related Fields 184 (2022), no. 3-4, 1139-1166. 82 (60)(arXiv:2106.07589 )


 
 \bibitem{AH} Amol Aggarwal and  Jiaoyang  Huang: {\em Edge statistics for Lozenge tilings of polygons, II: Airy line ensemble}  (arXiv:2108.12874)
 


  



 

\bibitem{AstDuse} Kari Astala, Erik Duse, Istvan Prause and Xiao Zhong: {\em Dimer Models and Conformal Structures}, (arXiv:2004.02599)


 
\bibitem{Bar} Yu. Baryshnikov. {\em GUEs and queues}. Probab. Theory Related Fields, 119(2):256-274, 2001.


\bibitem{BLR} Natana\"el Berestycki, Beno\^ it Laslier and Gourab Ray: {\em Dimers and Imaginary Geometry}. Ann. Probab. {\bf 48} (2020), no. 1, 1-52 (arXiv:1603.09740)


 
 \bibitem{BBNV} Dan Betea, J. Bouttier, P. Nejjar and M. Vuletic: {\em The free boundary Schur process and applications}. Ann. Henri Poincar\'e 19 (2018), no. 12, 3663-3742. (arXiv:1704.05809)




\bibitem{BD}
Alexei Borodin and Maurice Duits: {\em Limits of determinantal processes near a tacnode}, Ann. Inst. Henri Poincare (B), {\bf 47} , 243-258 (2011).

\bibitem{BF} Alexei Borodin and Patrik L. Ferrari : {\em Anisotropic growth of random surfaces in 2+1 dimensions}, Comm. Math. Phys, {\bf 325}, 603-684 (2014).

 \bibitem{BFP}
Alexei Borodin and Patrik L. Ferrari, Michael Pr\"ahofer, Tomohiro Sasamoto: \emph {Fluctuation properties of the TASEP with periodic initial configuration} J. Stat. Phys. {\bf 129} (2007) (arXiv:math-ph/0608056)
 
\bibitem{BGR}
 A. Borodin, V. Gorin and E. M. Rains, {\em $q$-Distributions on boxed plane partitions}, Selecta Math. 16 (2010),
731-789.

 \bibitem{BR} Alexei Borodin, Eric M. Rains: {\em Eynard-Mehta theorem, Schur process, and their Pfaffian analogs} J. Stat. Phys. 121 (2005), no. 3-4, 291-317.   (arXiv:math-ph/0409059)
 
  \bibitem{B } Alexei Borodin: {\em Determinantal point processes}, The Oxford handbook of random matrix theory, 231-249, Oxford Univ. Press, Oxford, 2011.




 







 

 

\bibitem{Buf} Alexey Bufetov and Alisa Knizel: {\em Asymptotics of random domino tilings of rectangular Aztec diamonds}, Ann. Inst. Henri Poincar\'e Probab. Stat. {\bf 54} no. 3, 1250-1290 (2018) (arXiv:1604.01491v2)
 
  

\bibitem{BuGo} Alexey Bufetov and Vadim Gorin: {\em Fourier transform on high-dimensional unitary groups with applications to random tilings}. Duke Math. J. {\bf 168}, 2559-2649 (2019)

\bibitem{Ciucu} Mihai Ciucu and Ilse Fischer {\em Lozenge tilings of hexagons with arbitrary dents}. Adv. in Appl. Math. 73, 1-22.  (2016)

\bibitem{CKP} Henry Cohn, Richard Kenyon and James Propp: {\em A variational principle for domino tilings}. J. Amer. Math. Soc. {\bf 14}, 297-346 (2001)



\bibitem{Cohn} H. Cohn, M. Larsen and J. Propp. {\em The shape of a typical boxed plane partition}, The New York Journal of Mathematics. {\bf 4} 137-165 (1998) 

 \bibitem{Defos} M. Defosseux: {\em Orbit measures, random matrix theory and interlaced determinantal processes}, Ann. Inst. H. Poincar Probab. Statist. {\bf 46},  209-249. (2010)
 
 \bibitem{Delv} Steven Delvaux: {\em The tacnode kernel: equality of Riemann-Hilbert and Airy resolvent formulas} (arxiv:1211.4845)

\bibitem{DelvKuiZh}S. Delvaux, A. Kuijlaars, L. Zhang: {\em Critical behavior of non-intersecting Brownian motions at a tacnode}, Comm. Pure Appl. Math. 64, 1305-1383 (2011)


\bibitem{DelvVeto} S. Delvaux, B. Vet\"o : {\em The hard edge tacnode process and the hard edge Pearcey process with non-intersecting squared Bessel paths}, Random Matrix Theory Appl. 4:155008 (2015)

 \bibitem{DFF}Maurice Duits: {\em The Gaussian free field in an interlacing particle system with two jump
rates}. 
Comm. Pure Appl. Math. 66 (2013), no. 4, 600?643 (arXiv:1105.4656)  
 
 \bibitem{D} Maurice Duits: {\em On global fluctuations for non-colliding processes},  Ann. Probab. 46 (2018), no. 3, 1279?1350 (arxiv: 1510.08248).
 
 \bibitem{DJM} Erik Duse, Kurt Johansson, Anthony Metcalfe {\em The Cusp-Airy Process}.  Electron. J. Probab. 21 (2016) (arXiv:1510.02057)

 \bibitem{Duse2} Erik Duse and Anthony Metcalfe:  
{\em Asymptotic geometry of discrete interlaced patterns: Part I.}
Internat. J. Math. {\bf 26} (2015),  1550093. 

\bibitem{Duse1} Erik Duse and Anthony Metcalfe:  
{\em Asymptotic geometry of discrete interlaced patterns: Part II.} Ann. Inst. Fourier (Grenoble) 70 (2020), no. 1, 375?436. (arXiv:1507.00467)

  \bibitem{Elkies}   N. Elkies, G. Kuperberg, M. Larsen, J. Propp, Alternating-sign matrices and domino
tilings. I. J. Algebraic Combin. 1 (1992), no. 2, 111-132

\bibitem{FerrariVetBr} Patrik L. Ferrari, B\'alint Vet\H o : {\em The hard-edge tacnode process for a Brownian motion}. Electron. J. Probab. 22, 1-32 (2017)

\bibitem{FerrariVet} Patrik L. Ferrari, B\'alint Vet\H o : {\em Fluctuations of the Arctic curve in the tilings of the Aztec diamond on restricted domains} 
Ann. Appl. Probab. 31 (2021), no. 1, 284-320 (arXiv:1909.10840)

\bibitem{Temp}M. Fisher, H. Temperley, {\em The dimer problem in statistical mechanics ? an exact result.}
Phil Mag. 6(1961), 1061-1063.

\bibitem{JPS} William Jokush, James Propp, Peter Shor, {\em Random domino tilings and the arctic circle theorem.}  (ArXiv: math.CO/9801068)

\bibitem{Jo01b}
Kurt~Johansson, \emph{Discrete orthogonal polynomial ensembles and the Plancherel measure}, Ann. of Math. {\bf 153}, 259-296. (2001) 
 

 
\bibitem{Jo02b}
Kurt~Johansson, \emph{Non-intersecting paths, random tilings and random
  matrices}, Probab. Theory Related Fields \textbf{123} (2002), 225--280.

\bibitem{Jo03b}
Kurt~Johansson, \emph{Discrete polynuclear growth and determinantal processes},
  Comm. Math. Phys. \textbf{242} (2003), 277--329.

\bibitem{Jo13} Kurt~Johansson, {\em Non-colliding Brownian motions and the extended tacnode process}.
Comm. Math. Phys. {\bf 319}, no. 1, 231-267 (2013)

 \bibitem{Jo05b}Kurt~Johansson,  \emph{Non-intersecting, simple, symmetric random walks and the extended Hahn kernel}, Ann. Inst. Fourier (Grenoble) {\bf 55}, 2129-2145. (2005) 
 
 \bibitem{Jo05c}Kurt~Johansson,  \emph{The arctic circle boundary and the Airy process}, Ann. Probab. {\bf 33}, 1-30. (2005)
 
 \bibitem{Jo16}Kurt~Johansson:  \emph{Edge Fluctuations of Limit Shapes}, Current developments in mathematics (Harvard Lectures, 2016), 47-110, Int. Press, Somerville, MA, 2018

 \bibitem{JN} Kurt Johansson and Eric Nordenstam: {\em Eigenvalues of GUE minors}, Electron. J. Probab. {\bf 11} , 1342-1371 (2006).
 
 
 

 
  \bibitem{Gorin} Vadim E. Gorin: \emph{Nonintersecting paths and the Hahn orthogonal polynomial ensemble},  Funct. Anal. Appl. {\bf 42}, 180-197 (2008).
  
   \bibitem{Gorin1} Vadim E. Gorin: \emph{Bulk universality for random lozenge tilings near straight boundaries and for tensor products},  Comm. Math. Phys. 354 (2017), no. 1, 317?344. (arXiv:1603.02707)
   
    \bibitem{Gorin2} Vadim E. Gorin: \emph{Lectures on Random Tilings} Cambridge Studies in Adv. Math., Cambridge University Press, August 2021. 
    
    
\bibitem{GorinPetrov} Vadim E. Gorin and L. Petrov: \emph{Universality of local statistics for noncolliding random walks}.  Ann. Probab. 47 (2019), no. 5, 2686?2753 (arXiv: 1608.3243)

\bibitem{Huang}Jiaoyang Huang: {\em Height Fluctuations of Random Lozenge Tilings Through Nonintersecting Random Walks} (arXiv:2011.01751)

  \bibitem{Kac} M. Kac and J. C. Ward:  {\em A combinatorial solution of the two-dimensional Ising
model}, Phys. Rev. {\bf 88}, 1332 - 1337 (1952)


 
 

\bibitem{Kast1}  P. W. Kasteleyn, {\em The statistics of dimers on a lattice. I. The number of dimer arrange-
ments on a quadratic lattice}, Physica 27 (1961), 1209?1225.

 \bibitem{Kast} Pieter W. Kasteleyn: {\em Graph theory and crystal physics}.  Graph Theory and Theoretical Physics pp. 43-110 Academic Press, London (1967).
 

\bibitem{Laslier} B. Laslier: {\em  Central limit theorem for lozenge tilings with curved limit shape}. (arXiv:2102.05544)


 \bibitem{Onsager} B. Kaufman and L. Onsager: {\em Crystal statistics. III. Short-range order in a binary
Ising lattice}, Phys. Rev {\bf 76}, 1244 -1252 (1949)

 \bibitem{K} Richard Kenyon: {\em Lectures on dimers} Statistical mechanics, 191-230, IAS/Park City Math. Ser., 16, Amer. Math. Soc., Providence, RI, 2009. arXiv: 0910.3129
 
  \bibitem{KO} Richard Kenyon and Andrei Okounkov: {\em Limit shapes and the complex Burgers equation}, Acta Math. {\bf 199}, no. 2, 263-302 (2007)
  
  \bibitem{Ke} Richard Kenyon: {\em Height fluctuations in the honeycomb dimer model}. Comm. Math. Phys. {\bf 281}, 675-709 (2008)
  
      
     \bibitem{KOS} Richard Kenyon, Andrei Okounkov and Scott Sheffield : {\em Dimers and Amoebae}, Annals of
Math. {\bf 163} , no.3, 1019-1056 (2006)
  
 \bibitem{Krat} Christian Krattenthaler: {\em Advanced determinantal calculus}, S\'eminaire Lotharingien de Combinatoire, European Math Society {\bf 42} (1999) (The Andrews Festschrift), paper B42q, 67 pp
 
\bibitem{Krat2} Christian Krattenthaler, {\em Descending plane partitions and rhombus tilings of a hexagon with a triangular hole},  European J. Combin. {\bf 27}  no. 7, 1138-1146 (2006)

\bibitem{McD} I. Macdonald. {\em Symmetric functions and Hall polynomials}. Oxford Mathematical Monographs. Clarendon Press (1995).
 
\bibitem{McMa} P. A. MacMahon, {\em Memoir on the theory of the partition of numbers, Part V. Partitions in two-dimensional
space}, Phil. Trans. R. S., 1911, A.

\bibitem{McMa1} P.A. MacMahon, Combinatory Analysis, Vol. 2, Cambridge University Press, 1916; reprinted by Chelsea, New York, 1960.
 


 \bibitem{Metc} Anthony Metcalfe: {\em Universality properties of Gelfand-Tsetlin patterns}, Probab. Theory Related Fields {\bf 155}(1-2) 303-346 (2013).
 
 
 \bibitem{Nov} Jonathan Novak, {\em Lozenge tilings and Hurwitz numbers}, Journal of Stat. Phys., 161 , 509-517 (2015) (arXiv:math/0309074)
 
 \bibitem{OR1} Andrei Okounkov and Nicolai Reshetikhin: {\em Correlation function of Schur process with application to local geometry of a random 3-dimensional Young diagram} J. of the American Math. Society {\bf 16}, 581-603 (2003)
 
 \bibitem{OR2} Andrei Okounkov and Nicolai Reshetikhin: {\em The birth of a random matrix}. Mosc. Math. J. {\bf 6} , 553-566, 588.(2006)
 


 
  \bibitem{Petrov1} Leonid Petrov: {\em Asymptotics of uniformly random lozenge tilings of polygons. Gaussian free field}, Ann. Probab. {\bf 43} 1-43 (2015).
 
  \bibitem{Petrov2} Leonid Petrov: {\em Asymptotics of random lozenge tilings via Gelfand-Tsetlin schemes}. Probab. Theory Related Fields {\bf 160}, 429-487 (2014) 
  
  \bibitem{PrSpohn} Michael Pr\"ahofer and Herbert Spohn: {\em Scale invariance of the PNG droplet and the Airy process} J. Stat. 108, 1071-1106 (2002)
  
 \bibitem{Romik}Dan Romik: {\em The Surprising Mathematics of Longest Increasing Subsequences}, Institute of Mathematical Statistics Textbooks, 4. Cambridge University Press, New York, 2015. xi+353 pp.
 
 \bibitem{Stanley}Richard Stanley: {\em Enumerative Combinatorics}. Cambridge Studies in Advanced Mathematics 62. Vols I and II. (2001)

  
\end{thebibliography}
\end{document}